\documentclass{jfm}
\usepackage[utf8]{inputenc}
\usepackage{caption}
\usepackage{subcaption}
\usepackage{epsfig}
\usepackage{graphicx}
\usepackage{amsmath}
\usepackage[version=4]{mhchem}
\usepackage{siunitx}
\usepackage{longtable,tabularx}
\usepackage{graphicx}
\usepackage{placeins}
\usepackage{makecell}
\usepackage{multicol}
\usepackage{xcolor}

\shorttitle{Response of the Separated Flow over an Airfoil to Actuator Bursts}
\shortauthor{X. An, D. R. Williams}

\title{Response of the Separated Flow over an Airfoil to Short-Time Actuator Bursts}

\author{Xuanhong An\aff{1}
  \corresp{\email{xan2@hawk.iit.edu}},
  \and David R. Williams\aff{1}}

\affiliation{\aff{1}Department Mechanical, Materials and Aerospace Engineering,
Illinois Institute of Technology, IL, 60616, USA}

\begin{document}

\maketitle

\begin{abstract}
The separated flow response to single and multiple burst mode actuation over a 2-D airfoil at $12^o$ angle of attack was studied experimentally. For the single-burst actuation case, surface pressure signals were correlated with the flowfield observations of the roll-up and convection of a large-scale vortex structure that follows the actuator burst input. A spatially localized region of high pressure occurs below and slightly upstream of a “kink” that forms in the shear layer, which is responsible for the lift reversal that occurs within $2.0t^+$ after the burst signal was triggered. Proper orthogonal decomposition of the single-burst flow field shows that the time-varying coefficients of the first two modes correlate with the negative of the lift coefficient and pitching moment coefficient. The dynamic mode decomposition (DMD) of the single-burst flow field data identified the modes related to the kinetic energy growth of the disturbance. The mode with the largest growth rate had a Strouhal number close to that associated with the separation bubble dynamics.  However, when multiple bursts are used to control the separation, the interactions between the bursts were observed which depend on the time intervals between the bursts. The convolution integral and DMD were performed on the multi-burst flow field datasets. The results indicate that the nonlinear burst-burst interaction only affects the reverse flow strength within the separation bubble, which is related to the main trend of the lift. On the other hand, the linear burst-burst interaction contributes to the high-frequency lift variation associated with the bursts.   
\end{abstract}

\begin{keywords}
Authors should not enter keywords on the manuscript, as these must be chosen by the author during the online submission process and will then be added during the typesetting process (see http://journals.cambridge.org/data/\linebreak[3]relatedlink/jfm-\linebreak[3]keywords.pdf for the full list)
\end{keywords}

\section{Introduction}
The application of  active flow control (AFC) methods to reattach  separated flows has attracted a lot of attention, due to its potential to enhance the performance of machines in a wide range of applications, such as, aircraft flight efficiency, pressure recovery in diffusers, road vehicle drag reduction and control in gusting flow, noise generation from propellers, and power output from wind turbines. In general, on nominally two-dimensional airfoils AFC methods are capable of performance enhancement in one of two ways. The first is to prevent flow separation at high angles of attack by using pre-set actuation.  Delaying the onset of flow separation enables lower takeoff and landing speeds for aircraft, and enables  better controllability at high angle of attack. When the AFC actuator runs continuously, then we refer to this approach as steady separation control. The second method deals with AFC when the flow or the vehicle has a flow separation that is time-varying.  When unsteadiness is involved then the control effect often needs to be synchronized with the changing external flow or changing vehicle trajectory.  We refer to the second method as unsteady separation control.  Unsteady separation control allows the "reattachment length" to be varied with time, so that forces and moments can be applied on the airfoils without changing the pitch attitude or using moving the control surfaces. Similarly, alleviating the unsteady flow separation due to some rapid maneuvers can also be achieved.  Unsteady separation control with AFC has proven to be challenging to implement, because of the nonlinear behavior of the separated flow and the significant time delays between the onset of actuation and the flow response.    

Understanding the dynamic behavior of separated flow response to actuator input will benefit the active flow control system design in at least two ways. First, it provides more insight into the flow physics behind the actuation effectiveness, which will benefit the actuator design for both steady and unsteady separation control. Second, a better understanding of the interaction between the separated flow response and the actuator input could benefit the development of low-order models of the aerodynamic loads to the time-varying actuator input.  Modeling the load response to actuator input is essential for the real-time controller design for active flow control systems, especially for the unsteady separation control. Some important earlier investigations into the behavior of separated flow response to actuator input are discussed next. 

\citet{raju2008dynamics} investigated the steady state behavior of separated flow over an NACA 4418 airfoil and identified three distinct time scales associated with the separated flow. The time scales are associated with the Kelvin-Helmholtz shear layer instability, the separation bubble, and the wake. They reported that continuous harmonic actuation at the separation bubble frequency gives the maximum lift recovery, but we expect any type of AFC actuation is likely to excited all three instabilities.  We also expect interactions between the different modes of instability will be important.  For example, by conducting experiments on circular cylinders with different diameters, \citet{prasad1996instability} reported that the shear layer instability frequency, $f_{SL}$, is related to the wake frequency, $f_{wake}$ as $f_{SL}=0.0235Re^{0.67}f_{wake}$. For an airfoil, the wake frequency is expressed as $St=\frac{f_{wake}\cdot c\cdot sin(\alpha)}{U_{\infty}}$ \citep{strouhal1878besondere, fage1927flow}, where $c$ is the is the chord length, $\alpha$ is the angle of attack and $U_{\infty}$ is the freestream speed. The value of the separation bubble frequency $f_{sep}$ scales as $f_{sep}\sim U_{\infty}/L_{sep}$ by \citet{raju2008dynamics}, where $L_{sep}$ is characteristic length of the separation region. 



Instead of using continuous harmonic excitation with the actuator, a different approach to study the dynamics of the separated region response to the actuation is to use an impulse-like disturbance to perturb the separated flow region, and then follow the development of the disturbance. When the system is linear and time invariant, the impulse response can be used in a convolution integral to predict the response from any arbitrary input.  The impulse-like disturbances excite a broad spectrum of modes within the separated flow, which can trigger multiple instabilities \citep{monnier2016comparison}. \citet{AG} conducted experiments on a stalled wing using this approach, and identified an initial reversal in the lift increment following a short burst (single-burst) input from a synthetic jet actuator prior to the lift increasing. The lift reversal typically occurs within the first 1.5 convective times, and it turns out to be a key feature of the transient response at the onset of actuation.  The lift reversal is  a characteristic of non-minimum phase behavior of a system. From a control theory perspective, this means the system will have an inherent time delay that will limit the bandwidth of control, and there will be an upper limit to how fast the lift can be controlled. In fact, it was shown by \citet{kerstens2011closed}, that this time delay was responsible for limiting the bandwidth in the gust alleviation experiments on a three-dimensional wing.

In addition, \citet{monnier2016comparison} reported that the circulatory force on a wing in response to an impulse-like pitching motion also follows the same lift reversal behavior as the AFC  actuators (e.g. synthetic and burst-blowing jets). Therefore, the similar inherent time delay (lift reversal) observed with the pneumatic actuators also exists when the wing is mechanically moved. Under this circumstance, faster actuators alone will not make the flow respond faster. In fact, \citet{williams2018alleviating} reported that the inherent time delay (lift reversal) is due to the nature of the separated flow and independent of the actuators. Thus, to increase the speed at which the forces can be controlled, a deeper understanding of the fluid dynamics responsible for the non-minimum phase (lift reversal) behavior of the separated flow system is required.  

The lift enhancement that follows the lift reversal \citep{amitay2006flow} for the single-burst actuation should also be related to the three instabilities associated with the separated flow region. The instabilities contained in the separated flow are the mechanisms by which the high-energy external flow from the freestream becomes transported into the separation region. These instabilities are also closely related to the lift enhancement maximization for continuous actuation. In fact, \citet{raju2008dynamics} suggested that the maximum lift gain, relative to the non-actuation baseline case, will be achieved by the harmonic actuation running at the vicinity of the subharmonics of the separation bubble frequency.     

However, it is unlikely that the separated flow response to a single burst is linear.  Linear systems have the additive property $f(x+y) = f(x)+f(y)$ and the homogeneous property $f(ax) = af(x)$.  The homogeneous property of linear systems is not followed by the separated flow.  For example, \cite{an2016modeling} showed that when the amplitude of the input pulse is doubled, the response amplitude is not doubled.  The nonlinear response behaves like a static nonlinearity with a square root dependence. 

Finite-length sequences of actuator bursts fall between continuous steady state actuation and single burst actuation.  By assembling a finite sequence of single-burst actuation signals with specific spacing in time between them, we can observe the transition from a pulse response to the steady state behavior of the separated flow.  

Amplitude modulation of the continuous burst actuation is often used for flow separation control, in which unsteady aerodynamic loads occur  \citep{WK} \citep{kerstens2011closed} \citep{williams2015dynamic}. But this further complicates the dynamic response of the separated flow to the actuation comparing to the single-burst actuation \citep{an2016modeling}. Therefore, studying on the separated flow response to the continuous actuation will benefit the actuation effectiveness as well as the low-order modeling approach for the real-time flow control applications. In the current work, we employed multi-burst actuation with specific burst-burst spacing in time to study the continuous actuation. Unlike the continuous burst actuation, the multi-burst actuation can be initialized from the baseline state and ended within a limited time. Thus, besides the fully developed burst-burst interaction (the same as in the continuous actuation case), it is able to capture both the initial transition during the first several bursts and the relaxation stage after the last burst.  

In this paper, a study of the flow physics behind the single-burst, multi-burst actuation is carried out. Detailed particle image velocimetry (PIV), pressure and force measurements were conducted. Dynamic mode decomposition (DMD) was performed on the flow field data to extract their dynamic characteristics. The remaining of this paper is organized as follows, the detailed experimental setup is described in section \ref{sec:exp}. The investigation of lift reversal, POD analysis and instability analysis following the single-burst actuation is carried out in section \ref{Sec:single}. The burst-burst interaction in the multi-burst actuation is studied in section \ref{sec:multi}. The conclusions are given in section \ref{Sec:conc}.


 
\section{Experimental Setup}\label{sec:exp}

\begin{figure}
\centering
\includegraphics[width=.7\textwidth]{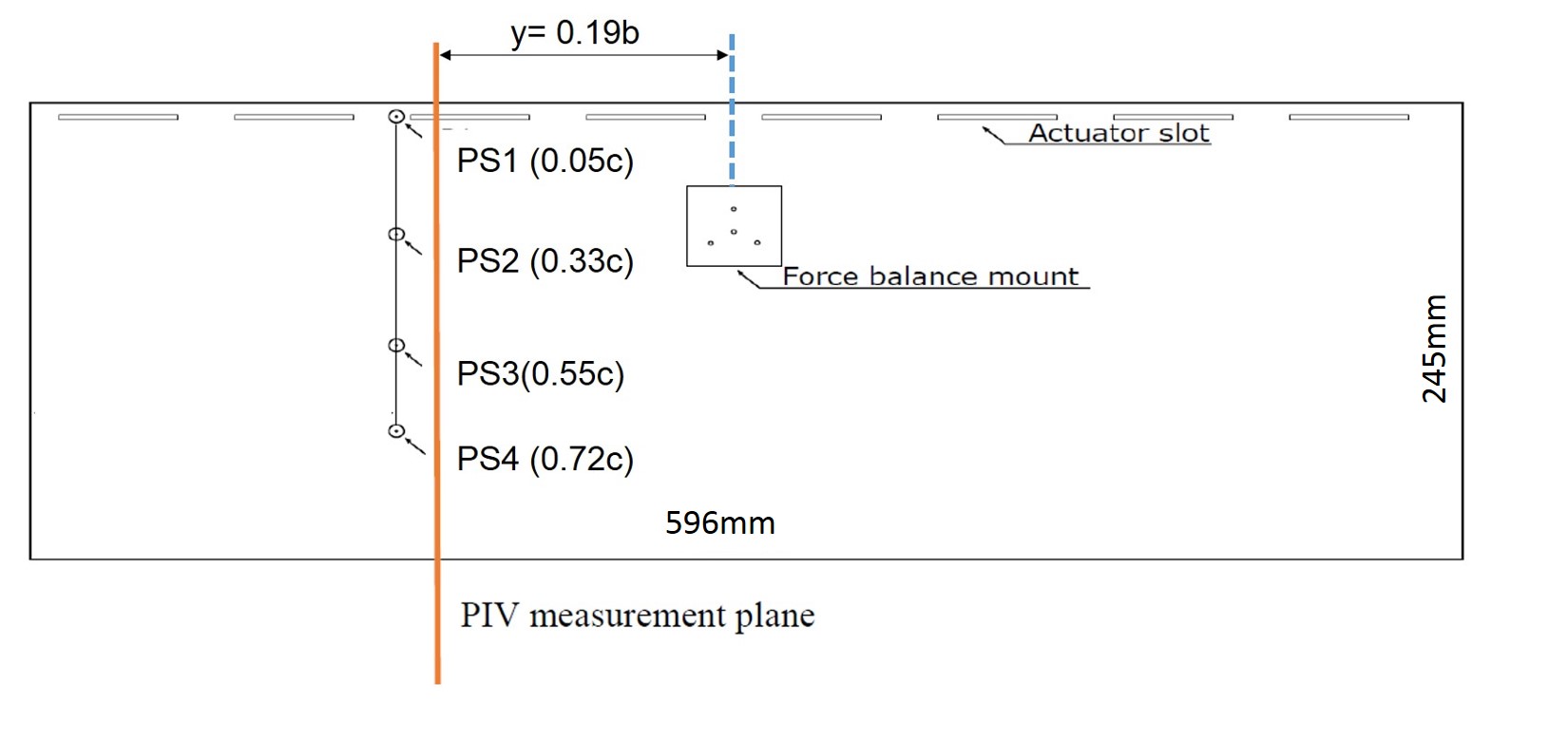}
\caption{Plan view of the wing with NACA 0009 profile. Pressure sensor, force balance, and actuator locations are shown.}
\label{fig:wing}
\end{figure}

The experiments were conducted in the Andrew Fejer Unsteady Flow Wind Tunnel at Illinois Institute of Technology. The wind tunnel has cross-section dimensions $600mm \times 600mm$. The right-hand coordinate system is defined with the origin at the leading edge of the wing and the x-axis in the flow direction, the y-axis pointing upward, and the z-axis pointing in the direction of the left side of the wing. A nominally two-dimensional NACA0009 wing with a wingspan $b = 596mm$ and chord length $c = 245mm$ was used as the test article that is shown in figure \ref{fig:wing}. The freestream speed was $U_{\infty}=3m/s$, corresponding to a convective time $t_{convect}=c/U_{\infty}=0.082 s$.  Dimensionless time $t^+$ is normalized by the convective time so that $t^+=t/t_{convect}$.  The chord-based Reynolds number is $Re_c = 49,000$. The angle of attack of the wing was fixed at $\alpha=12^o$, at which the flow is fully separated on the suction side of the airfoil.

Eight piezoelectric (zero net mass) actuators were installed in the leading edge of the wing. The slots of the actuators were located 0.05c from the leading edge with an exit angle of 30 degrees from the tangent to the surface on the suction side of the wing. The dimension of each actuator orifice slot is $2mm \times 40mm$. 

Surface pressure measurements were made with All-Sensors D2-P4V Mini transducers built into four chord-wise locations on the wing.  The pressure range for these sensors is +/- 1 inch of water. The four pressure sensors are shown in figure \ref{fig:wing} as PS1 – PS4, and the corresponding pressure coefficient measurements will be denoted as $C_P 1$ - $C_P 4$ in the rest of this paper.
Forces and moments were measured with an ATI, Inc. model Nano-17 force balance located inside the model at $30\%$ of the chord, which is also the center of gravity of the wing. The reference point of the pitching moment is located at $25\%$ of the chord.   

\begin{figure}
\centering
\includegraphics[width=.8\textwidth]{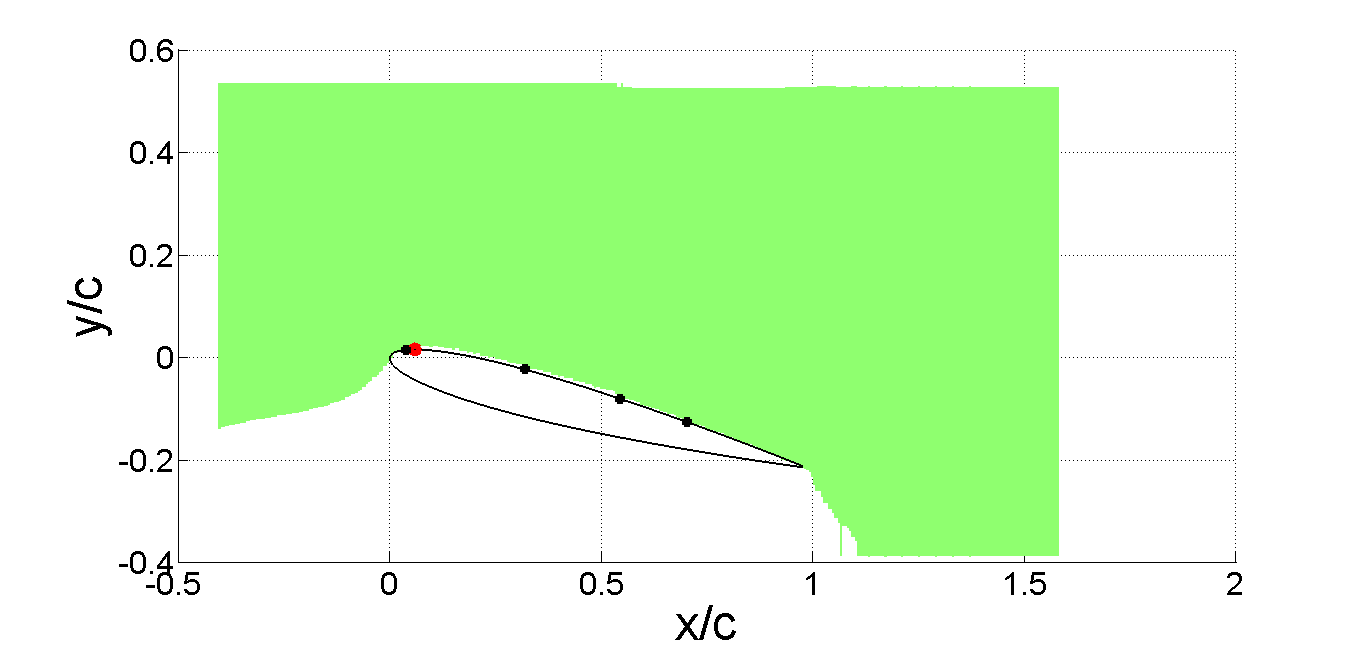}
\caption{Side view of the wing and the PIV measurement window.  Pressure sensor locations are shown by the black dots, and the actuator locations are shown by the red dot. }
\label{fig:PIV_window}
\end{figure}

Particle Image Velocimeter (PIV) flow field measurements were obtained in the x,y plane located at z=0.19b away from the centerline (indicated by the orange line in figure \ref{fig:wing}). The PIV data window in the x,y plane is shown in figure \ref{fig:PIV_window} with green color.  The small red circle denotes the streamwise location of the actuators and the black dots are the locations of the pressure sensors. The time interval between the phase-averaged PIV measurements is $0.005s$  $(0.0625t^+)$, which resulted in 800 phases covering 4s $(50t^+)$. The phase averaging was done by averaging 100 flow field images for each phase.  The initial (reference) phase corresponded to the beginning of the actuator burst signal.

\begin{figure}
	\centering
	\begin{subfigure}{0.45\textwidth}
	        \includegraphics[width=3.0in]{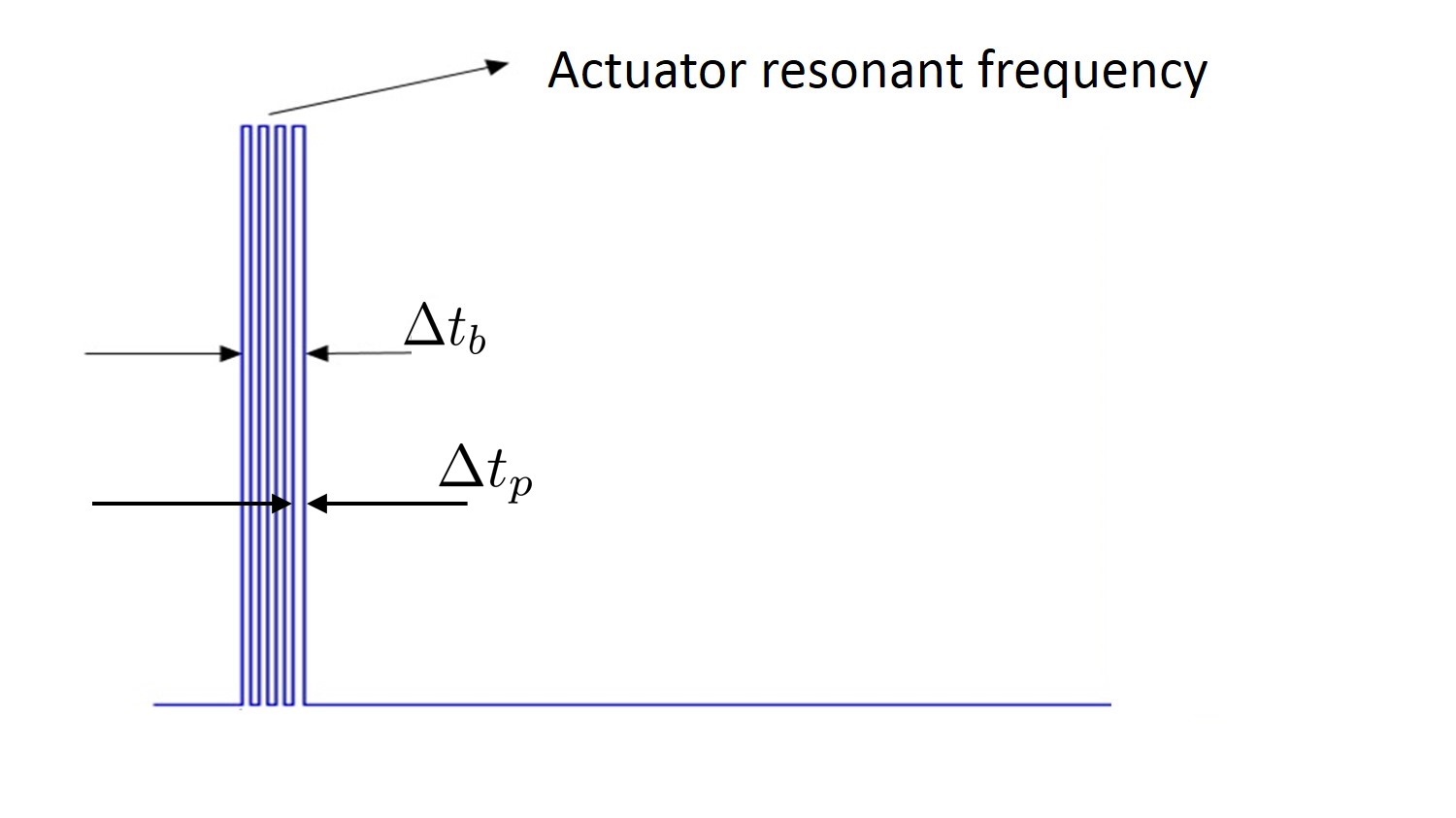}
	          \caption{Single-burst actuation consists of four$\Delta t_p$ pulses.}
	          \label{fig:input_single}
	\end{subfigure}
~
	\begin{subfigure}{0.45\textwidth}
	        \includegraphics[width=2.2in]{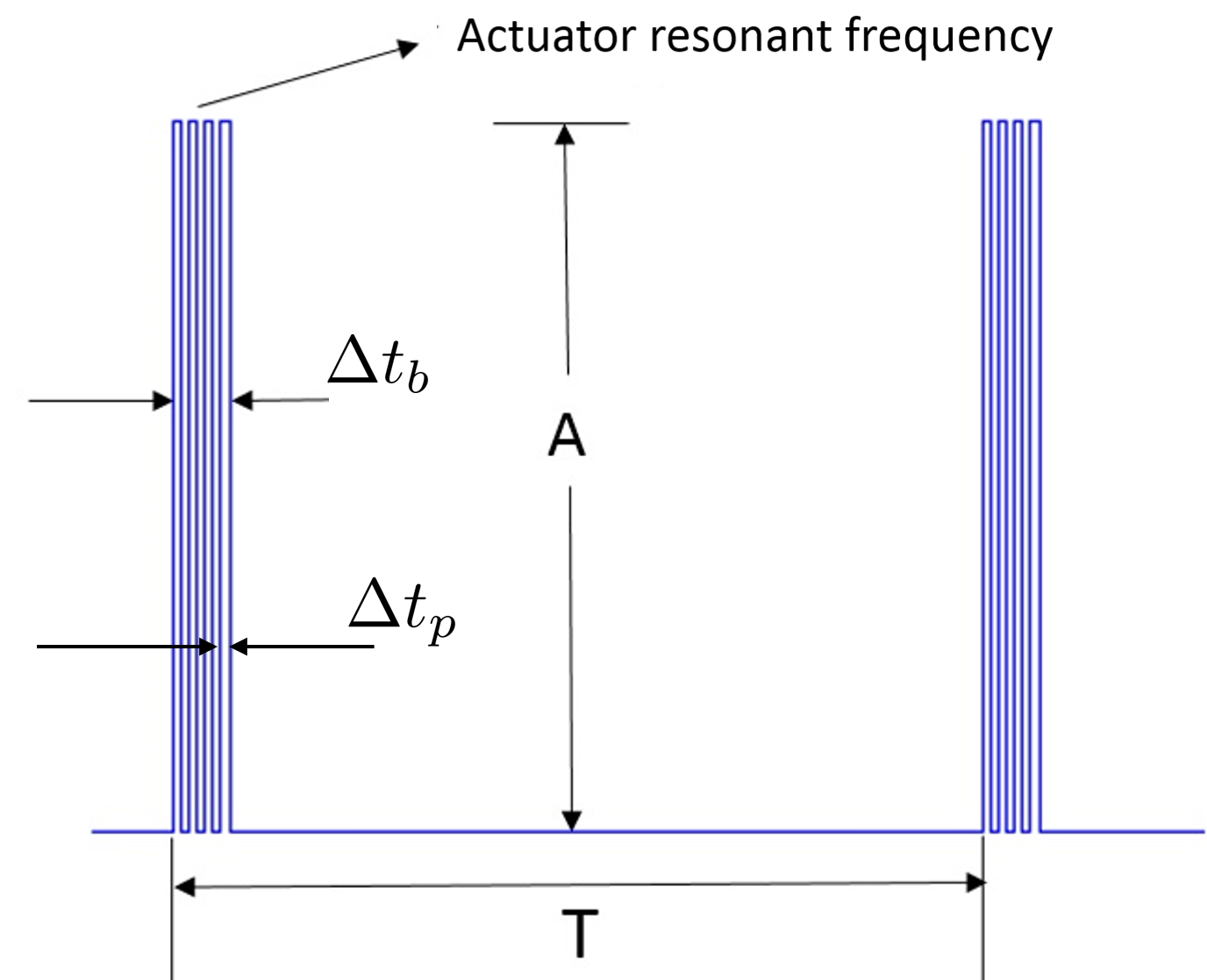}
	          \caption{Multi-burst actuation}
	          \label{fig:input_multi}
	\end{subfigure}	

    \caption{Input signal to the actuators.}
    \label{fig:input_sig}		
\end{figure}

\FloatBarrier
To produce the maximum exit jet velocity, the zero net mass actuators are operated at their mechanical resonance frequency, $f_r=400Hz$ with a pulse width of $\Delta t_p=0.03125t^+$, and 60 Volts ($V$) amplitude. A second square wave signal was superposed on the 400 Hz carrier signal to create the `burst signal'.  The burst signal width is $\Delta t_b=0.125t^+$. Therefore, the actual input signal to the actuators is a short burst signal containing 4 high-frequency (400Hz) pulses, the amplitude of the input signal A$=60V$ for all the cases in the current research (figure \ref{fig:input_sig}). The  peak exit jet velocity measured with a hot-wire anemometer at the actuator exit is $4.9m/s$ corresponding to the peak  $C_{\mu}=\frac{\rho {V_{jet}}^2 A_{jet}}{0.5\rho U^2 cb}=0.01$, where $V_{jet}$ is the peak velocity of the actuation jet, $A_{jet}$ is the opening area of the actuators, $\rho$ is the air density, $U_{\infty}$ is the freestream velocity, $c$ is the chord length of the wing, and $b$ is the wingspan.  

\FloatBarrier

\section{Single-burst Actuation}\label{Sec:single}	
\begin{figure}
	\centering
    \begin{subfigure}{0.3\textwidth}
    	        \includegraphics[width=2in]{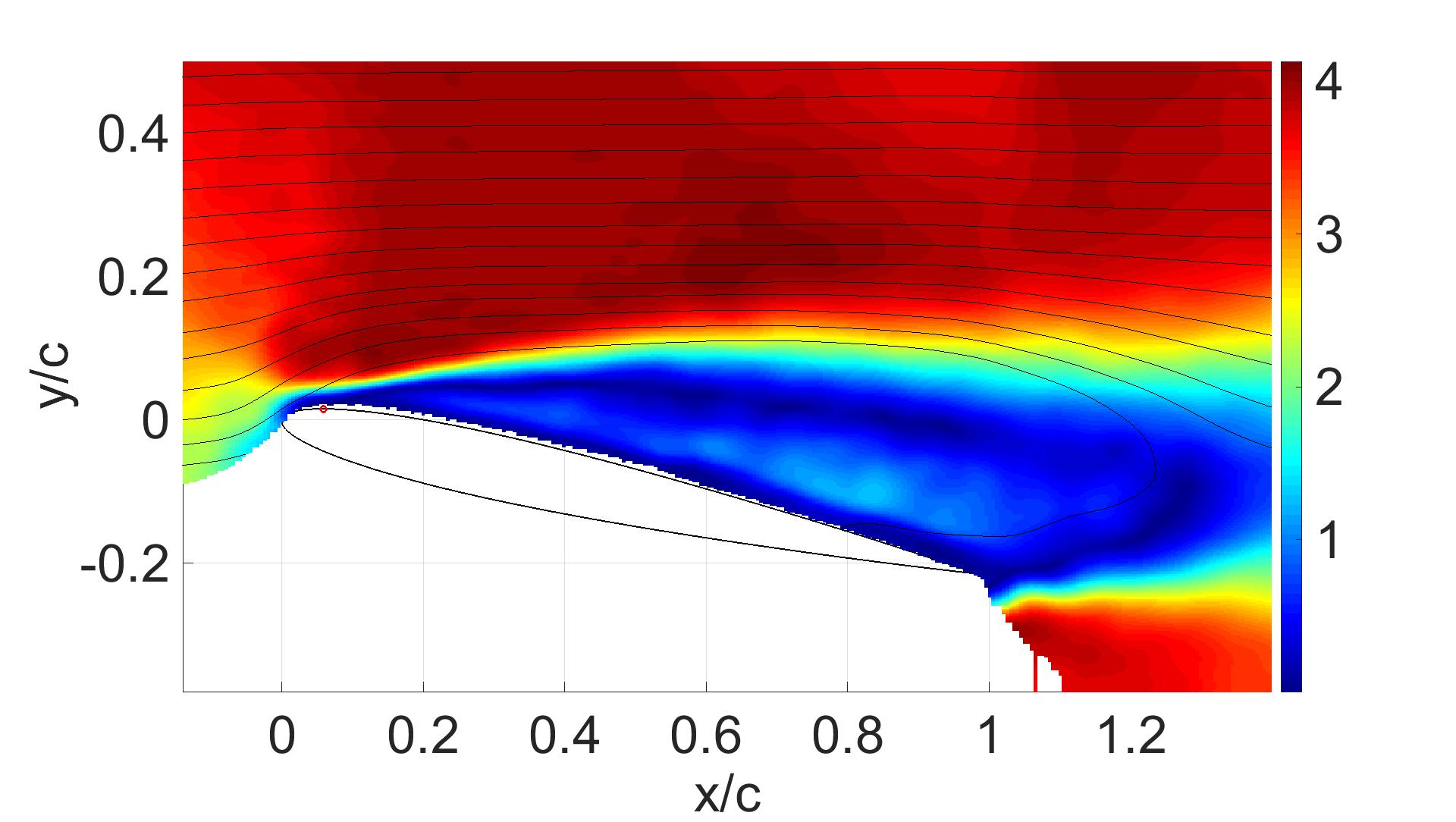}
    			\caption{$0t^+$}
    	          \label{fig:flow_field_0t+}
    	\end{subfigure}	
    	~
    	\begin{subfigure}{0.3\textwidth}
    	    	  \includegraphics[width=2in]{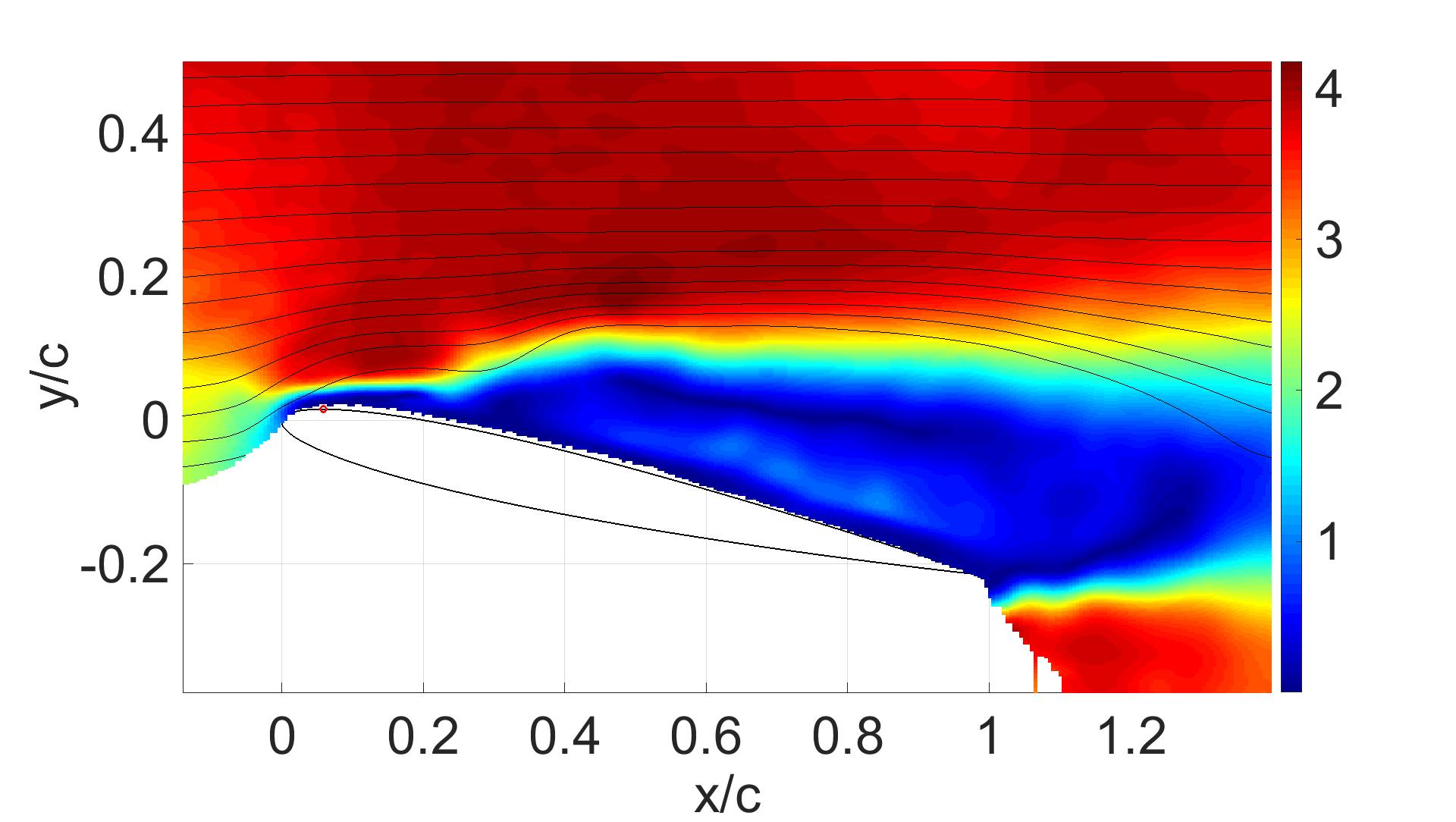}
    	    	  \caption{$0.5t^+$}
    	    	   \label{fig:flow_field_0.5t+}
    	\end{subfigure}
    	~
    	\begin{subfigure}{0.3\textwidth}
    	        \includegraphics[width=2in]{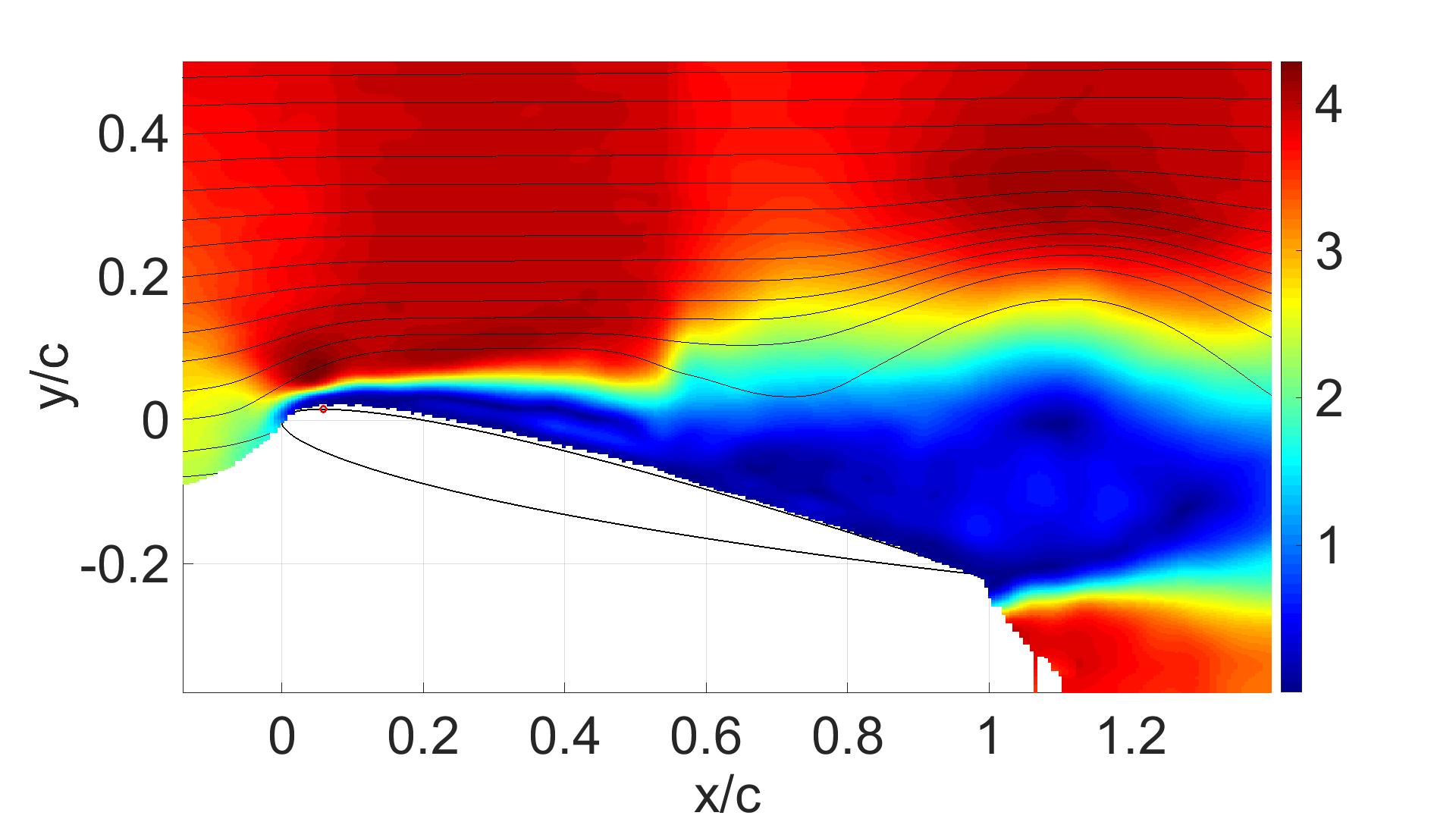}
    	          \caption{$1.4t^+$}
    	          \label{fig:flow_field_1.4t+}
    	\end{subfigure}    	
 
    \begin{subfigure}{0.3\textwidth}
    	        \includegraphics[width=2in]{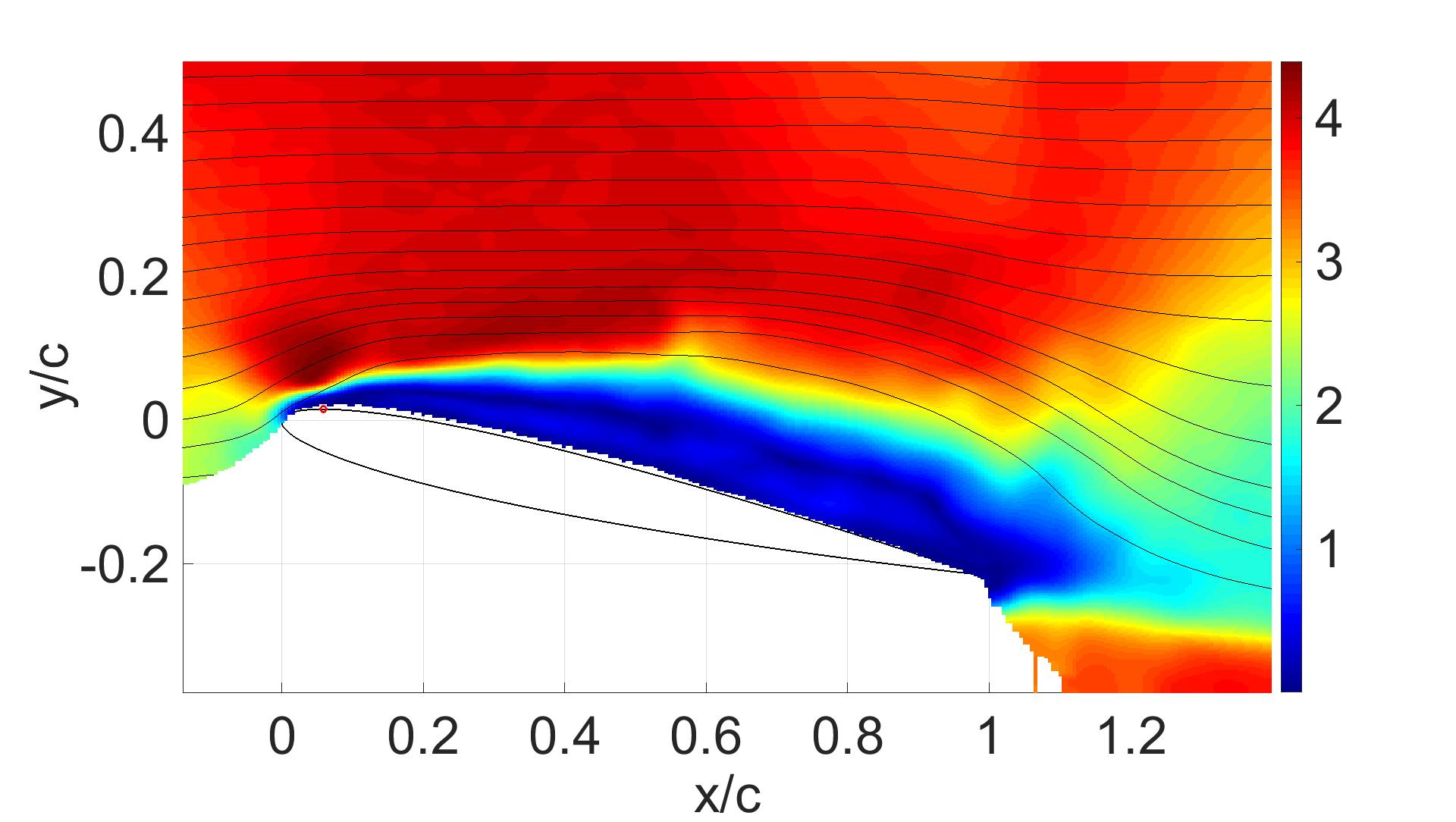}
    			\caption{$2.8t^+$}
    	          \label{fig:flow_field_2.8t+}
    	\end{subfigure}	
    	~
    	\begin{subfigure}{0.3\textwidth}
    	    	  \includegraphics[width=2in]{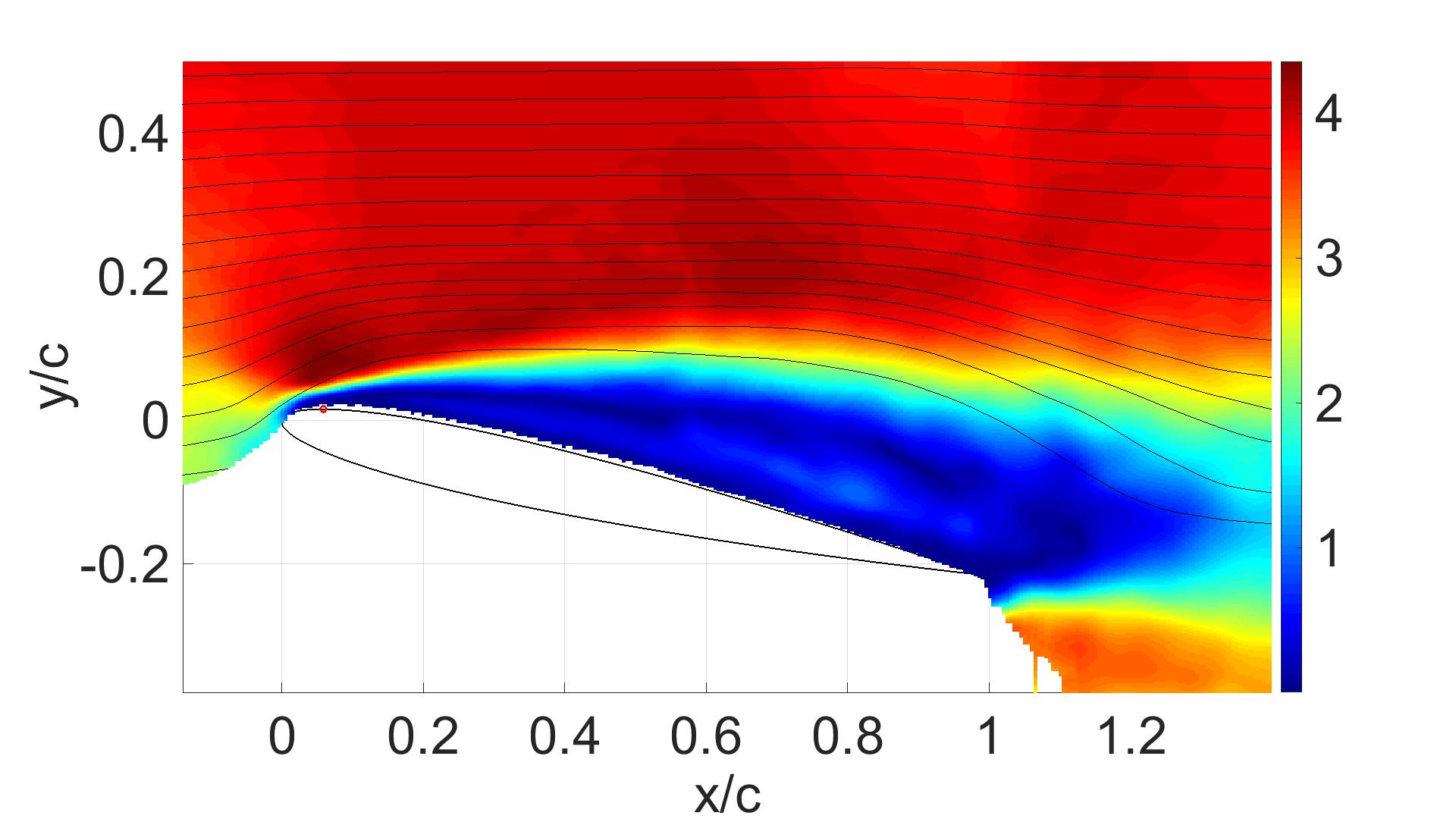}
    	    	  \caption{$4t^+$}
    	    	   \label{fig:flow_field_4t+}
    	\end{subfigure}
    	~
    	\begin{subfigure}{0.3\textwidth}
    	        \includegraphics[width=2in]{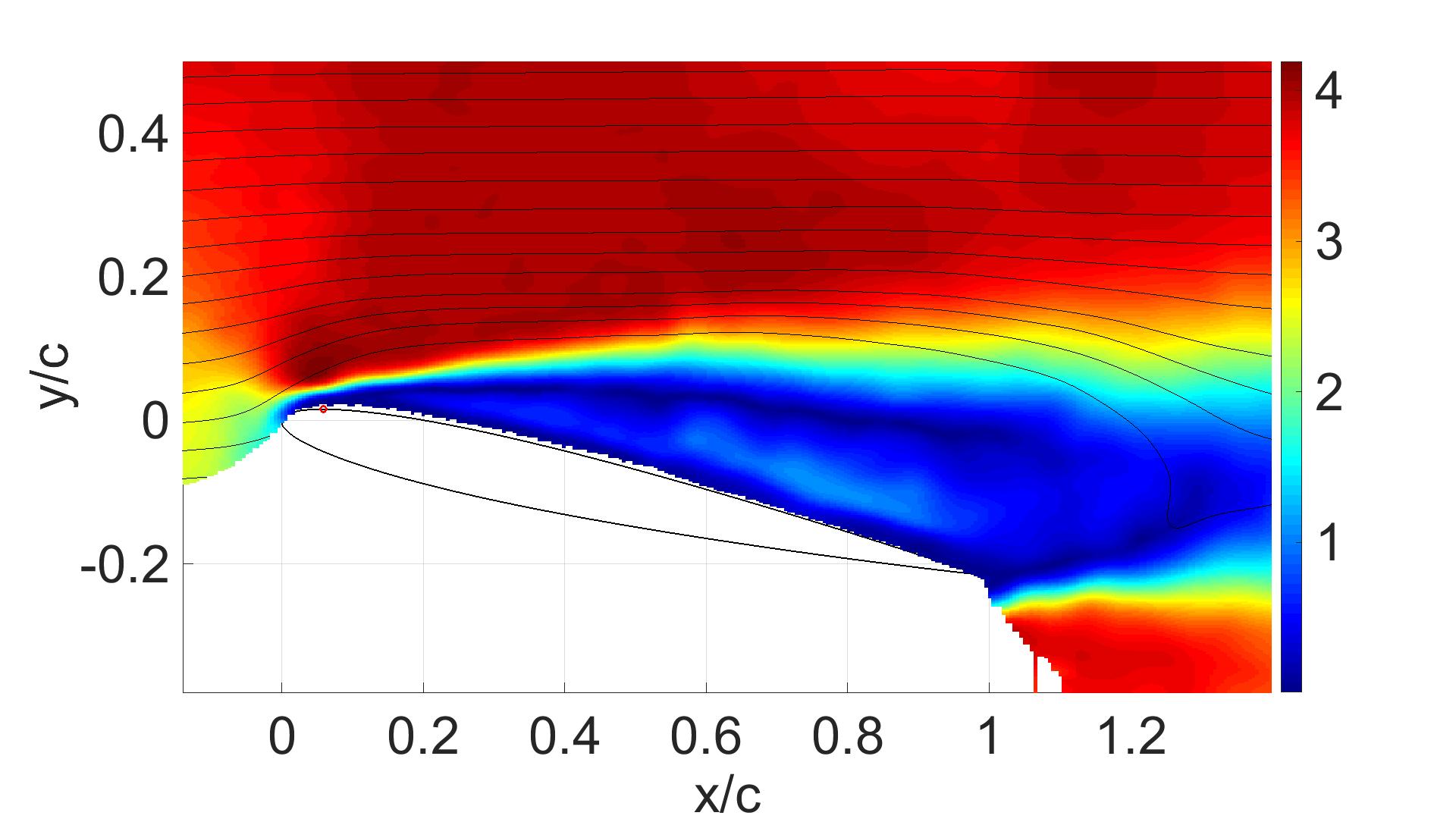}
    	          \caption{$20t^+$}
    	          \label{fig:flow_field_20t+}
    	\end{subfigure}

    \caption{Phase-averaged velocity magnitude $(\sqrt{U^2+V^2})$ and streamline time sequence plots following a single-burst actuation that was triggered at $0t^+$. The color indicates the velocity magnitude and the black lines are showing the streamline. The red circle on the leading edge denotes the streamwise location of the actuators.} 
    \label{fig:flow_field} 
\end{figure}

\begin{figure}
	\centering
    \begin{subfigure}{0.3\textwidth}
    	        \includegraphics[width=1.7in]{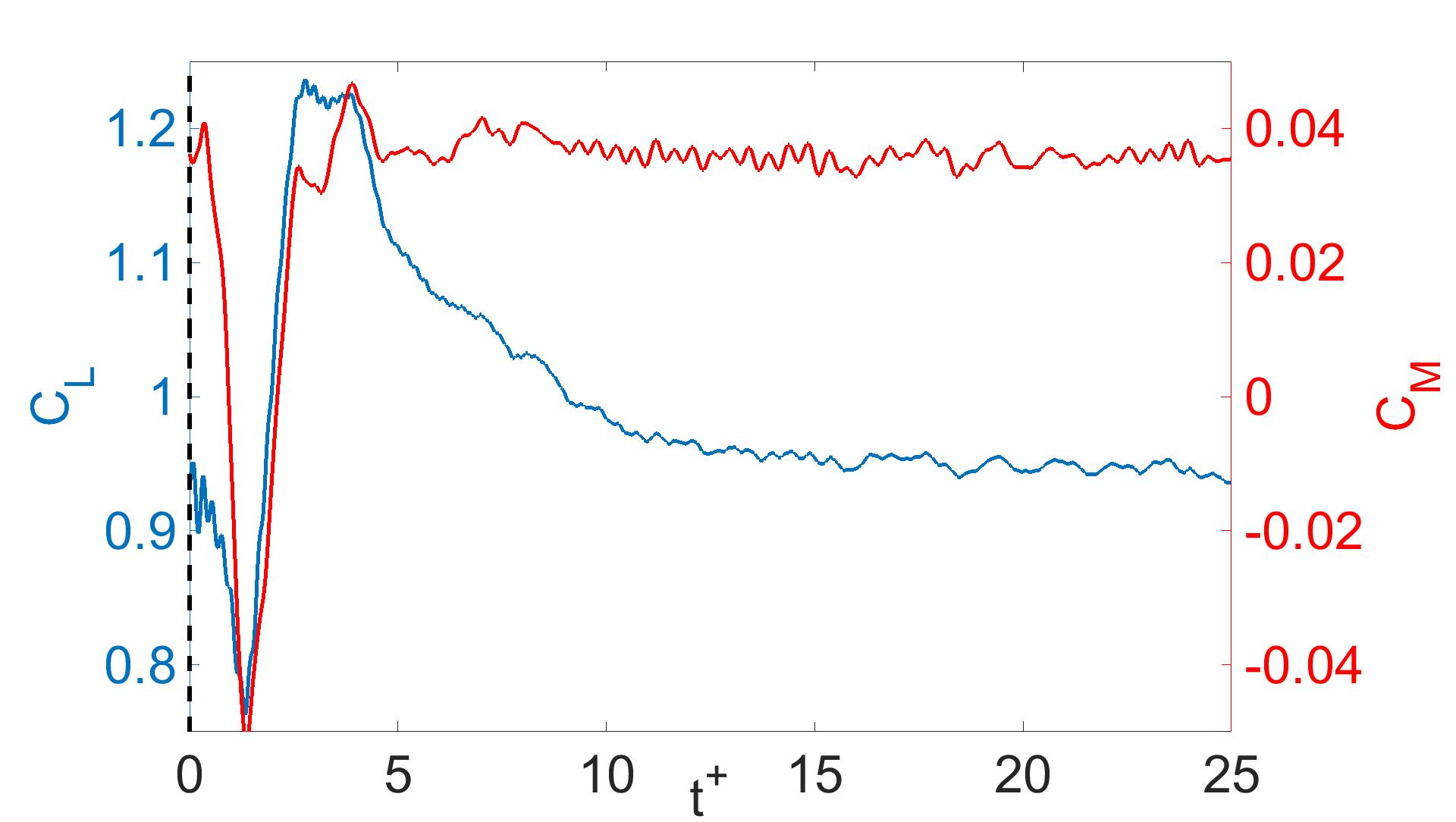}
    			\caption{$0t^+$}
    	          \label{fig:single_CL_CM_0t+}
    	\end{subfigure}	
    	~
    	\begin{subfigure}{0.3\textwidth}
    	    	  \includegraphics[width=1.7in]{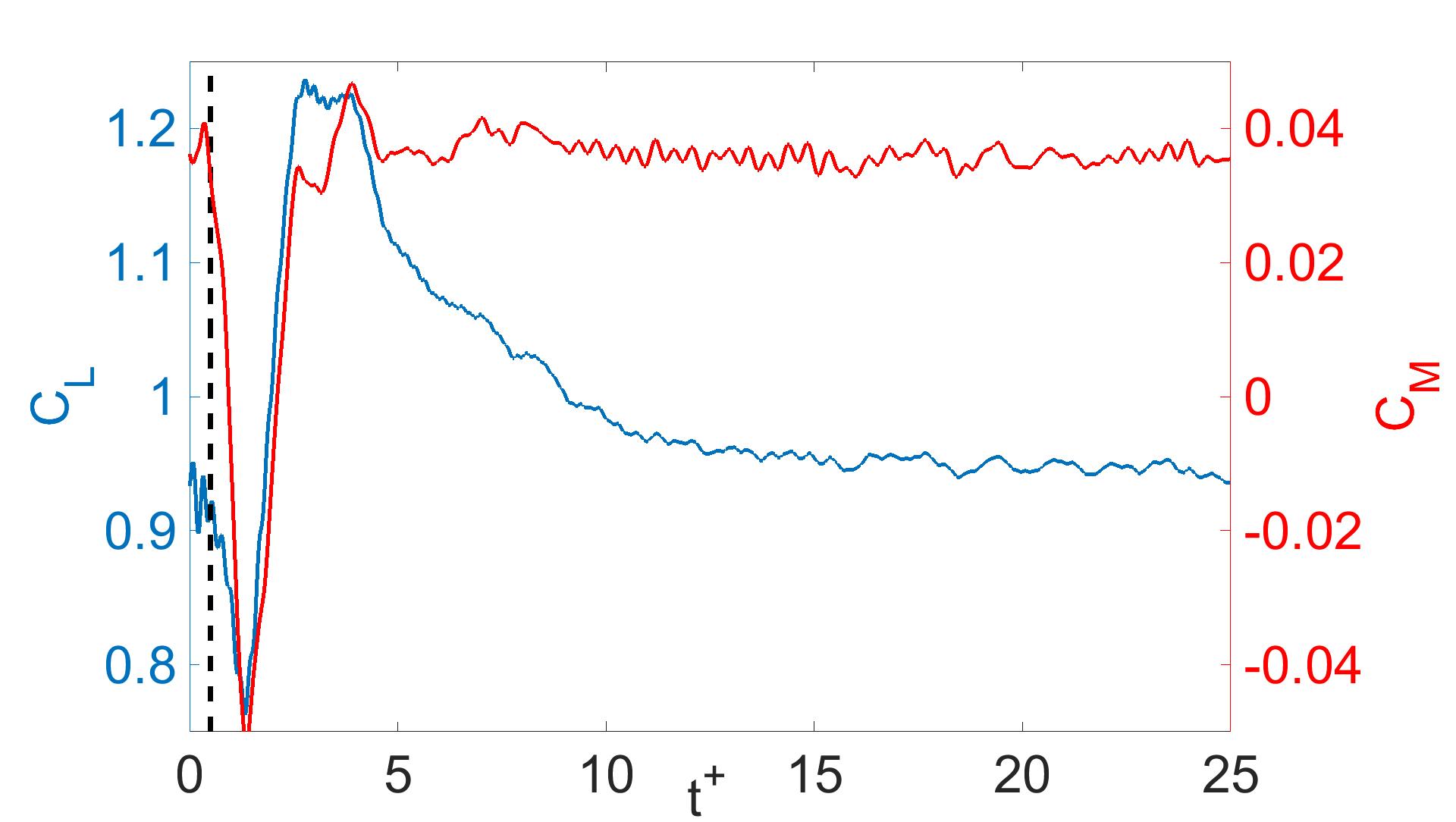}
    	    	  \caption{$0.5t^+$}
    	    	   \label{fig:single_CL_CM_0.5t+}
    	\end{subfigure}
    	~
    	\begin{subfigure}{0.3\textwidth}
    	        \includegraphics[width=1.7in]{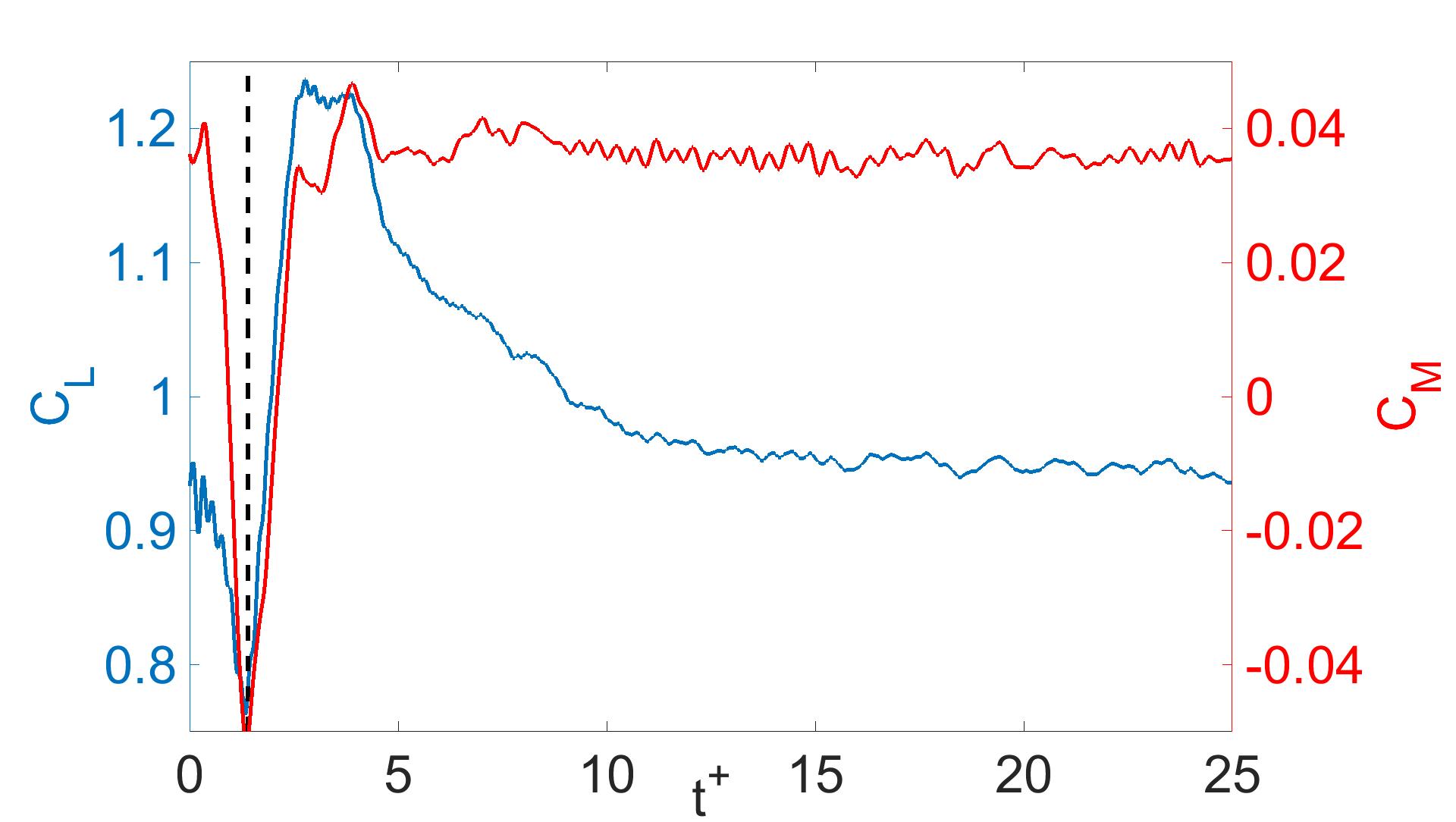}
    	          \caption{$1.4t^+$}
    	          \label{fig:single_CL_CM_1.4t+}
    	\end{subfigure}    	
 
    \begin{subfigure}{0.3\textwidth}
    	        \includegraphics[width=1.7in]{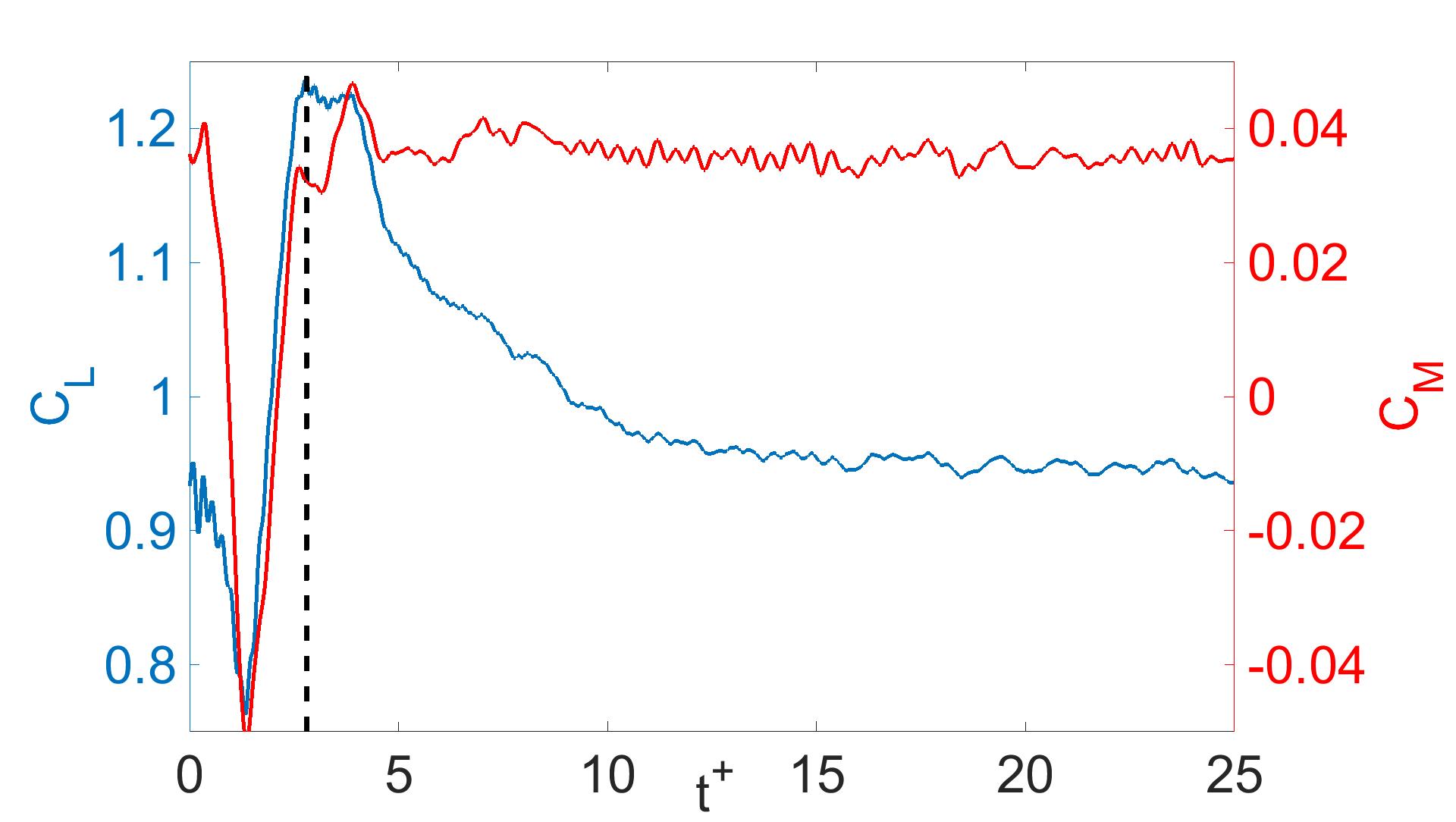}
    			\caption{$2.8t^+$}
    	          \label{fig:single_CL_CM_2.8t+}
    	\end{subfigure}	
    	~
    	\begin{subfigure}{0.3\textwidth}
    	    	  \includegraphics[width=1.7in]{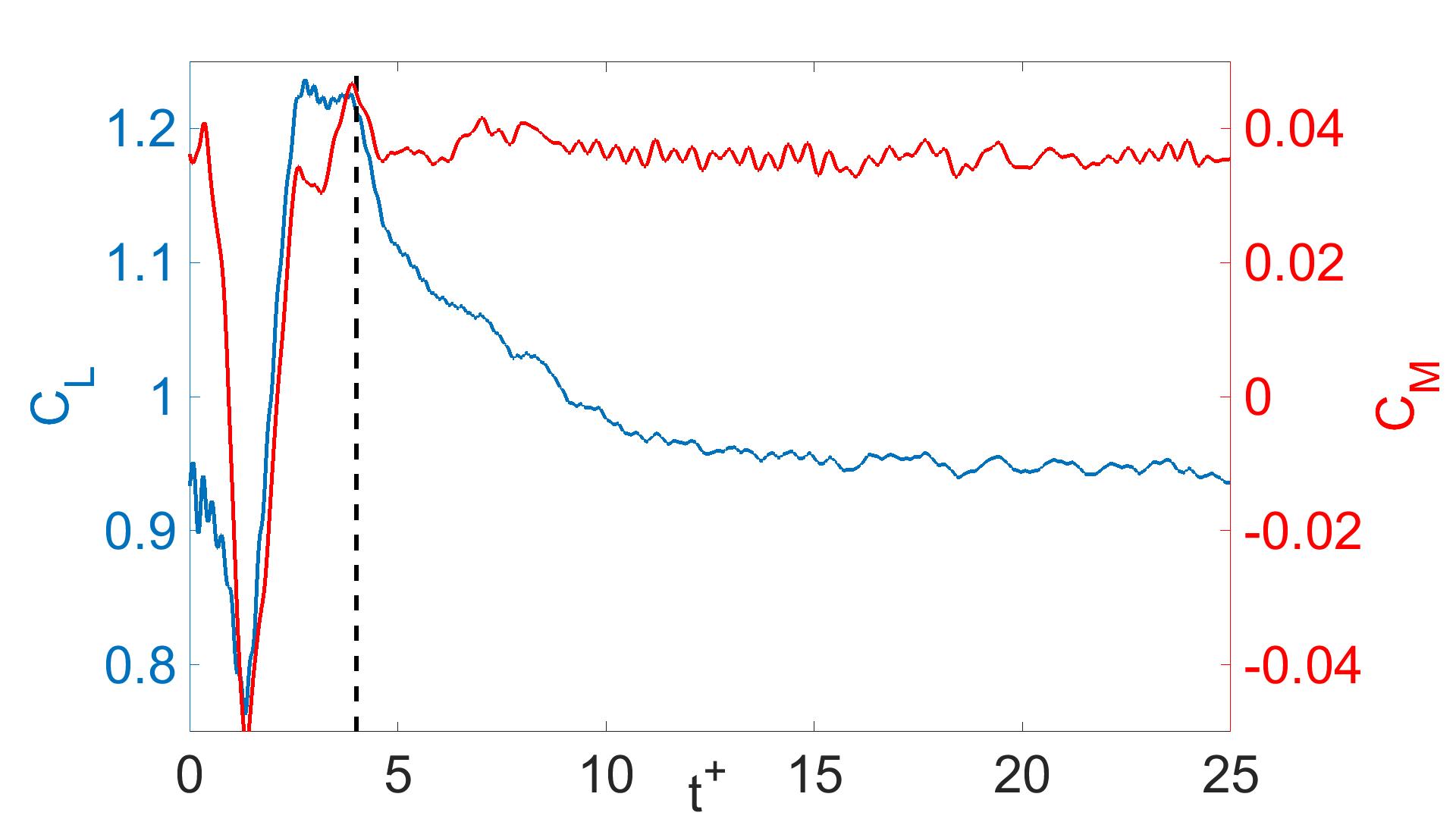}
    	    	  \caption{$4t^+$}
    	    	   \label{fig:single_CL_CM_4t+}
    	\end{subfigure}
    	~
    	\begin{subfigure}{0.3\textwidth}
    	        \includegraphics[width=1.7in]{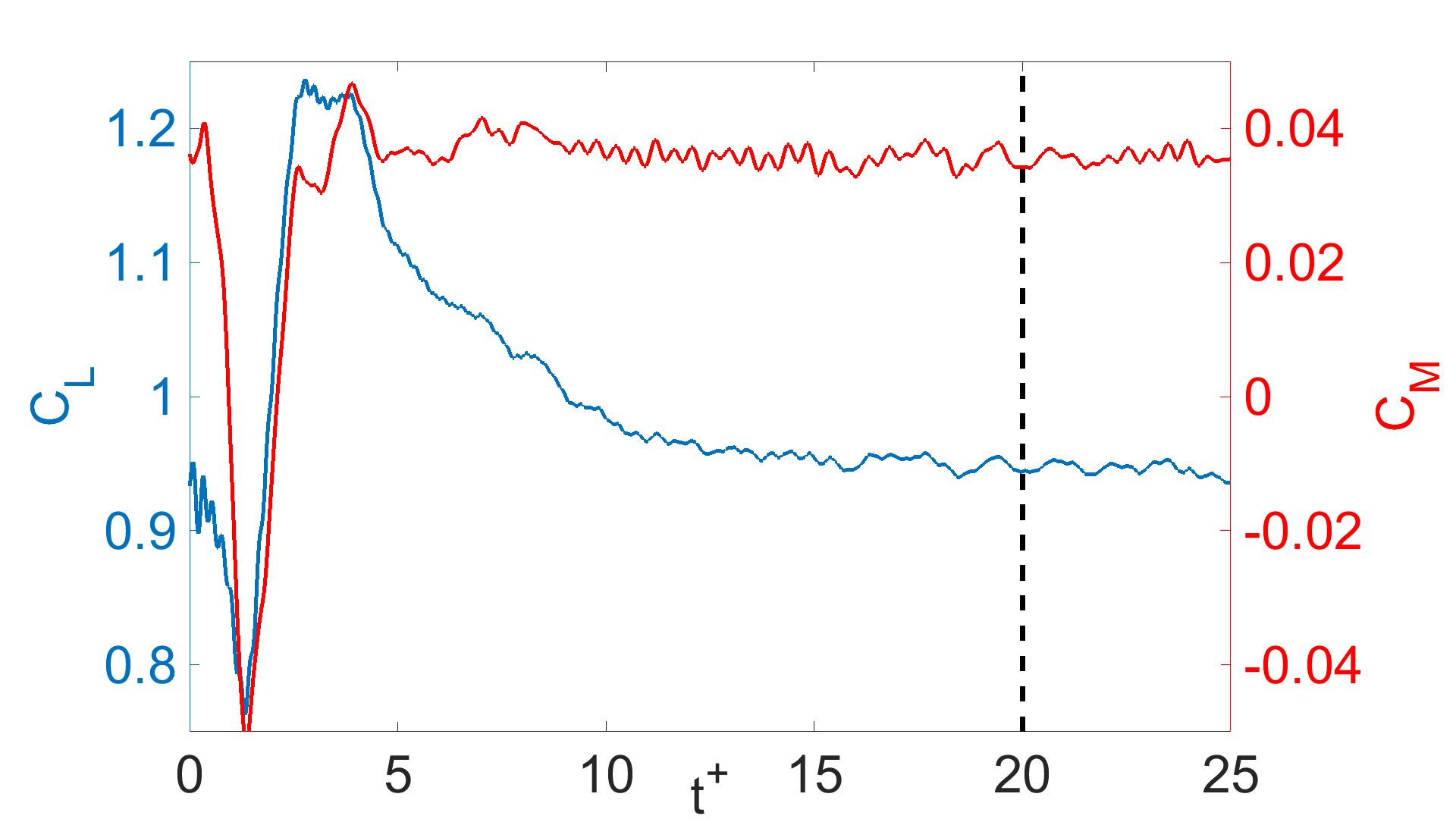}
    	          \caption{$20t^+$}
    	          \label{fig:single_CL_CM_20t+}
    	\end{subfigure}

    \caption{Phase-averaged $C_L$ (blue) and $C_M$ (red) variation following a single-burst actuation that was triggered at $0t^+$. The solid blue line denotes $C_L$, the solid red line denotes $C_M$ and the dashed black line indicates the time instant correlated to each flow field in figure \ref{fig:flow_field}.} 
    \label{fig:single_CL_CM} 
\end{figure}

The single-burst actuation consists of a sequence high-frequency pulses as shown in the diagram in figure \ref{fig:input_single}. The corresponding velocity field and streamline response to the single-burst actuation is shown at different instants in figure \ref{fig:flow_field}. The color corresponds to velocity magnitude.  Figure \ref{fig:single_CL_CM} shows the  lift coefficient $C_L=\frac{L}{0.5\rho U^2 cb}$ (blue) and pitching moment coefficient $C_M=\frac{M}{0.5\rho U^2 c^2b}$ (red) response, where $L$ is the lift force and $M$ is the pitching moment. Positive $M$ corresponds to a nose-up pitching moment. The vertical dashed black lines in figure \ref{fig:single_CL_CM} correspond to the instantaneous flow fields shown in figure \ref{fig:flow_field}. Both $C_L$ and $C_M$ decrease immediately following the single-burst signal, which can be seen from the time series $C_L$, $C_M$ data plotted in figure \ref{fig:single_CL_CM_0t+} to figure \ref{fig:single_CL_CM_1.4t+}. Similar lift reversal phenomenon was identified first by \citet{amitay2002controlled}. Since then the effect has been observed by numerous investigators, \citep{GW, brzozowski2010transient, woo2009transitory}, and is now an established feature of the separated flow dynamics. 

Referring to the velocity field prior to the burst in figure, the baseline flow on the suction side is fully separated at $\alpha=12^o$ angle of attack (figure \ref{fig:flow_field_0t+}). The single-burst actuation was initiated at $0t^+$ and lasted for $0.12t^+$. The beginning of the reattachment process can be seen at  $0.5t^+$ at $x/c=0.2$ from the leading edge (figure \ref{fig:flow_field_0.5t+}). The reattachment produces a “kink” in the shear layer that divides the new `reattached' flow from the `old' separated flow region. The reattached region grows with time as the kink convects downstream from the leading edge towards the trailing edge as shown in figure \ref{fig:flow_field_0.5t+} to figure \ref{fig:flow_field_2.8t+}.  The maximum flow reattachment occurs when the kinked region of the shear layer reaches the trailing edge at $2.8t^+$ (figure \ref{fig:flow_field_2.8t+}). The lift coefficient is also a maximum at this time, which agrees with Rival's observations \citep{rival2014characteristic}.  After $4t^+$, the lift coefficient begins to decrease as the flow field gradually relaxes to its original baseline state.  This relaxation process is exhibited in figure \ref{fig:flow_field_4t+} and figure \ref{fig:flow_field_20t+}, and it takes about $10t^+$. \citet{brzozowski2010transient} described much of the same behavior when using combustion-based pulsed actuators on an NACA 4415 cambered wing.

Figure \ref{fig:single_CL_CM_1.4t+} shows that both $C_L$ and $C_M$ reach their minimum at $1.4t^+$ before they start their climb to the maximum increments. The $C_L$ reaches its maximum at $2.8t^+$ (figure \ref{fig:single_CL_CM_2.8t+}), which is consistent with the flow field measurement shown in figure \ref{fig:flow_field_2.8t+} where the flow reattachment length also reaches its maximum. The maximum $C_L$ increment is about 30\% of its undisturbed baseline value. 

In contrast to the $C_L$ behavior there is no significant increase in $C_M$ above the baseline (figure \ref{fig:single_CL_CM}).  At $4t^+$ later, as can be seen in figure \ref{fig:single_CL_CM_4t+} and figure \ref{fig:single_CL_CM_20t+} both $C_L$ and $C_M$ start to return to their baseline undisturbed condition, which is also consistent with the flow field (figure \ref{fig:flow_field_4t+} and figure \ref{fig:flow_field_20t+}). A more detailed discussion about the $C_L$, $C_M$ reversal and increment will be provided in the following section by using pressure measurement and vortex structure.         
\FloatBarrier

\FloatBarrier

\subsection{The mechanism of $C_L$, $C_M$ reversal}

The mechanism behind the lift and pitching moment reversals is important to understand, because it is the reason for the non-minimum phase behavior that limits control bandwidth.  By examining the vortex structure and surface pressure evolution, we intend to gain some insight into the physics of the lift and pitching moment reversal. The methodology here is that we will first relate the lift and pitching moment to the pressure, and then link the pressure to the flow field evolution.

\begin{figure}

\centering
    \begin{subfigure}{0.4\textwidth}
    \centering
    
    	        \includegraphics[width=1\textwidth]{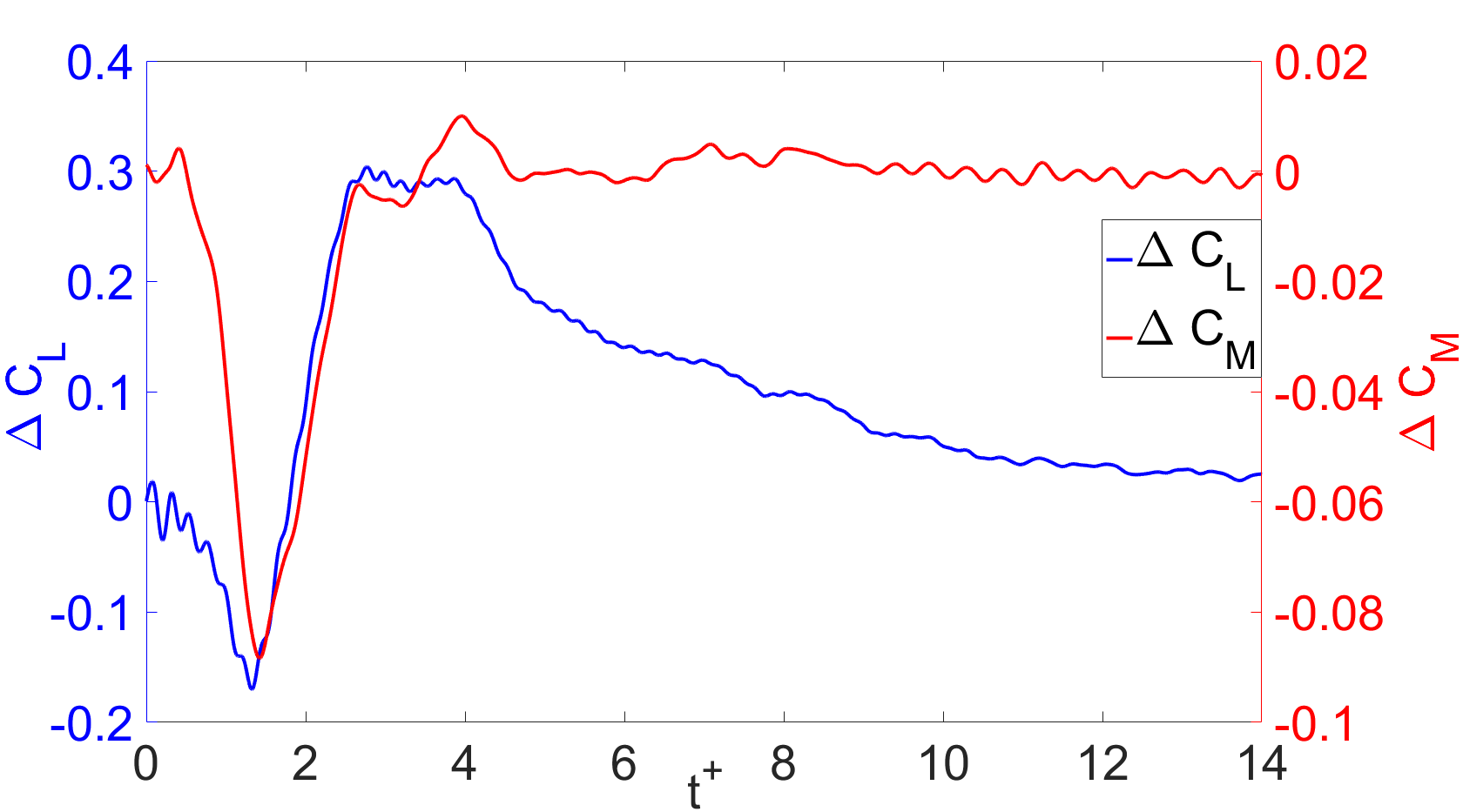}
    			\caption{$\Delta C_L$, and $\Delta C_M$ following the single-burst actuation.}
    	          \label{fig:single_DCL_DCM}
    	\end{subfigure}	
~    	
    	\begin{subfigure}{0.4\textwidth}
    	\centering
    	    	  \includegraphics[width=1\textwidth]{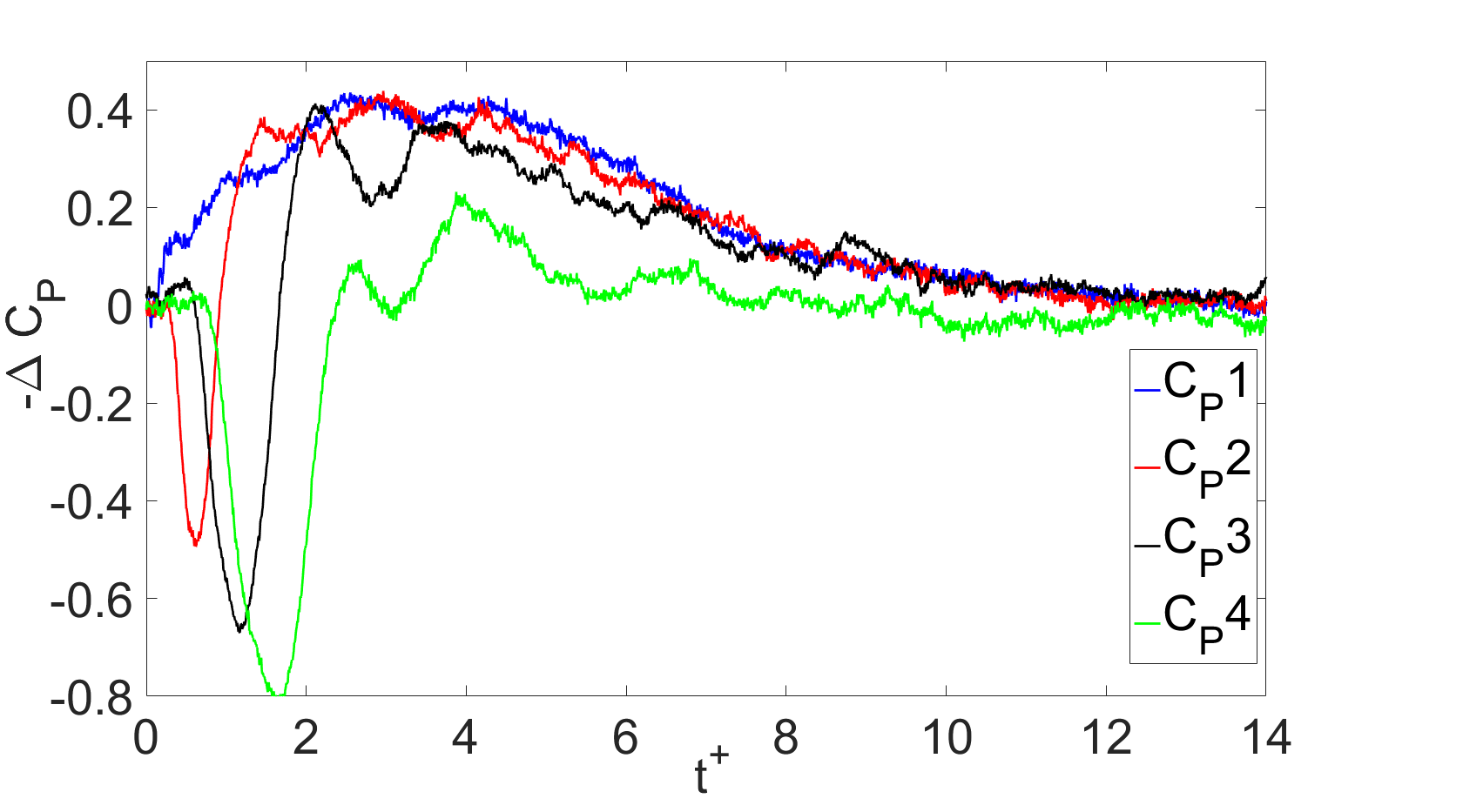}
    	    	  \caption{$\Delta C_P$ following the single-burst actuation.}
    	    	   \label{fig:single_pressure}
    	\end{subfigure}

\caption{Lift coefficient increment and pitching moment coefficient increment measured by the force transducer, and pressure coefficient increments measured by the four pressure sensors.}
\label{fig:CL_CM_pressure}
\end{figure}

The time-varying increments $\Delta C_L$, $\Delta C_M$ and $\Delta C_P$ are shown in figure \ref{fig:CL_CM_pressure}. Here, $"\Delta"$ denotes the disturbed value relative to the baseline. Note that the vertical axis on figure \ref{fig:single_pressure} is $"-\Delta C_P"$ for a easier comparison with $\Delta C_L$ and $\Delta C_M$. The pressure response to the short burst actuation was investigated at four chordwise locations. It can be seen in figure \ref{fig:CL_CM_pressure} that the pressure at locations PS2, PS3, and PS4 follow a similar trend following a pulse disturbance from the actuator. There is an approximately $0.6 t^+$ constant time delay  between the minima in the pressure sensor signals due to the convection of the disturbance. The first pressure signal on PS1 does not follow the same pattern as the other pressure sensors, because it is upstream of the actuator. The pressure data variation will be discussed in more detail later associated with the vortex structure. 

\begin{figure}
\centering
    	\begin{subfigure}{0.45\textwidth}
    	\centering
		\includegraphics[width=1\textwidth]{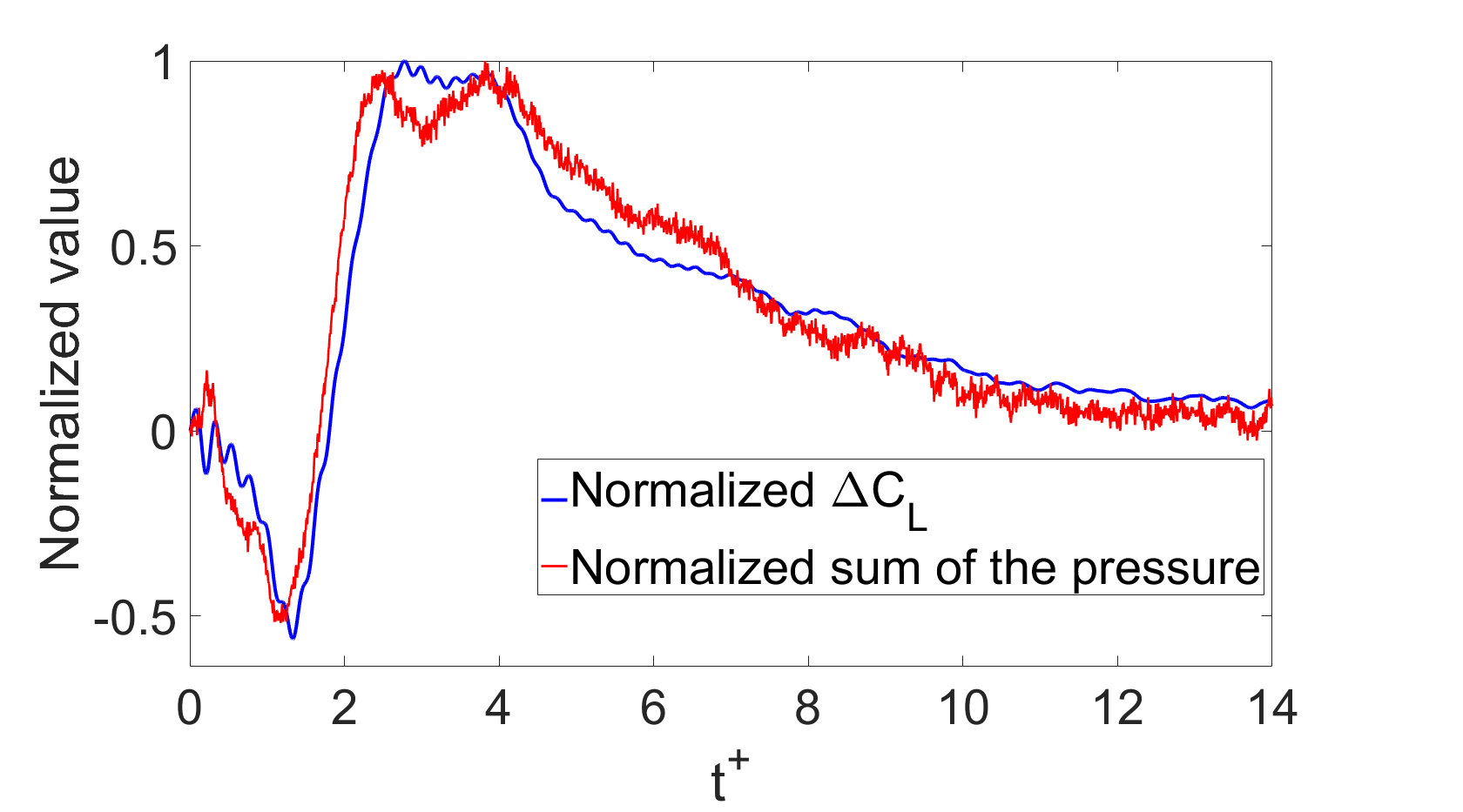}
			\caption{Comparison of $\Delta C_L$ normalized by the maximum lift increment, and the sum of the all 4 pressure measurements normalized by the maximum.}
	    \label{fig:DCL_single_pressure}
	    \end{subfigure}
~
    	\begin{subfigure}{0.45\textwidth}
    	\centering
		\includegraphics[width=1\textwidth]{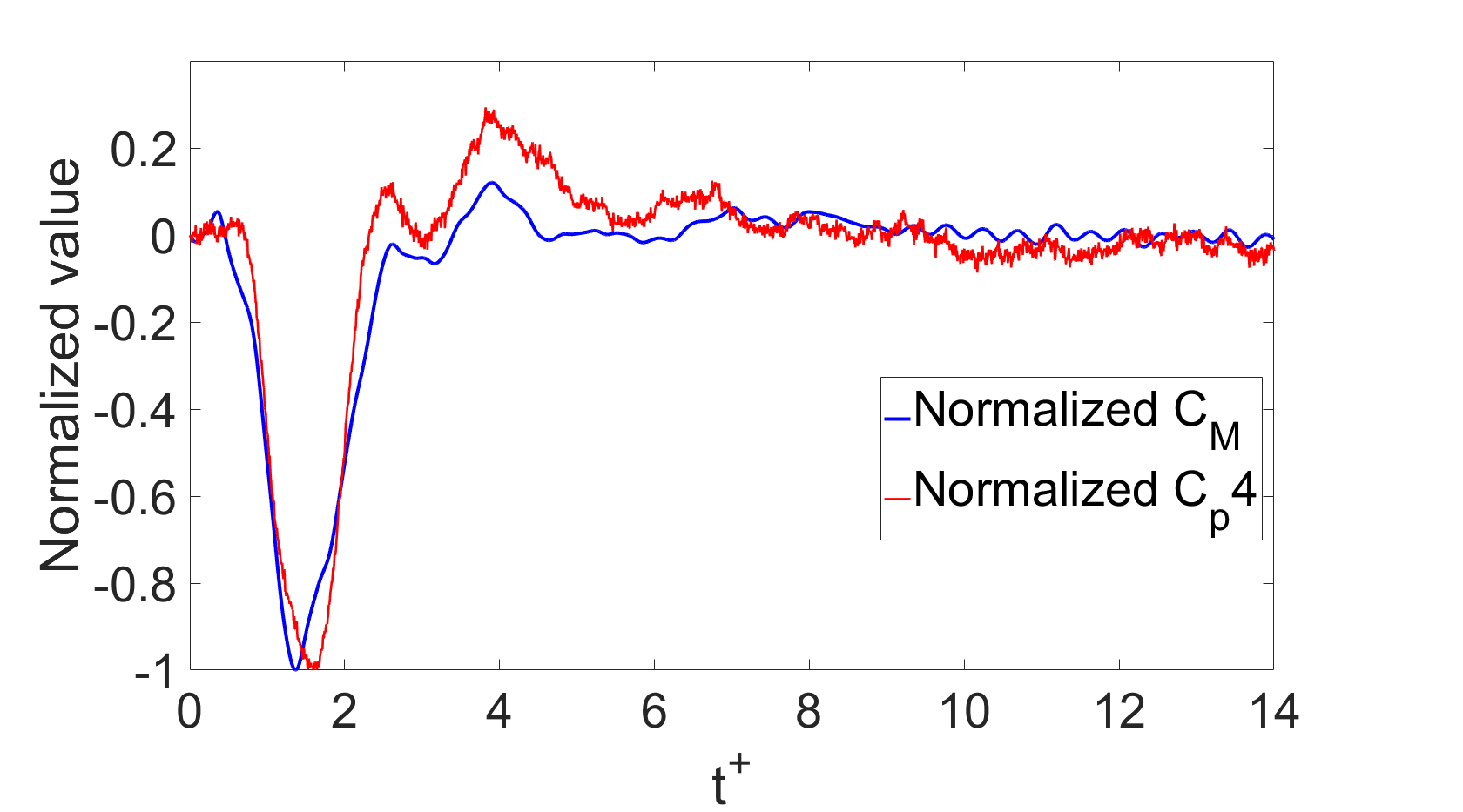}
			\caption{Comparison of the normalized $\Delta C_M$ and the normalized $C_P4$.}
	    \label{fig:DCM_single_pressure}
	    \end{subfigure}
\caption{Lift and moment coefficient increments normalized by their respective maxima.}
\end{figure}

The trend of $\Delta C_L$ data follows approximately the sum of all the pressure measurements. A comparison  of $\Delta C_L$ normalized by the maximum lift coefficient increment $\Delta C_{L max}$ (blue) and the normalized value of the sum of the pressure measurements is plotted in figure \ref{fig:DCL_single_pressure} (red). The trend of the pitching moment closely tracks $C_P4$ (figure \ref{fig:DCM_single_pressure}), which is close to the trailing edge.  This is because the moment arm of PS4 is the largest relative to the reference point of the moment measurement. 

Up to now, we can conclude that the $C_L$ and $C_M$ reversals are a consequence of the surface pressure reversal following the initial actuator input. Next, we will relate the time-varying surface pressure to the vortex structure, so that we obtain a complete picture of how the flow field evolution contributes to the $C_L$, $C_M$ reversal.

\FloatBarrier

\FloatBarrier

\begin{figure}
\centering
\includegraphics[width=0.5
\textwidth]{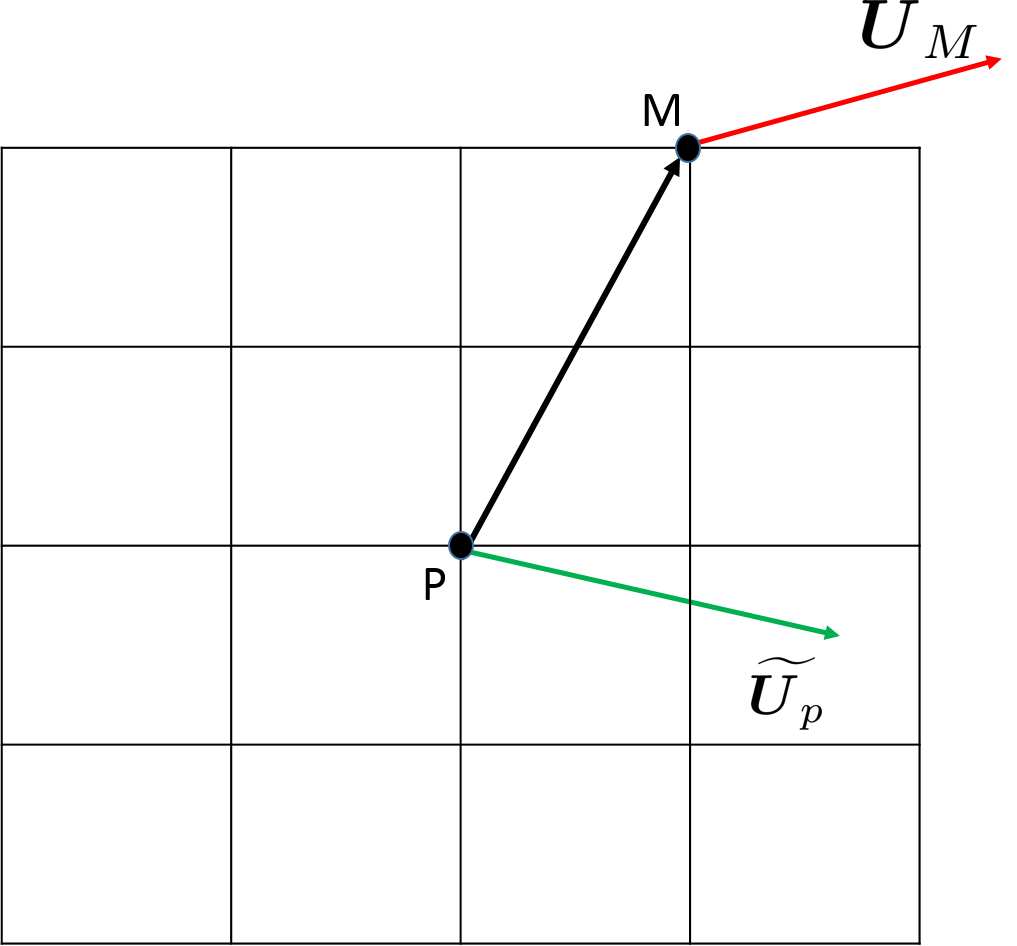}
\caption{The sketch of the $\Gamma$ computational area.}
\label{fig:Gamma2}
\end{figure}

To investigate the vortex structure associated with the lift reversal, the PIV data was analyzed using a method following \citet{graftieaux2001combining}. A Galilean invariant vortex strength $\Gamma$ in the flow was calculated that uses the local swirling velocity and the spatial vector relative to the center point of the computational region. To reduce the noise in the measured flow field, a local averaging method was used. 
\begin{equation}
\label{eq:Gamma}
\Gamma(p)=\frac{1}{N}\sum_S[\boldsymbol{P}\boldsymbol{M}\wedge(\boldsymbol{U}_M-\widetilde{\boldsymbol{U}_p})]\cdot \boldsymbol{Z}
\end{equation}
where $\boldsymbol{P}\boldsymbol{M}$ is the spatial vector from the center point $P$ of the computational area to each individual point $M$ surrounding $P$ in the computational area $S$. $N$ is the number of points in the surrounding area. $\boldsymbol{U}_M$ is the velocity at the point $M$. $\widetilde{\boldsymbol{U}_p}$ is the mean velocity in the area $S$ and $\boldsymbol{Z}$ is the unit vector normal to the measurement plane. In the 2-D case, Eq. \ref{eq:Gamma} becomes
\begin{equation}
\label{eq:Gamma_2D}
\Gamma(p)=\frac{1}{N}\sum_S[\boldsymbol{P}\boldsymbol{M}\wedge(\boldsymbol{U}_M-\widetilde{\boldsymbol{U}_p})]
\end{equation}
The computational region of $\Gamma$ at the spatial point $P$ is sketched in figure \ref{fig:Gamma2}.  
\FloatBarrier

To better visualize the vortex structure, the Galilean invariant $\Lambda_2$ criterion \citep{jeong1995identification} was used to identify the vortex boundary. The vortex strength $\Gamma$ is only shown inside the vortex boundary defined by $\Lambda_2$. Taking the gradient of Navier-Stokes equations, the symmetric part without unsteady and viscous effects is

\begin{equation}
\label{eq:lambda}
\Omega^2+S^2=-\frac{1}{\rho}\bigtriangledown(\bigtriangledown P)
\end{equation}
where $S=(J+J^T)/2$, $\Omega=(J-J^T)/2$ and $J$ is the velocity gradient tensor. The tensor $\Omega^2+S^2$ is the
corrected Hessian of pressure. The vortex core can be defined as a connected region with two negative eigenvalues $(\Lambda_2<0)$ of tensor $\Omega^2+S^2$. In our case, $(\Lambda_2<-100)$ is used to reduce the number of the small vortices that are caused by the turbulence and measurement noise.

\begin{figure}
\centering
    \includegraphics[width=0.2\textwidth]{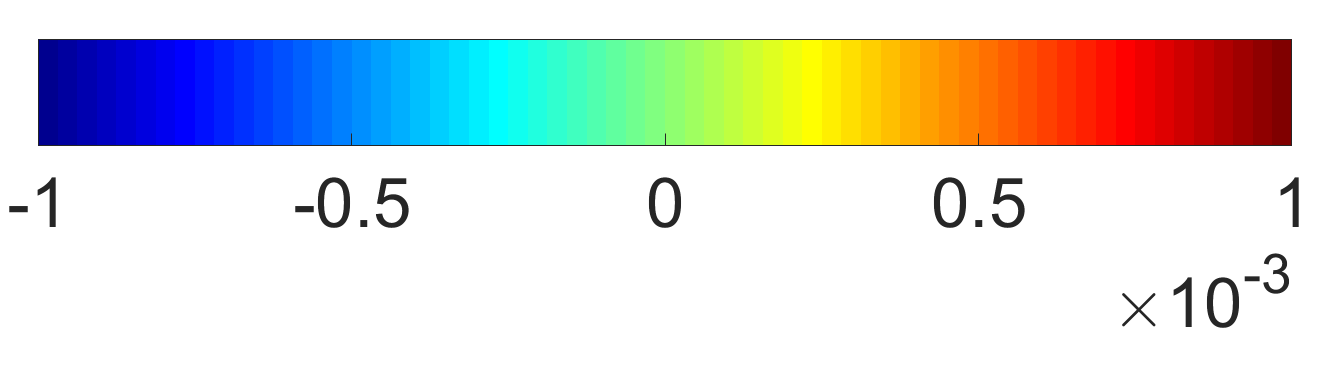}
\begin{multicols}{2}
	\centering
	\begin{subfigure}{0.45\textwidth}
	        \includegraphics[width=2.8in]{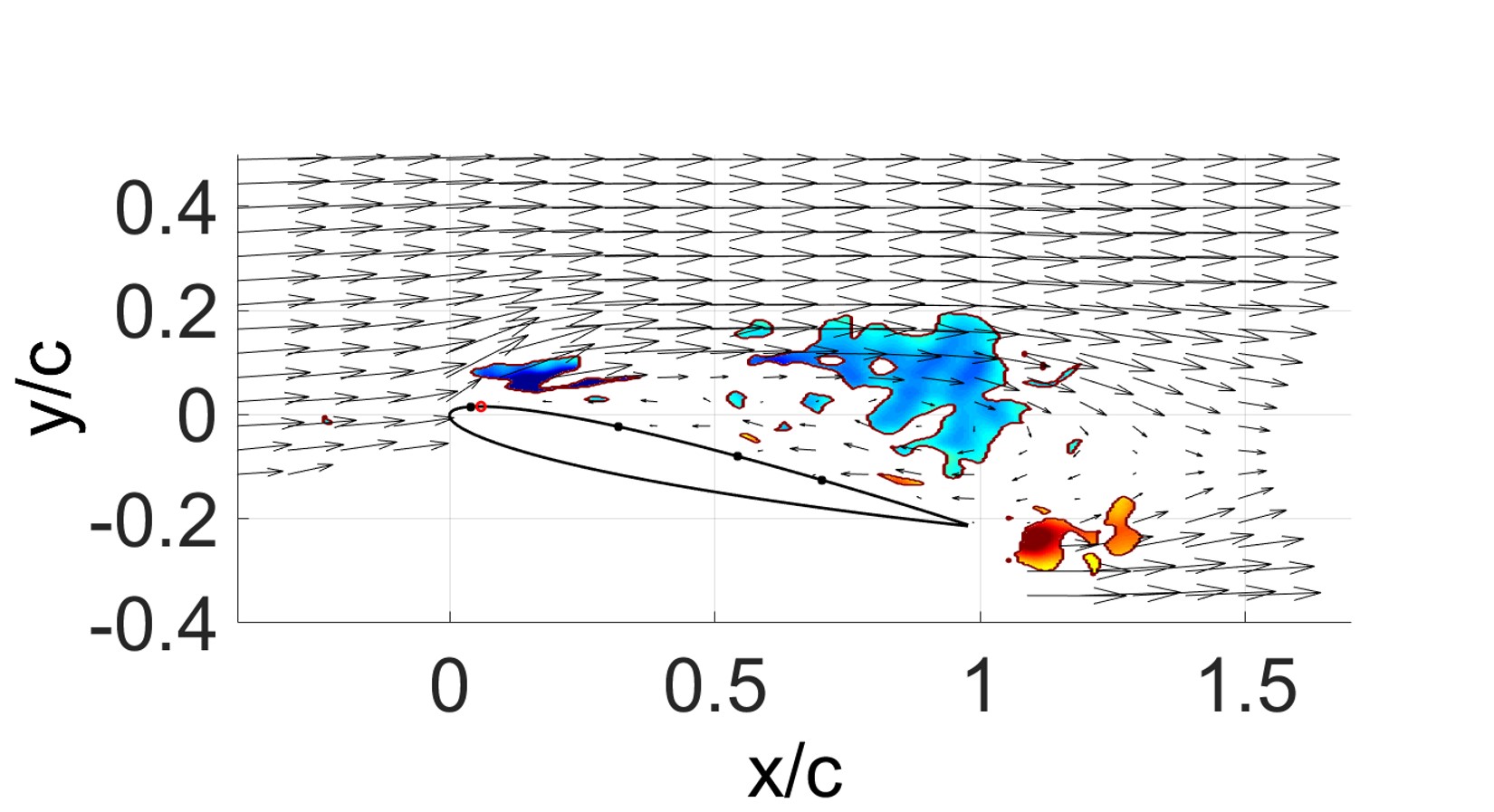}
			\caption{$0t^+$}
	          \label{fig:Gamma2_0}
	\end{subfigure}	
	
	\begin{subfigure}{0.45\textwidth}
	        \includegraphics[width=2.8in]{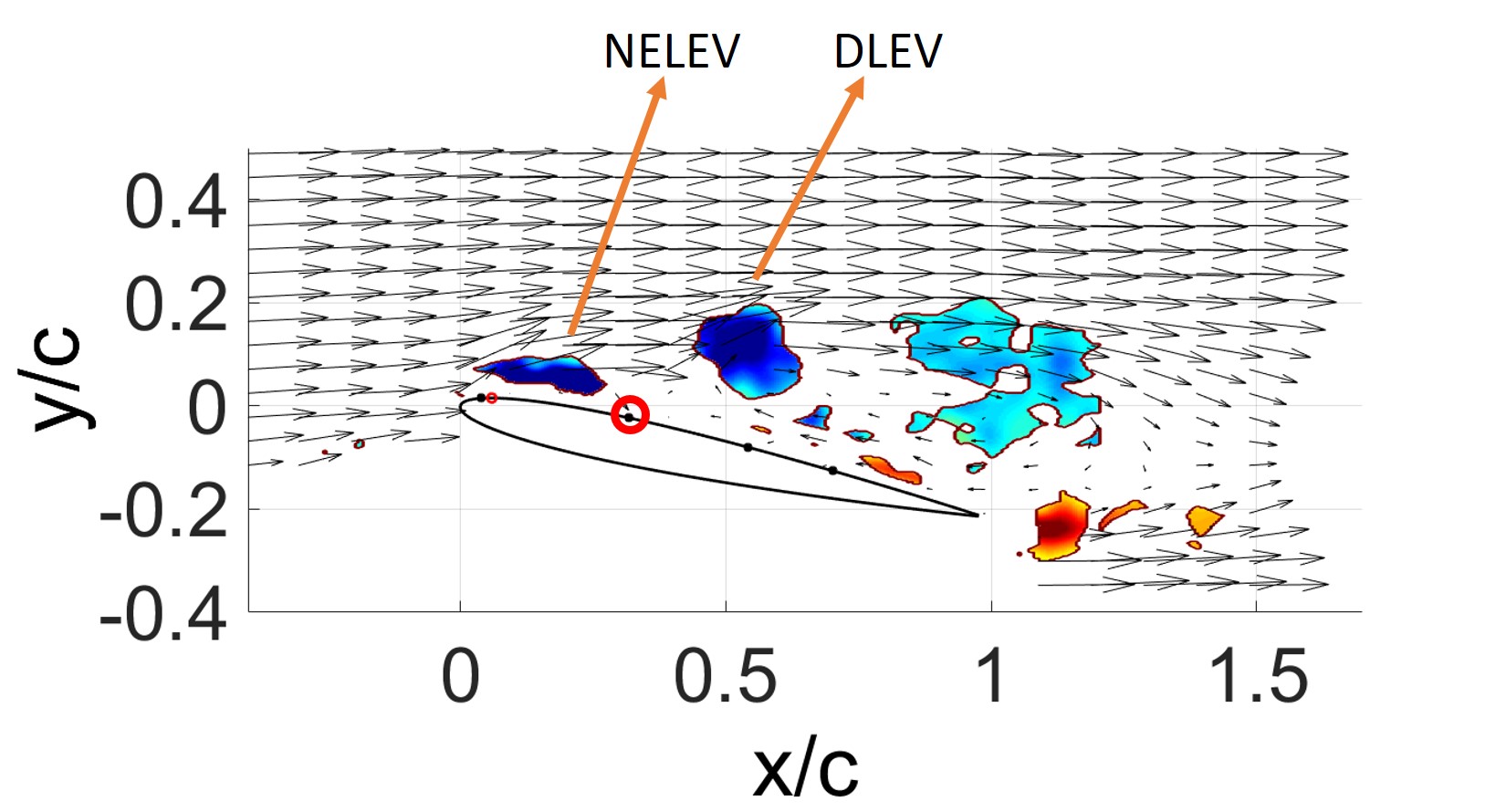}
	          \caption{$0.7t^+$}
	          \label{fig:Gamma2_0p7}
	\end{subfigure}	
	
	\begin{subfigure}{0.45\textwidth}
	        \includegraphics[width=2.8in]{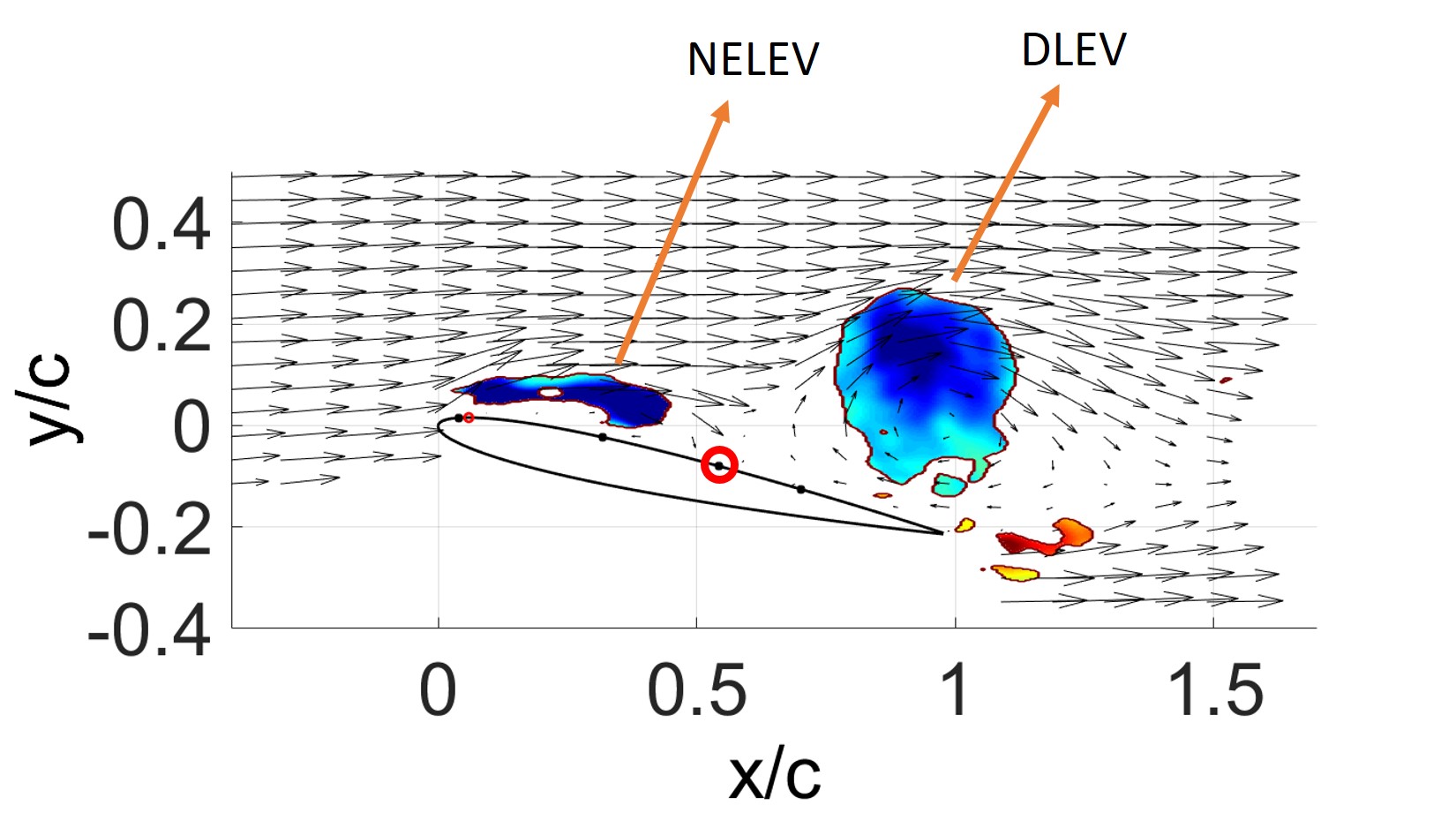}
	          \caption{$1.2t^+$}
	          \label{fig:Gamma2_1p2}
	\end{subfigure}	
	
	\begin{subfigure}{0.45\textwidth}
	        \includegraphics[width=2.8in]{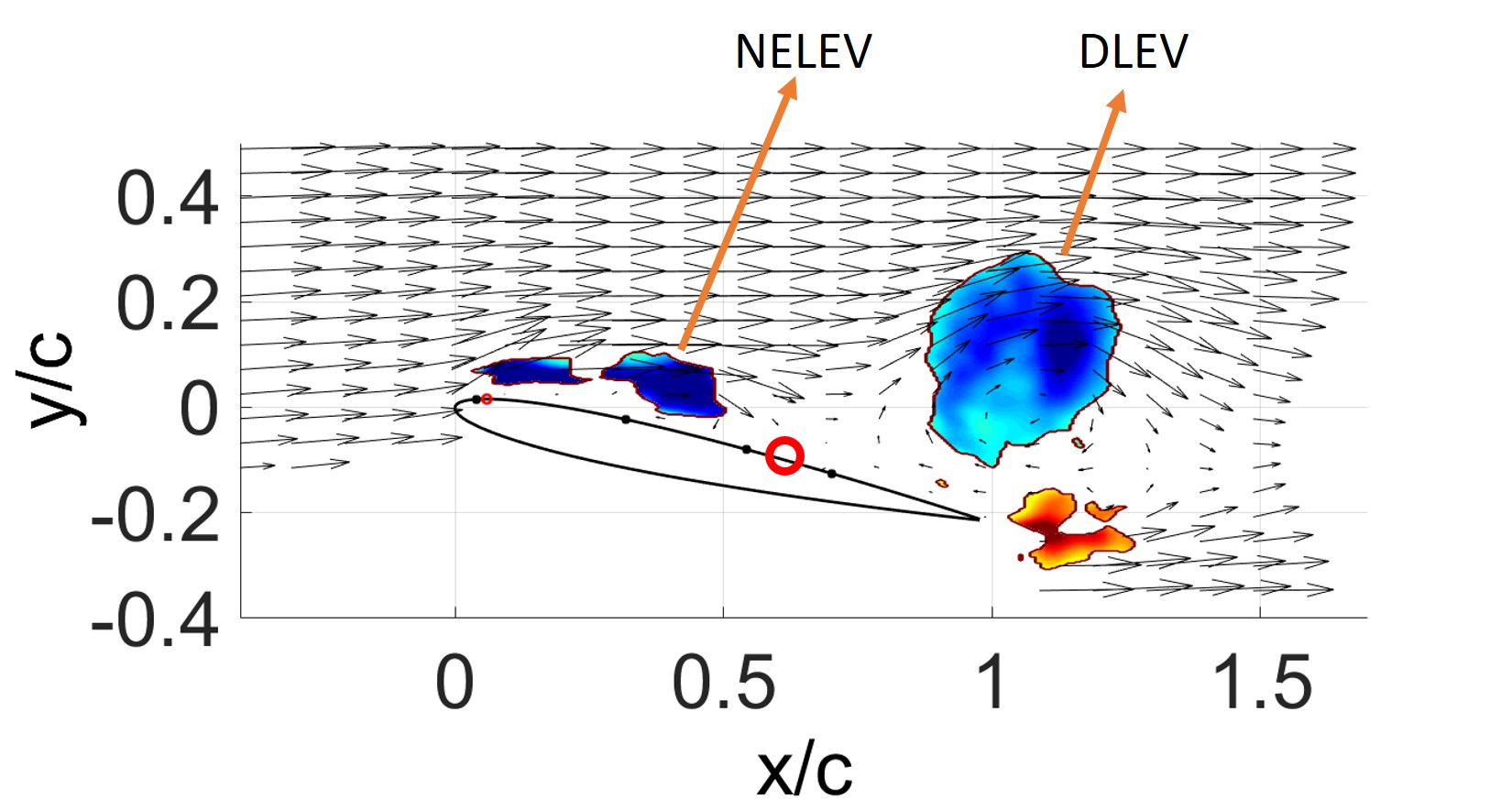}
	          \caption{$1.4t^+$}
	          \label{fig:Gamma2_1p4}
	\end{subfigure}	
	
	\begin{subfigure}{0.45\textwidth}
	        \includegraphics[width=2.8in]{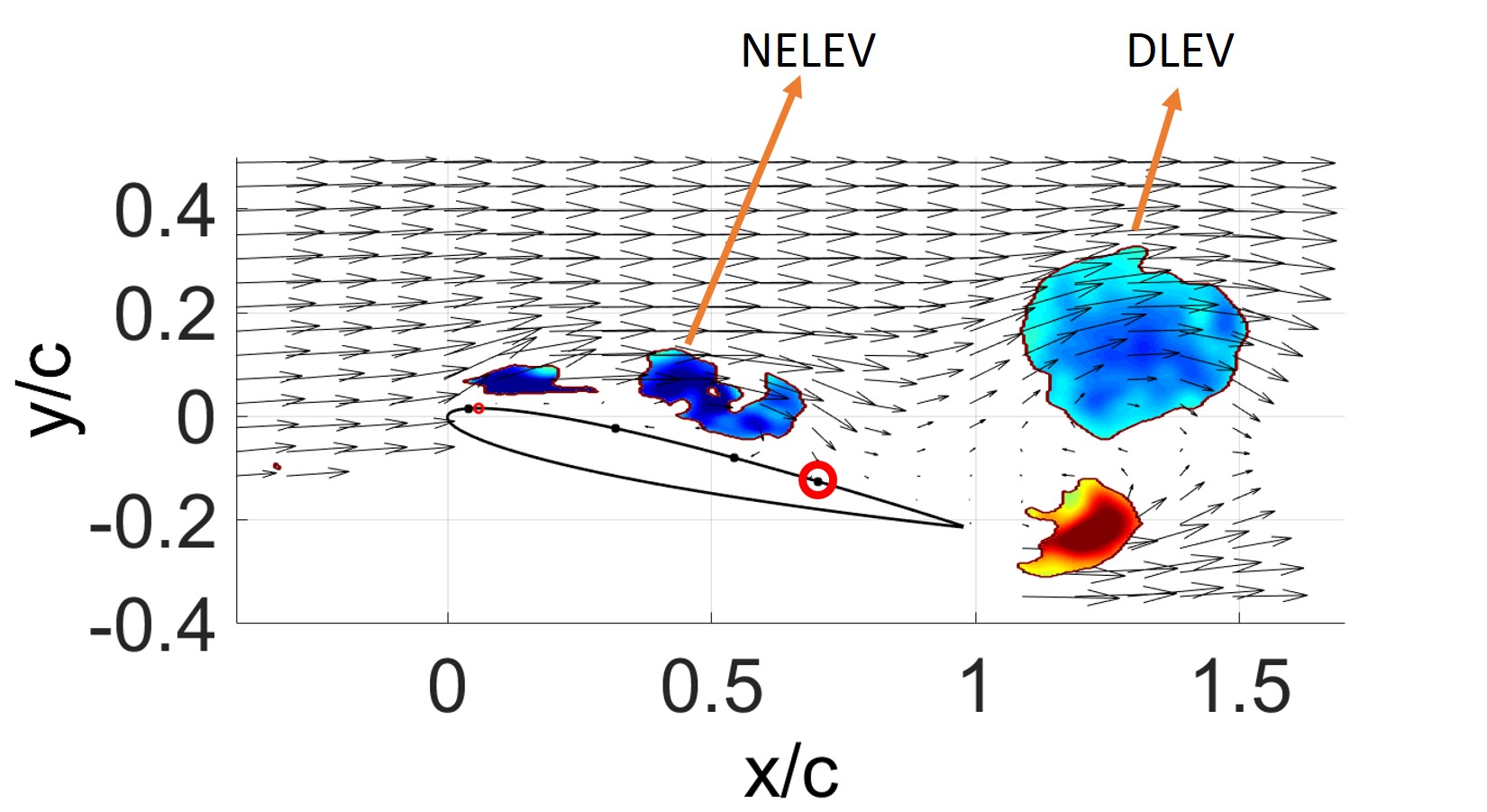}
	          \caption{$1.7t^+$}
	          \label{fig:Gamma2_1p7}
	\end{subfigure}
	
		\begin{subfigure}{0.45\textwidth}
		        \includegraphics[width=2.8in]{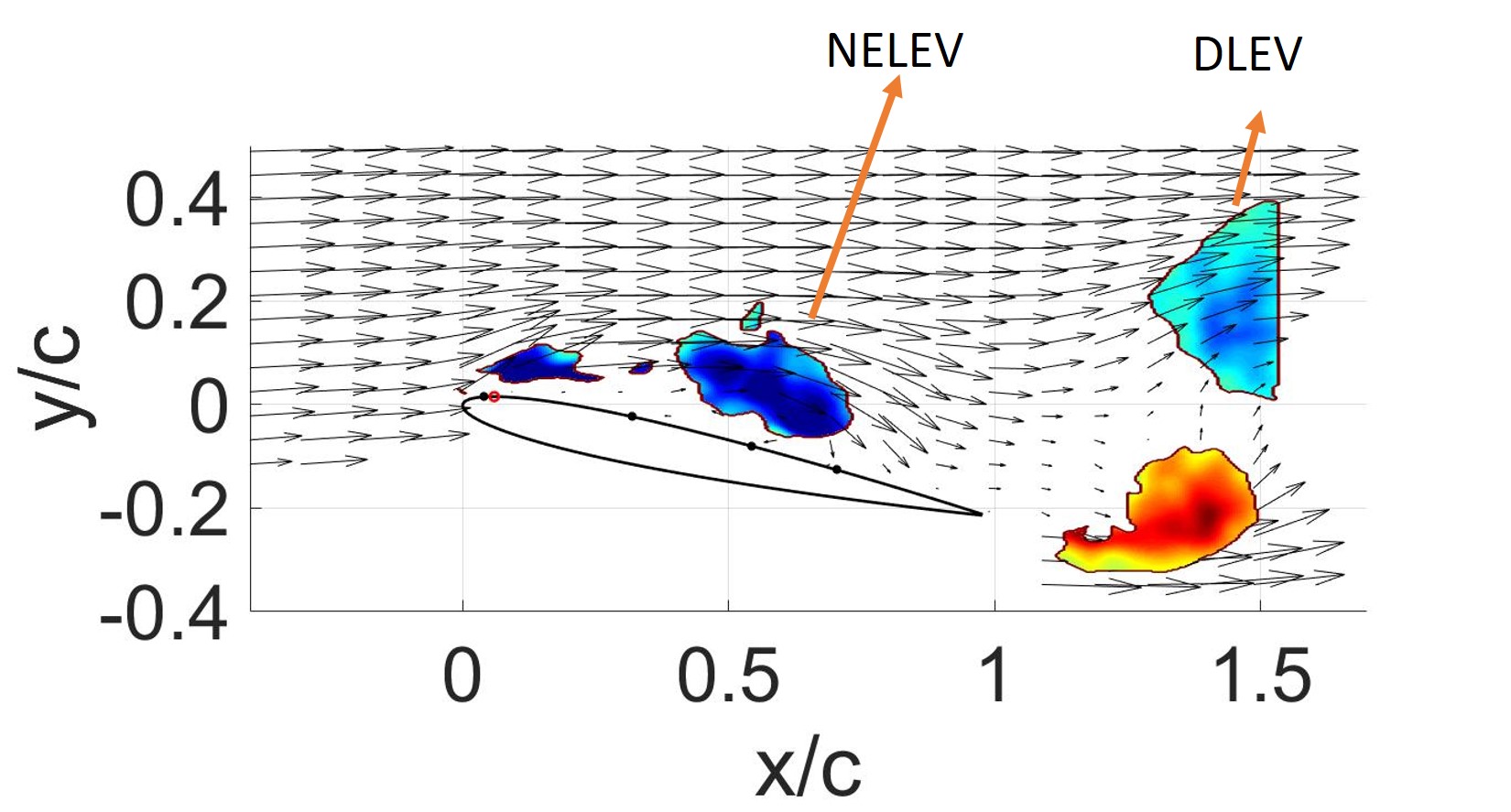}
		          \caption{$2t^+$}
		          \label{fig:Gamma2_2}
		\end{subfigure}
	
	\begin{subfigure}{0.45\textwidth}
	        \includegraphics[width=2.8in]{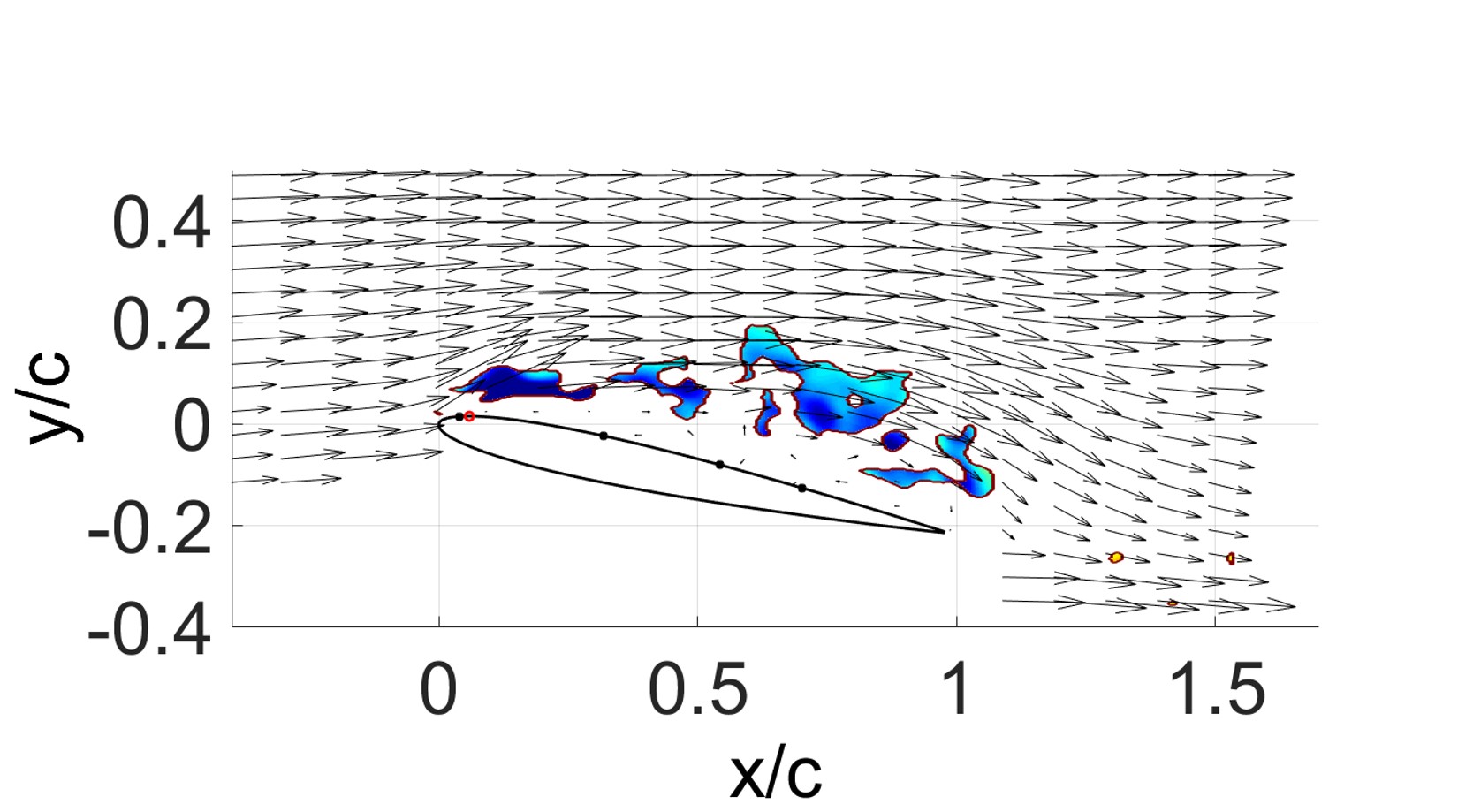}
	          \caption{$2.8t^+$}
	          \label{fig:Gamma2_2p8}
	\end{subfigure}		
	
	\begin{subfigure}{0.45\textwidth}
	        \includegraphics[width=2.8in]{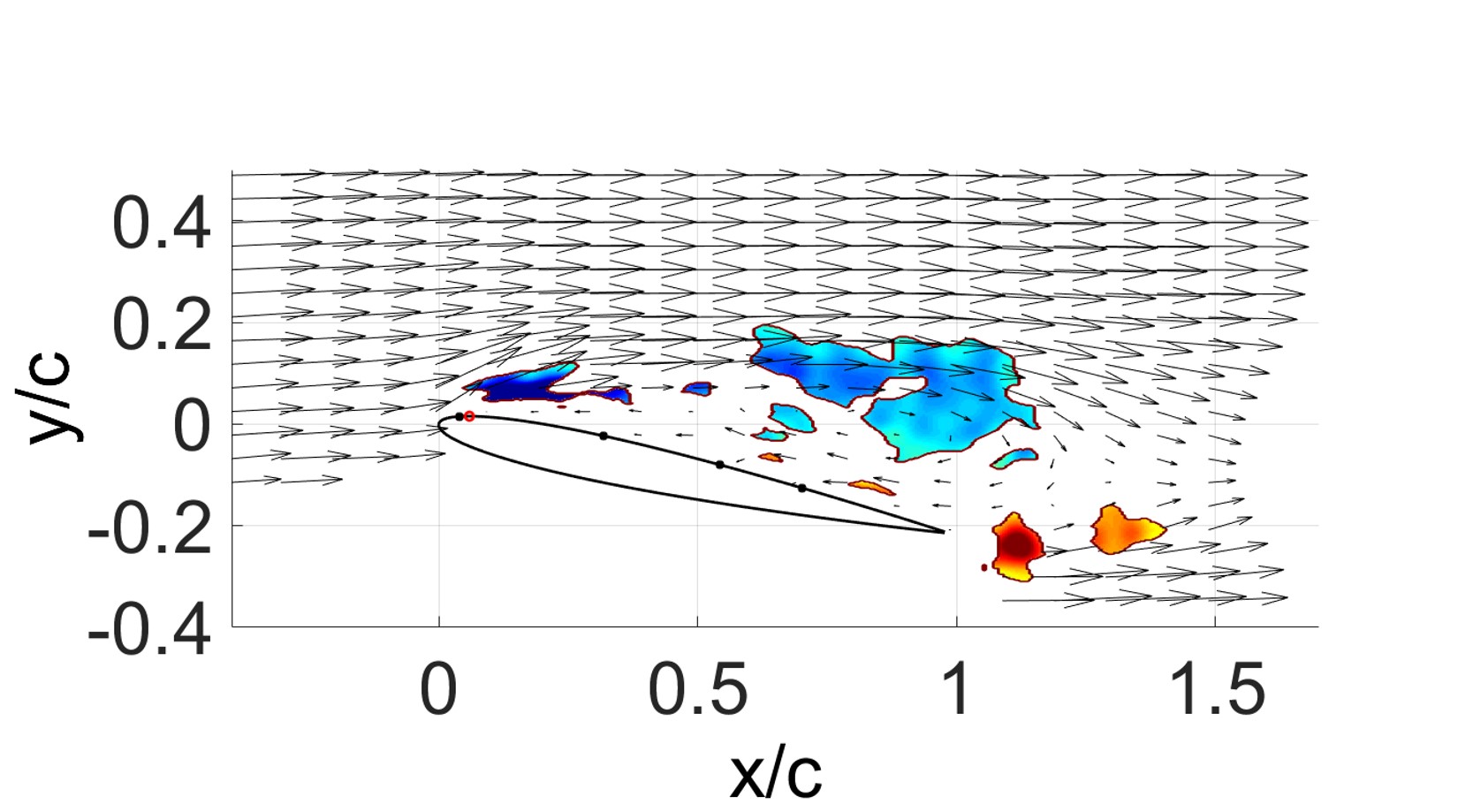}
		         \caption{$20t^+$}
              \label{fig:Gamma2_20}
	\end{subfigure}				


		

\end{multicols}

    \caption{Vorticity maps and velocity vectors following a single burst from the actuator. The red circles denote the location where the downwash impacts the wing. The red dot close to the leading-edge indicates the location of the actuators, and the black dots show the location of the pressure sensors. The plots from a to h represent baseline $(0t^+)$, maximum pressure on PS2 $(0.7t^+)$, maximum pressure on PS3 $(1.2t^+)$, minimum $C_L$ $(1.4t^+)$, maximum pressure on PS4 $(1.7t^+)$, vortices shed from the wing $(2t^+)$, maximum $C_L$ $(2.8t^+)$ and the baseline $(20t^+)$.}
    \label{fig:Gamma}		
\end{figure}

The vortex structures in figure \ref{fig:Gamma} are plotted at the `critical' instants that include the minimum $C_L$, maximum $C_L$, the downwash (which will be discussed in more detail later) acting on each pressure sensor and the original baseline flow. 
The vortex structure at $0t^+$ (figure \ref{fig:Gamma2_0}) shows a group of clockwise rotating vortices (blue) above the airfoil denoting the separated boundary layer, and some counterclockwise rotating vortices (red) at the vicinity of the trailing edge indicating the trailing-edge vortex (TEV). As shown in figure \ref{fig:Gamma2_0p7}, the single-burst actuation cuts the shear layer into two parts. The upstream part forms a new leading-edge vortex that bonds to the leading edge. We call this vortex the newly established leading-edge vortex (NELEV). The downstream portion of the shear layer rolls up into another vortex which eventually detaches from the airfoil.  The previously discussed `kink' in the shear layer is at the interface between the two vortex structures.  The size of the downstream vortex grows continuously before it completely detaches from the airfoil (figure \ref{fig:Gamma2_1p2} to \ref{fig:Gamma2_1p7}). This vortex is referred to as the detached leading-edge vortex (DLEV). At $2.8t^+$ (figure \ref{fig:Gamma2_2p8}), both the DLEV and the counter-clockwise rotating TEV have detached from the wing. The separation bubble becomes smaller than the baseline separated flow, and the resulting flow leaves the trailing edge smoothly. The reattachment point reaches the trailing edge at this time, and the lift increment reaches its maximum as previously shown in figure \ref{fig:flow_field}d and figure \ref{fig:single_CL_CM}d. Finally, as expected, at $20t^+$ after the actuation (figure \ref{fig:Gamma2_20}) the flow has returned to the original baseline state.

\FloatBarrier

The NELEV vortex structure in figure \ref{fig:Gamma} induces a downward flow that correlates with the pressure reversal at each pressure sensor following the burst actuation. The location, where the downwash acts on the suction side surface of the wing is shown in figure \ref{fig:Gamma} b-e with red circles. At $0.7t^+$ after the firing of the burst (figure \ref{fig:Gamma2_0p7}), the downwash flow is acting on the pressure sensor 2, and this pressure sensor is showing maximum pressure reading (figure \ref{fig:CL_CM_pressure}). The peak in the pressure reversal occurs at pressure sensors 3 and 4 (figure \ref{fig:Gamma2_1p2} and figure \ref{fig:Gamma2_1p7}) at slightly later times as the NELEV convects downstream towards the trailing edge of the airfoil. 

Therefore, by combining the results of the $\Delta C_L$ and $\Delta C_M$ measurements, the surface pressure measurements (figure \ref{fig:CL_CM_pressure} and figure \ref{fig:DCL_single_pressure}) and the vortex structure inside the flow field (figure \ref{fig:Gamma}), we conclude that the NELEV induces a downwash velocity that acts on the suction side of the airfoil and moves downstream. This downwash contributes to a local pressure reversal and as a consequence, the local pressure reversal leads to the lift and pitching moment reversal following the single-burst actuation. It is also worth pointing out that it takes about $2t^+$ (figure \ref{fig:Gamma2_2}) for the leading edge vortex to detach and convect downstream into the wake, which leads to the $\Delta C_L$ increase. But it takes about $10t^+$ for the "re-separation" process (figure \ref{fig:flow_field}e to figure \ref{fig:flow_field}f) to occur, which is responsible for the $C_L$ returning to its initial condition. 


\subsection{Proper Orthogonal Decomposition (POD) of the flow field}

Next we apply the Proper Orthogonal Decomposition (POD) to the transient case of the single burst actuation to gain additional insight into the modes that are responsible for the lift and pitching moment reversal. 
In a similar experiment, \citet{monnier2016comparison} reported that the temporal coefficient of the second Proper Orthogonal Decomposition (POD) mode correlates with the negative of the lift coefficient variation following a single-burst input. The POD method can reduce a large number of interdependent variables to a much smaller number of independent modes, while retaining as much as possible the variation in the original variables \citep{kerschen2005method}.

	\begin{equation} \label{eq:51}
	v(x,t)=\sum\limits_{i=1}^{\infty}a_i(t)\phi_i(x)
	\end{equation}	

In Eq. \ref{eq:51} $a_i$ is the time-dependent temporal coefficient and $\phi_i(x)$ is the POD basis function.
Singular Value Decomposition (SVD) \citep{kerschen2005method} is performed on both horizontal and vertical velocity components obtained from the PIV measurements. 

For any given $(m\times n)$ matrix $X$
    \begin{equation} \label{eq:52}
    X=USV^T
    \end{equation}

where $U$ is an $(m\times m)$ orthonormal matrix containing the left singular vectors, $S$ is a $(m\times n)$ pseudo-diagonal and semi-positive definite matrix with diagonal elements $\delta_i$, and $V$ is an $(n\times n)$ orthonormal matrix containing the right singular vectors. There are physical meanings for each term from the SVD. The matrix $U$ represents the spatial distribution of velocity within each POD mode, $V$ contains the temporal coefficients for each mode, and the pseudo-diagonal elements of $S$ denote the energy level for the modes, in which the energy is descending with increasing of the mode number.        

\begin{figure}
	\centering
    \includegraphics[width=0.8\textwidth]{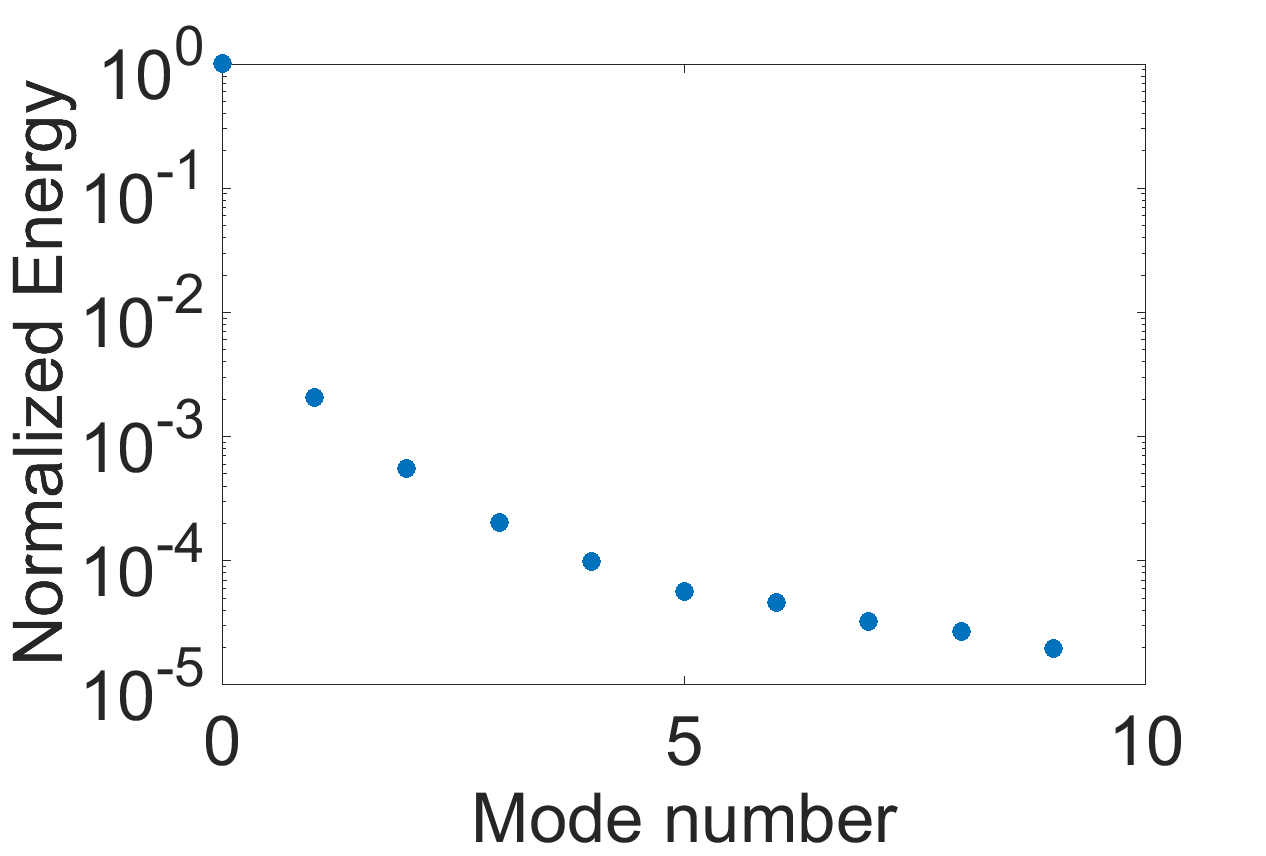}
    \caption{Energy distribution versus POD mode number.} 
    \label{fig:POD_energy} 
\end{figure}

The energy contained in each mode is normalized by the first mode's energy and plotted in figure \ref{fig:POD_energy}. It shows that about 84\% of the disturbed flow energy (with mode 0 subtracted from the flow field) is contained in mode 1 and mode 2, which means that mode 1 and mode 2 can reconstruct a flow field containing most of the energetic structures in the actual disturbed flow field.

\begin{figure}
	\centering

	\begin{subfigure}{0.45\textwidth}
	        \includegraphics[width=2.6in]{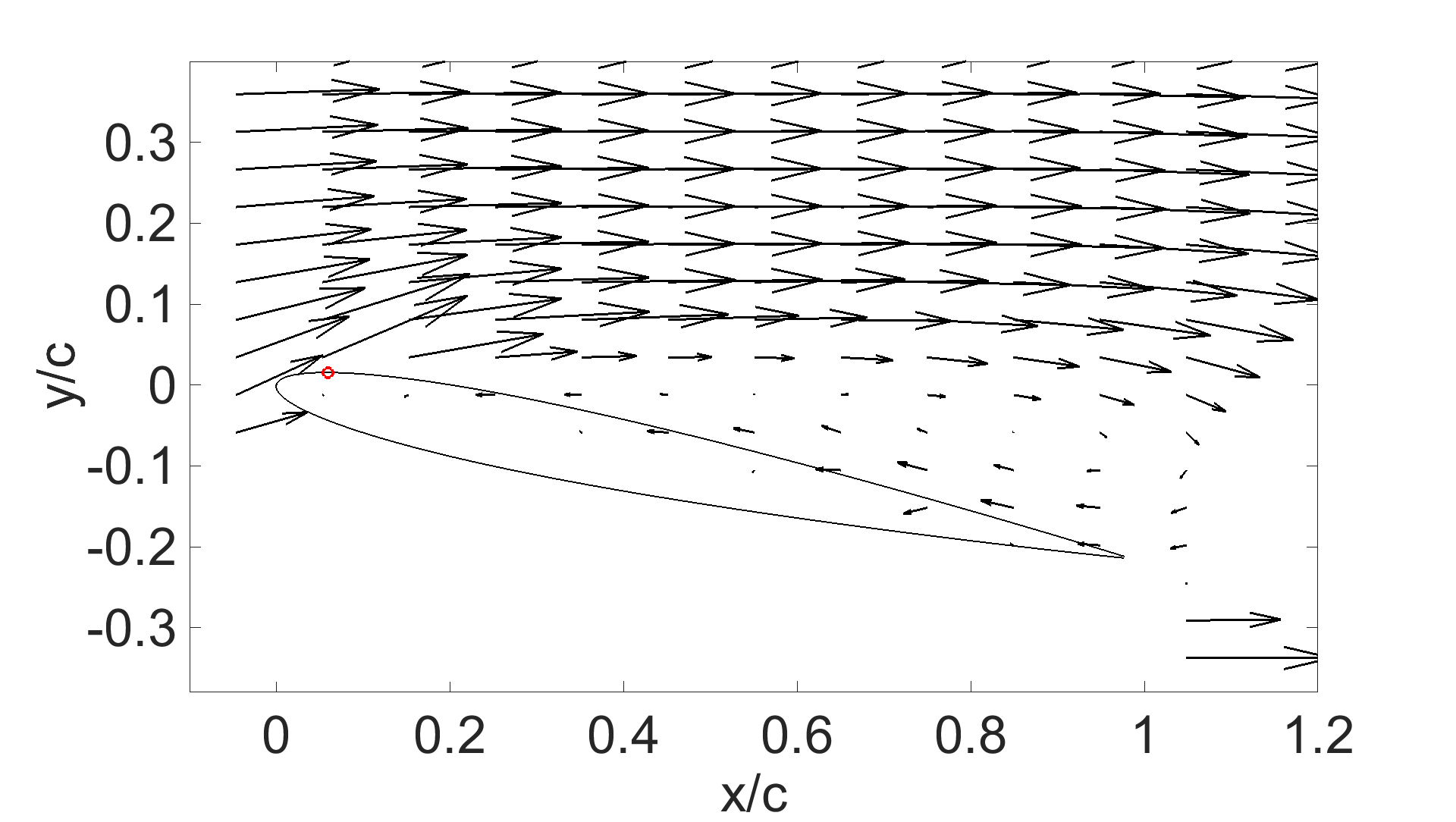}
			\caption{Mode 0}
	          \label{fig:POD_S_0}
	\end{subfigure}	
	~
	\begin{subfigure}{0.45\textwidth}
	        \includegraphics[width=2.6in]{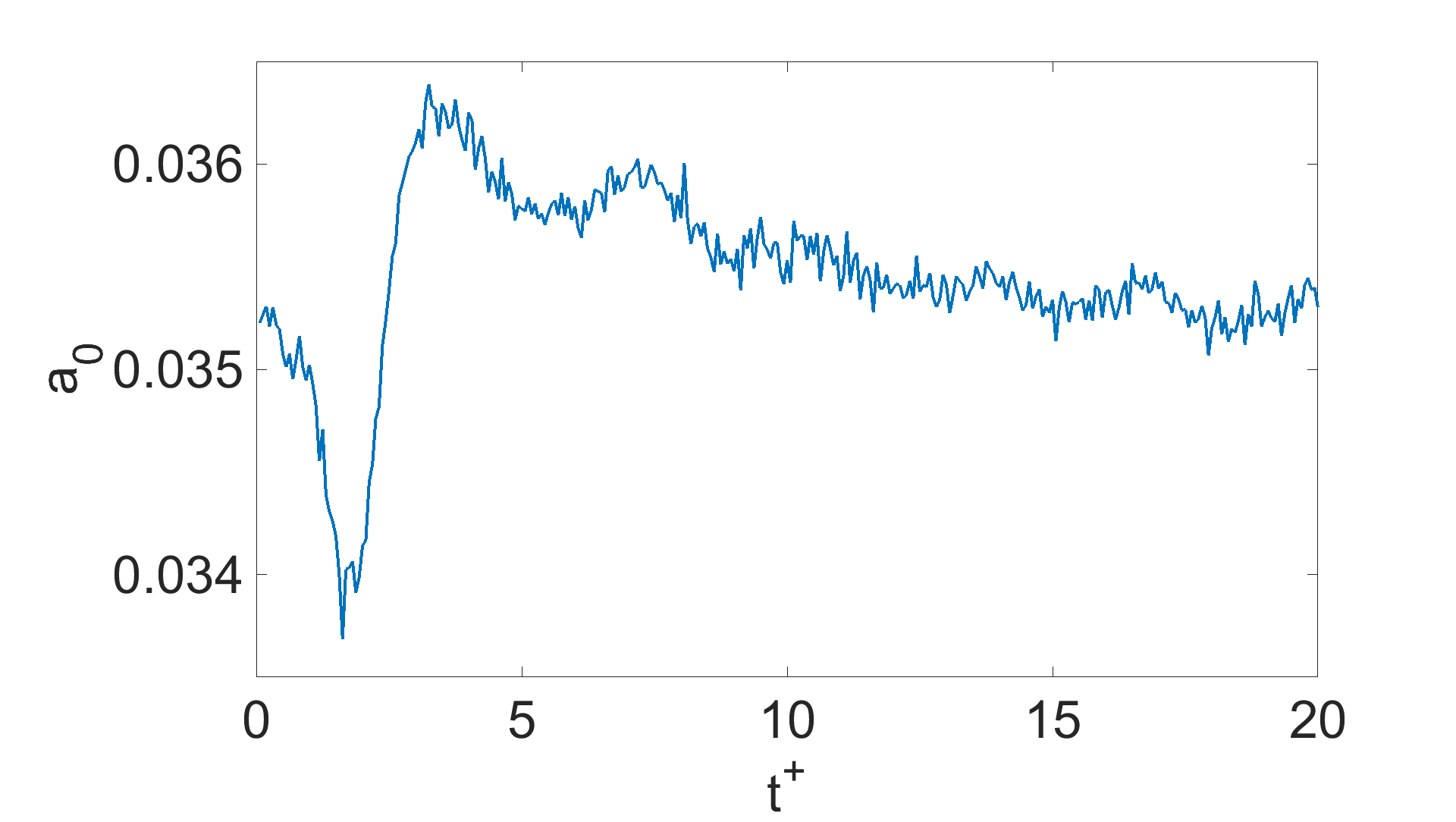}
	          \caption{Temporal coefficient of Mode 0}
	          \label{fig:POD_T_0}
	\end{subfigure}	
	
	\begin{subfigure}{0.45\textwidth}
	        \includegraphics[width=2.6in]{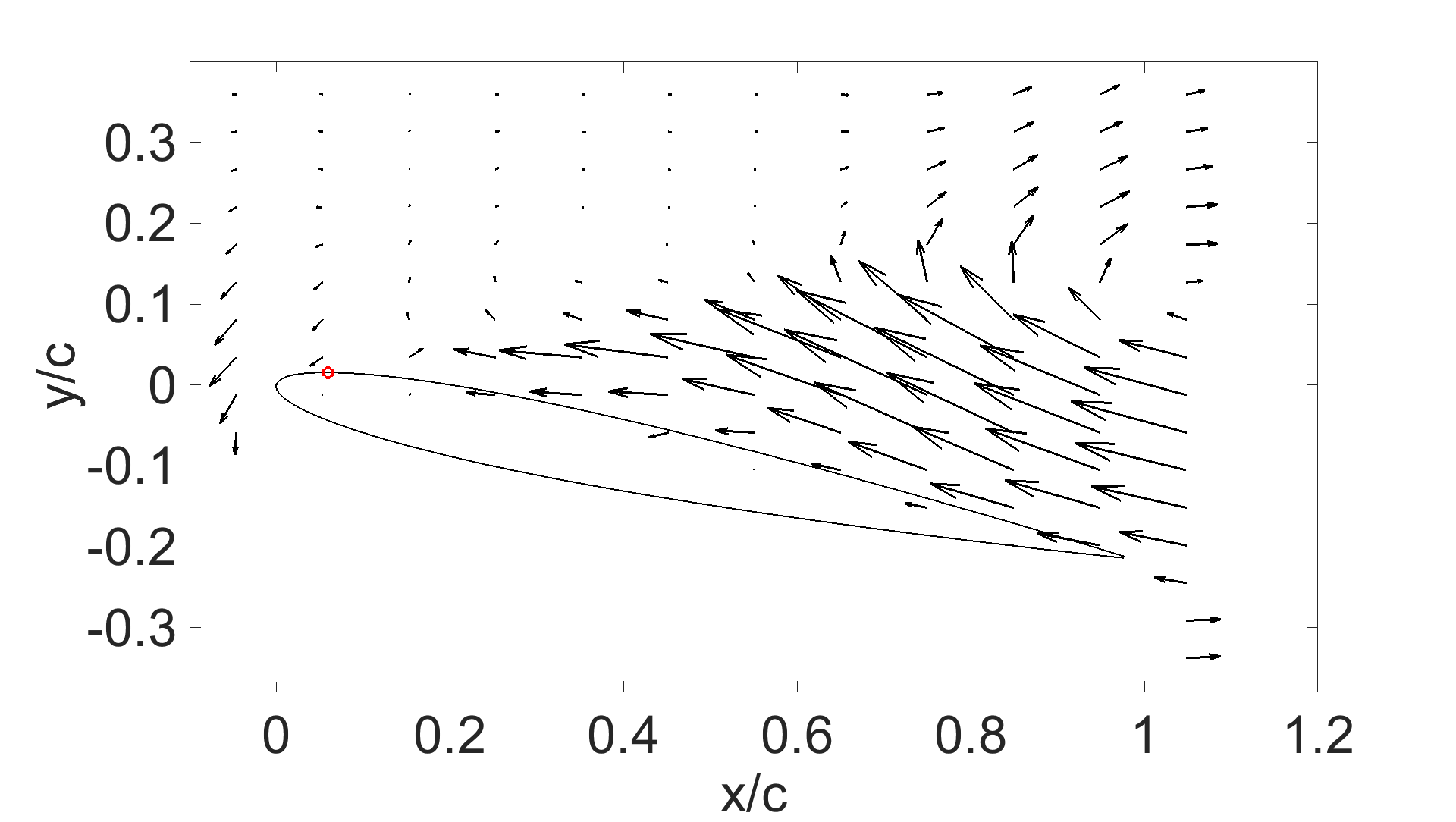}
			\caption{Mode 1}
	          \label{fig:POD_S_1}
	\end{subfigure}	
	~
	\begin{subfigure}{0.45\textwidth}
	        \includegraphics[width=2.6in]{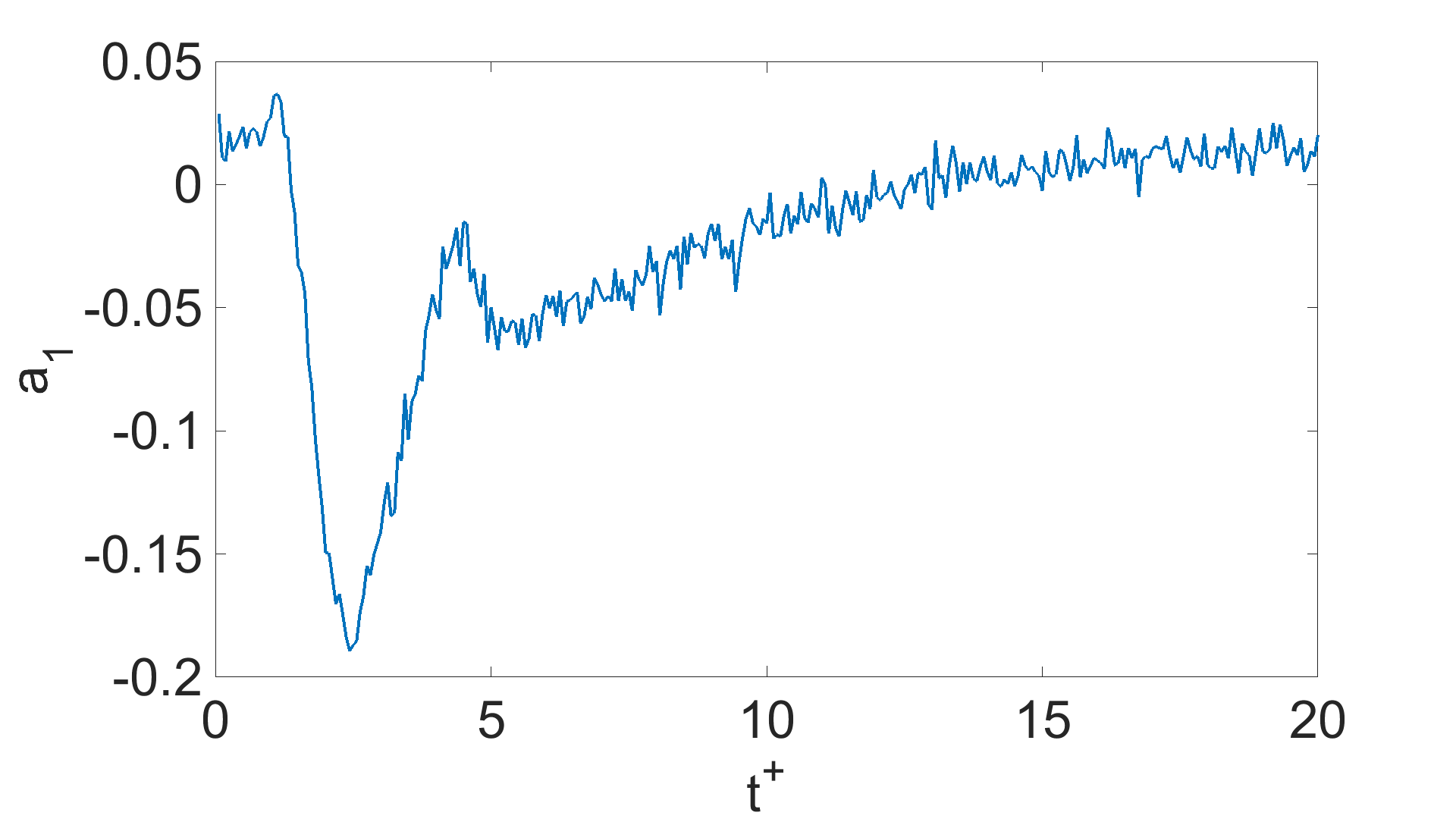}
	          \caption{Temporal coefficient of Mode 1}
	          \label{fig:POD_T_1}
	\end{subfigure}	

	\begin{subfigure}{0.45\textwidth}
	        \includegraphics[width=2.6in]{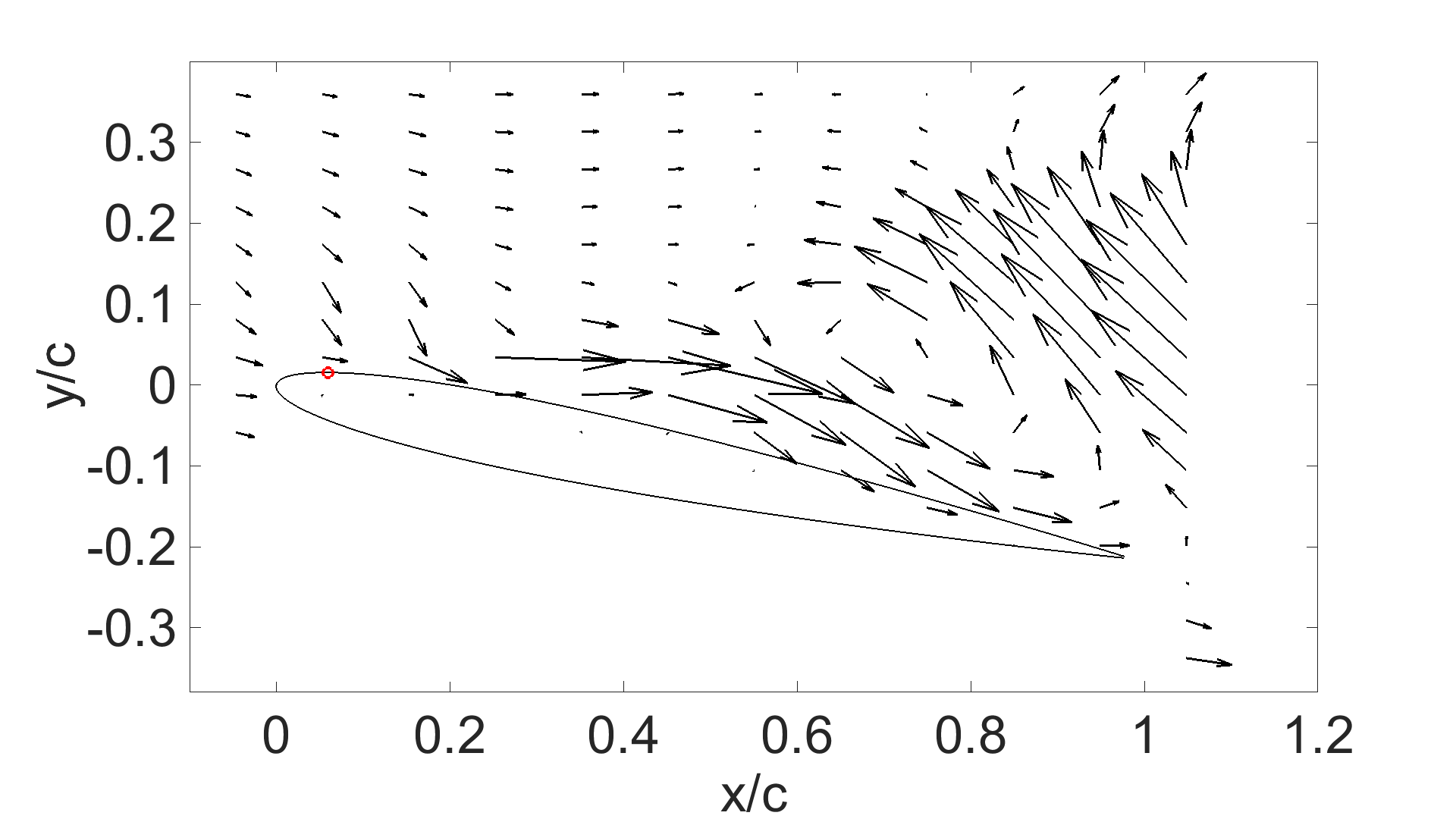}
			\caption{Mode 2}
	          \label{fig:POD_S_2}
	\end{subfigure}	
	~
	\begin{subfigure}{0.45\textwidth}
	        \includegraphics[width=2.6in]{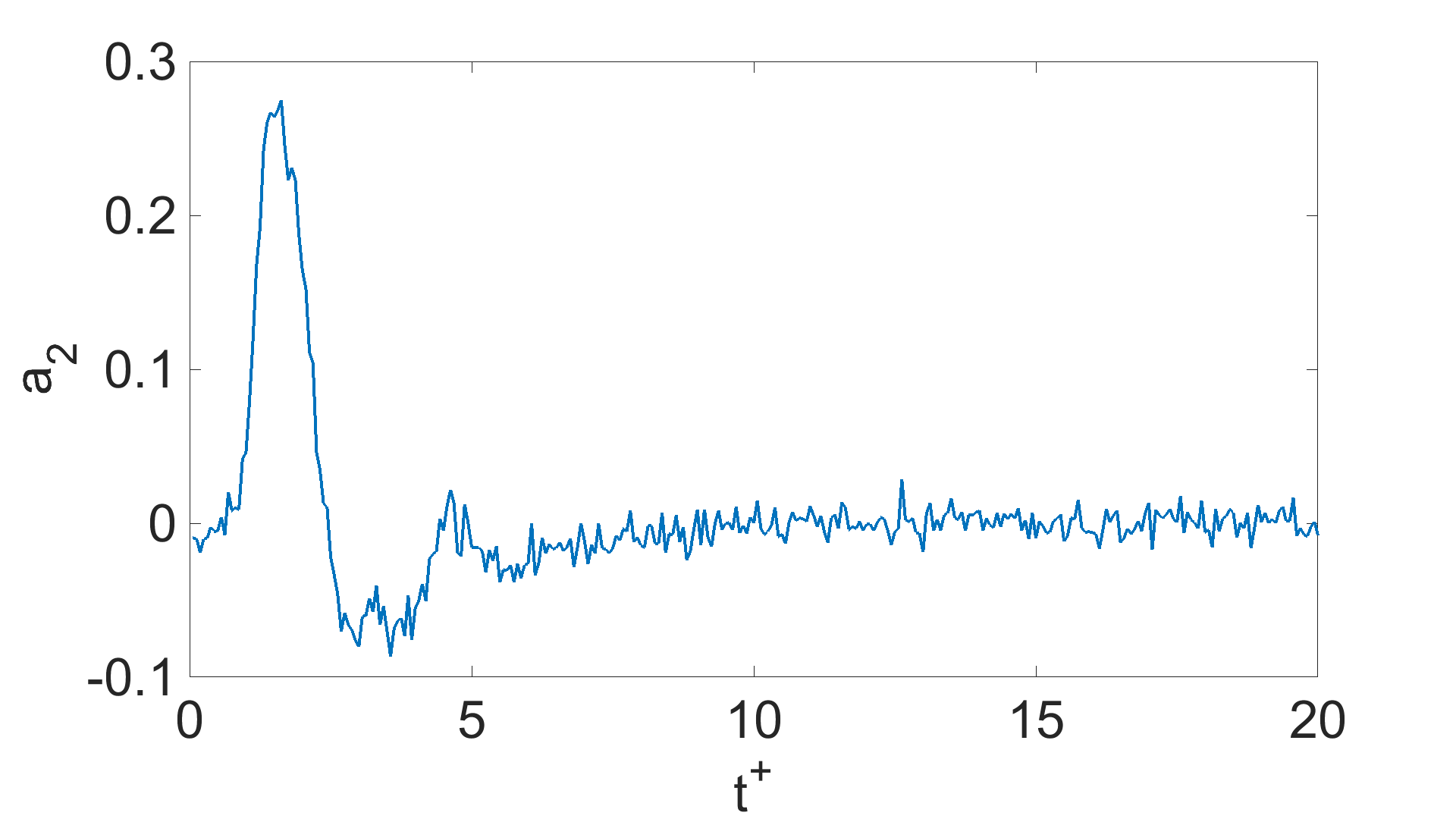}
	          \caption{Temporal coefficient of Mode 2}
	          \label{fig:POD_T_2}
	\end{subfigure}

				
   	   	
    \caption{Comparison of the first four spatial POD modes and their corresponding temporal coefficients.}
    \label{fig:POD}		
\end{figure}

The corresponding basis functions and the temporal coefficients for the first three modes are plotted in figure \ref{fig:POD}. Mode 0 (figure \ref{fig:POD_S_0}) shows the baseline separated flow. And figure \ref{fig:POD_T_0} shows that this mode almost remains unchanged with time. The temporal coefficient of this mode is always above 0 which means the velocity vector direction in figure \ref{fig:POD_S_0} will not be flipped. Mode 1 shown in figure \ref{fig:POD_S_1} is related to a clockwise rotating vortex above the trailing edge and the backward flow that causes the boundary to reattach. The temporal coefficient (figure \ref{fig:POD_T_1}) of this mode goes negative after the burst is fired, which implies that the velocity vectors in the spatial mode flip their directions. Mode 2 shown in figure \ref{fig:POD_S_2} indicates a counterclockwise rotational flow pattern upstream of the trailing edge and a clockwise rotational vortex which bonds to the mid chord on the suction side. This flow structure produces a downwash impinging on the upper surface of the wing close to the pressure sensor 4. The temporal coefficient of this mode (figure \ref{fig:POD_T_2}) goes above 0 following the burst, which preserve the direction of the velocity vectors in figure \ref{fig:POD_S_2}. This implies that mode 2 is related to the $C_L$ and $C_M$ reversal. Next, we will exhibit the relation between the $C_L$ and $C_M$ and the POD modes.      


 \FloatBarrier

\FloatBarrier

\begin{figure}
	\centering
	
		\begin{subfigure}{0.45\textwidth}
		        \includegraphics[width=2.6in]{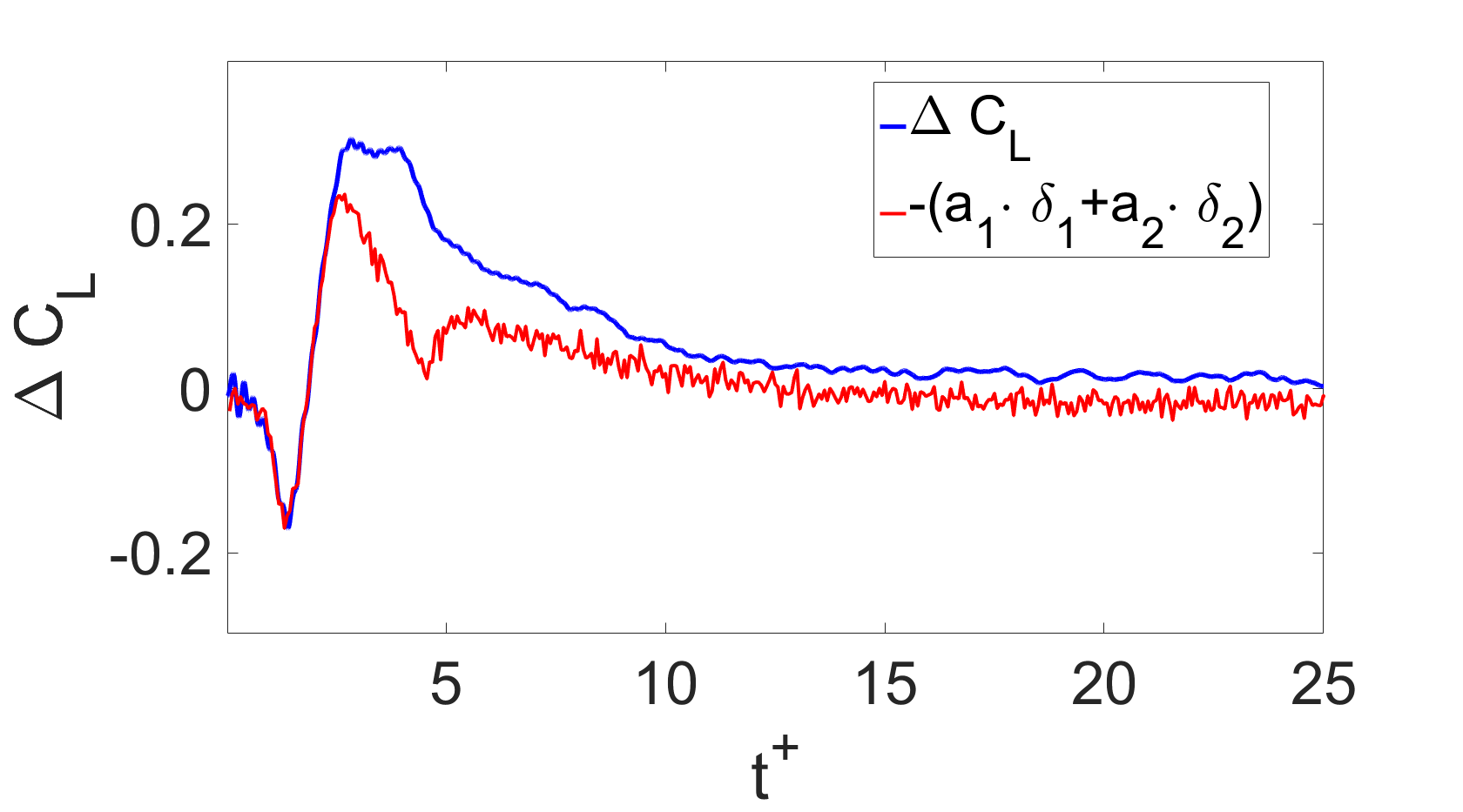}
		          \caption{Sing-burst actuation $\Delta C_L$}
		          \label{fig:DCL_POD}
		\end{subfigure}
~		
		\begin{subfigure}{0.45\textwidth}
		        \includegraphics[width=2.6in]{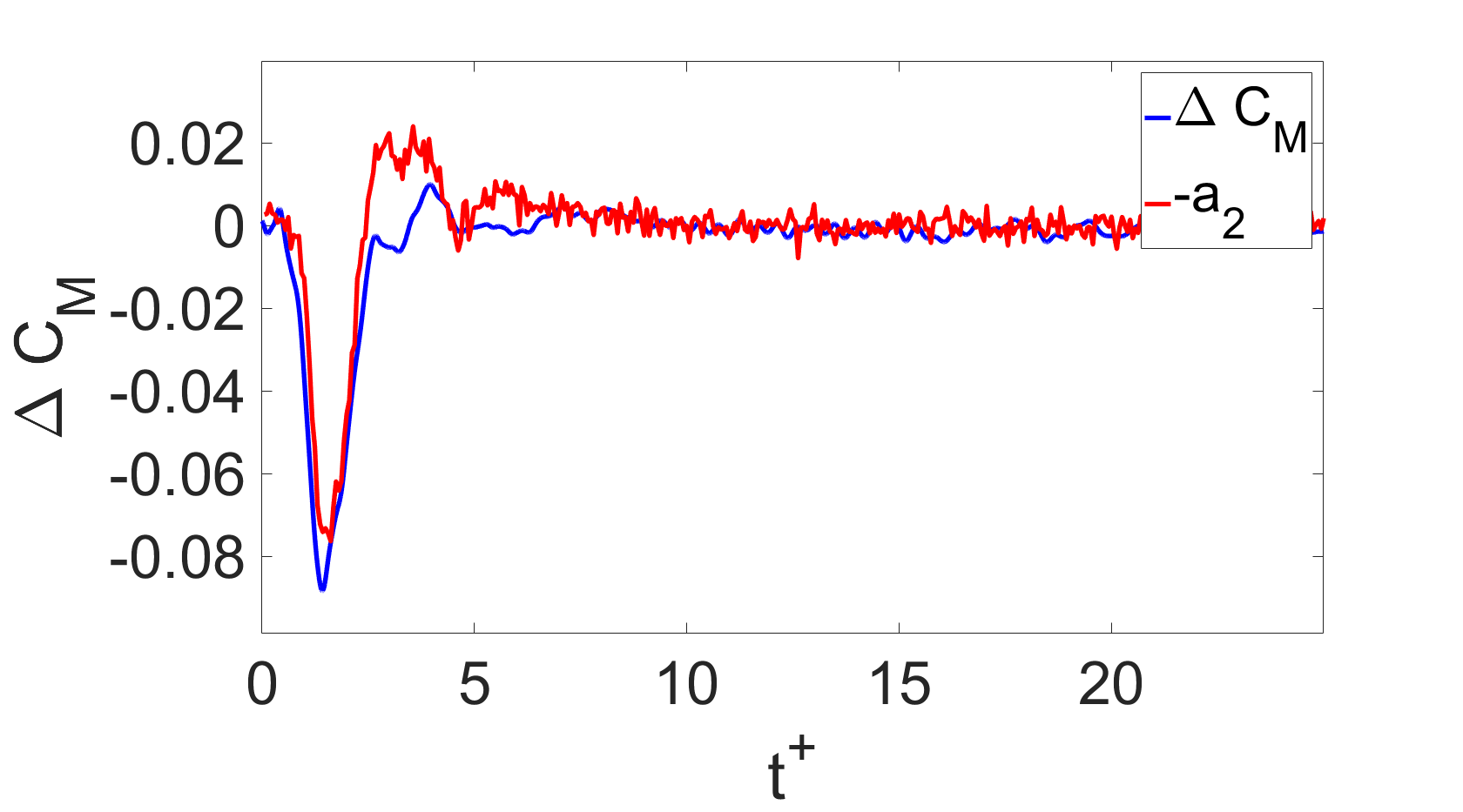}
		          \caption{Single-burst actuation $\Delta C_M$}
		          \label{fig:DCM_POD}
		\end{subfigure}


    \caption{Comparison of aerodynamic loads and time varying coefficient of POD modes, (a) lift coefficient, (b) pitching moment coefficient.}
    \label{fig:CL_CM_POD}		
\end{figure}

The sum of the temporal coefficients of mode 1 and mode 2 is plotted in figure \ref{fig:CL_CM_POD}a. Comparing the $\Delta C_L$ (baseline $C_L$ subtracted from the transient $C_L$) to $a_1\cdot \delta_1+a_2\cdot \delta_2$, the combined temporal coefficient tracks the negative $\Delta C_L$ quite well (especially for the lift reversal), other than the second mode alone, reported by \citet{monnier2016comparison}. 
The negative temporal coefficient of mode 2 is closely tracking the measured $\Delta C_M$, since this mode represents the downwash acting on pressure sensor 4. This can be seen in figure \ref{fig:CL_CM_POD}b.  
\FloatBarrier

\subsection{Stability Analysis of the flow field}

According to \citet{raju2008dynamics} there are three instability mechanisms within the separated flow that govern its response to disturbances; namely, the Kelvin-Helmholtz shear layer, the separation bubble, and the wake instability. The actuator injects a small-amplitude, spatially-localized disturbance into the separating shear layer.  In the single pulse case, the disturbance would have a broad spectrum. Based on the findings by \citet{raju2008dynamics} we surmise that the initial growth of the disturbance will follow linear dynamics that result from one or more instabilities. Nonlinear interactions and saturation of the disturbance will ultimately limit the maximum amplitude of the disturbance, which is reflected in the lift response. 

To investigate the possibility of linear instability mechanisms controlling disturbance development within the separated flow, the single-burst disturbance is studied utilizing the dynamic mode decomposition (DMD) \citep{rowley2009spectral} \citep{schmid2010dynamic} based on the kinetic energy within the PIV window. The single-burst actuation triggers a transient disturbance within the flow field, which is not a typical application of DMD \citep{rowley2009spectral}. Therefore, in order to determine the DMD modes corresponding to the kinetic energy growth within the flow field, DMD is performed only on the "near-equilibrium" region \citep{chen2012variants} of the snapshot kinetic energy ($KE_{snap}$). The "near-equilibrium" here refers to a partial regime of the transient state connecting one equilibrium state (the initial undisturbed separated flow)  to another equilibrium state (nonlinear saturation), where the oscillation amplitude of this transient state is growing or decaying exponentially in time.    

\begin{figure}
\centering
\includegraphics[width=.8\textwidth]{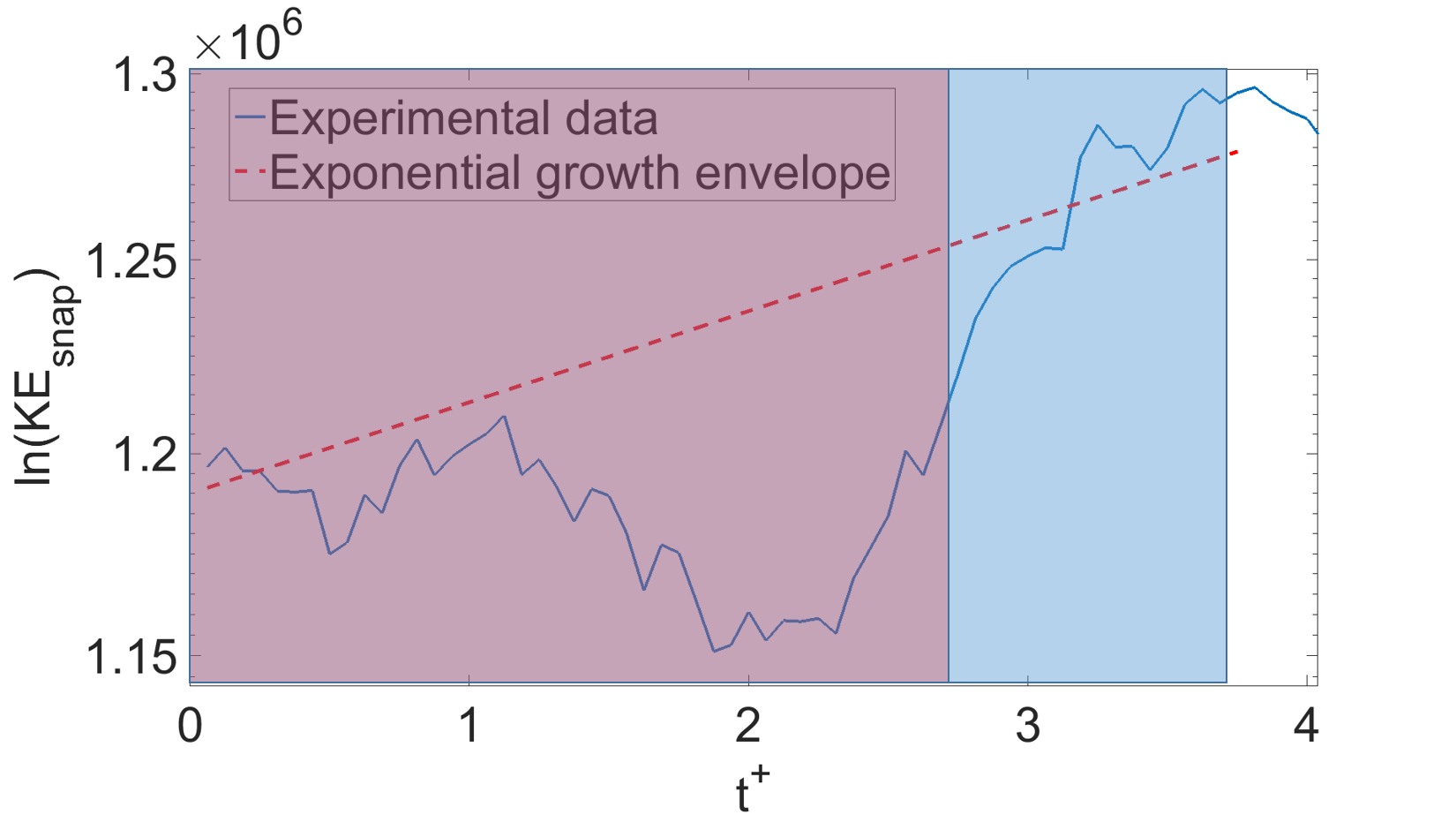}
\caption{Snapshot kinetic energy, The blue window indicates the end of the exponential growth (near-equilibrium) region, and DMD is performed within the red window.}
\label{fig:KE}
\end{figure}

The definition of the kinetic energy density ($KE$) and the snapshot kinetic energy ($KE_{snap}$) in the PIV measurement window are given as follows,
\begin{equation}
\label{eq:KE}
KE(i,j,k)=\frac{M}{2}(U(i,j,k)^2+V(i,j,k)^2)
\end{equation}
where $KE$ is the kinetic energy density, $U$ is the horizontal velocity and $V$ is the vertical velocity. The index $i,j$ denote the coordinates in the 2-D measurement window, $k$ is the snapshot number or discretized time series and $M$ is the fluid density. Note that $M$ is a constant number and does not change through out the entire space and time, therefore we let $M$ equals to 1 without losing any generality.     
\begin{equation}
\label{eq:KE_snap}
KE_{snap}(k)=\sum\limits_{i=1,j=1}^{m,n}KE(i,j,k)
\end{equation}
$KE_{snap}(k)$ is the snapshot kinetic energy, which is the total kinetic energy contained in the entire measurement window at each PIV snapshot $k$. The indices $m=400$ and $n=400$ denote the total number of the spatial points in horizontal and vertical direction, respectively. Here we just let each area represented by $i$, $j$ to be 1 since all the grid points are equally spaced and does not change through out the entire space and time. Note that DMD analysis is performed on the kinetic energy density, $KE$. The snapshot kinetic energy $KE_{snap}$ is only used to determine the "near-equilibrium" region. From figure \ref{fig:KE}, one can tell that the oscillation amplitude of the snapshot kinetic energy grows exponentially from $0t^+$ right after the burst signal until $3.3t^+$. Therefore, the DMD was performed on the kinetic energy density from $0t^+$ to $2.7t^+$. 

\begin{figure}
\centering
\includegraphics[width=.6\textwidth]{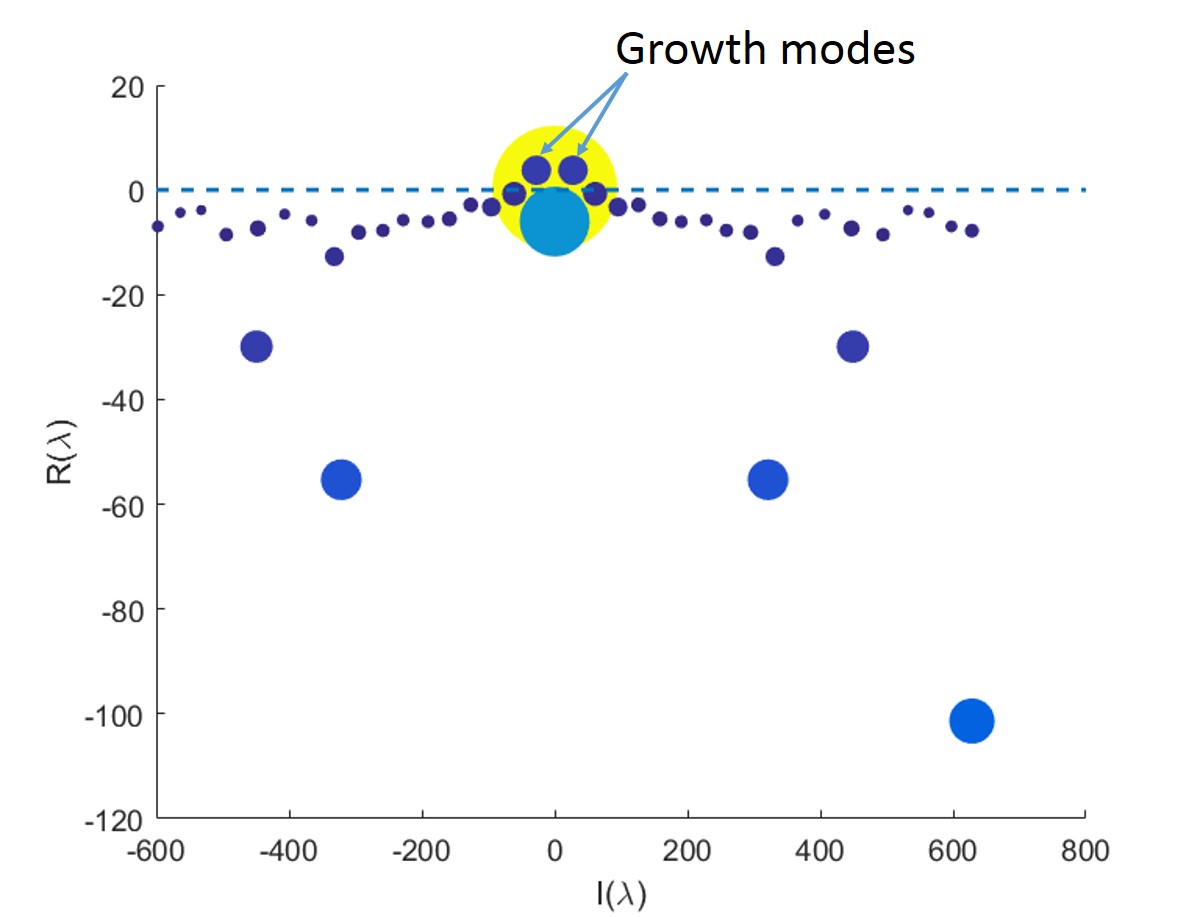}
\caption{Eigenvalues computed from DMD on the kinetic energy density, mode 7 and mode 8 are marked by the arrows.}
\label{fig:eigenvalue_zoom}
\end{figure}

\begin{figure}
\centering
\includegraphics[width=.8\textwidth]{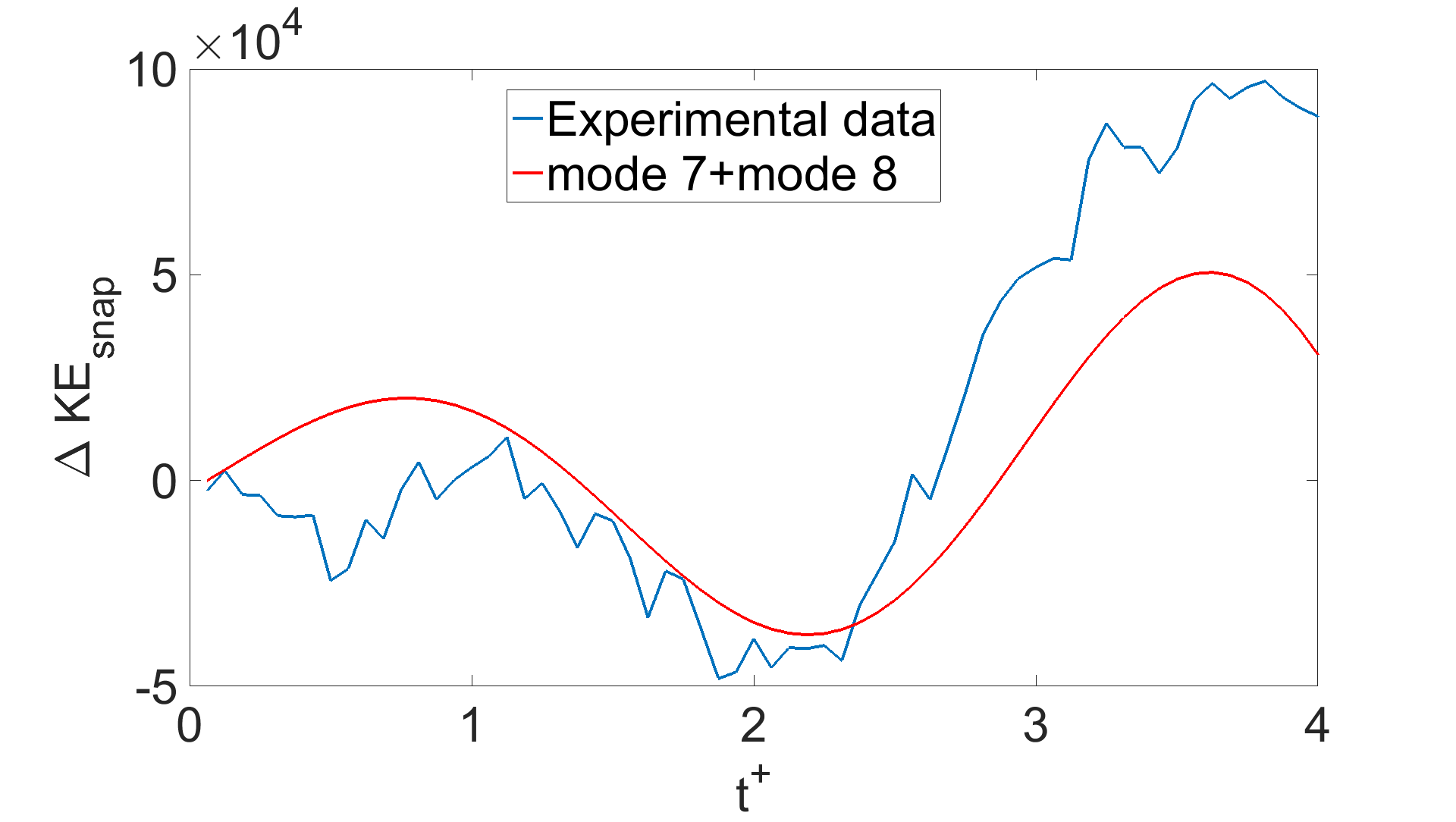}
\caption{Comparison of the experimental $\Delta KE_{snap}$ and the reconstructed $\Delta KE_{snap}$ by mode 7 and mode 8. Here "$\Delta$" indicates the disturbed value in which the non-actuated baseline value has been subtracted from the original data.}
\label{fig:KE_recon}
\end{figure}

\begin{figure}
	\centering
	
		\begin{subfigure}{0.48\textwidth}

		        \includegraphics[width=2.9in]{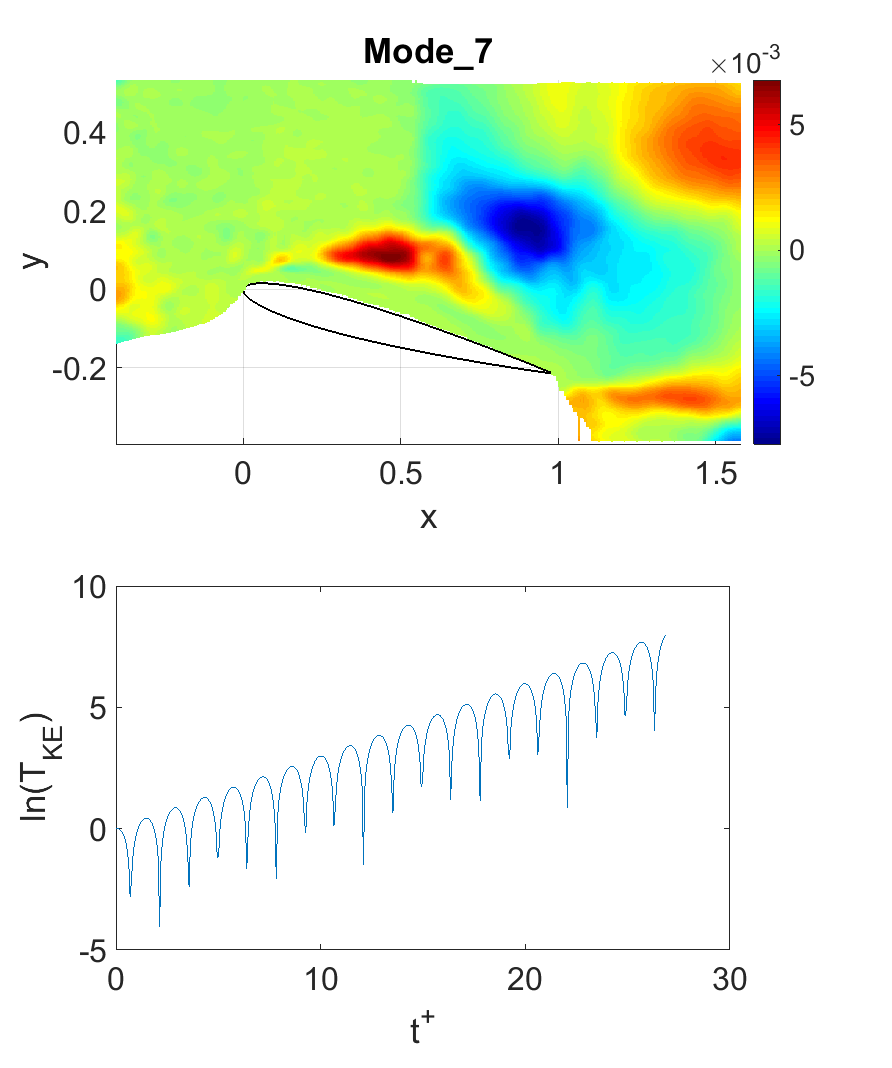}
		          \caption{ }
		          \label{fig:KE_mode_7}
		\end{subfigure}	
~		

	

    \caption{The growth mode, mode 7, the temporal coefficient $T_{KE}$ is plotted in log scale.}
    \label{fig:DMD_mode}		
\end{figure}

\FloatBarrier
The DMD modes are sorted in descending order of the kinetic energy contained in each spatial mode. The eigenvalues of the first 22 DMD modes are shown in figure \ref{fig:eigenvalue_zoom}. The horizontal axis is the imaginary part of the complex eigenvalues and the vertical axis is the real part of the complex eigenvalues. There is only one pair of complex conjugate modes (mode 7 and mode 8) associated with the positive real components of the eigenvalues ($3.75\pm27.60i$). These two modes are able to capture almost 60\% of the disturbance kinetic energy in the flow field (figure \ref{fig:KE_recon}), which indicates that these two modes are responsible for the kinetic energy growth. The corresponding frequency of this pair of modes is 4.4Hz ($F^+\approx0.36$), where $F^+=fc/U_{\infty}$, $f$ is the frequency in 'Hz', $c$ is the chord length in 'meters' and $U_{\infty}$ is the freestream speed in 'm/s'. The real components of the spatial and temporal evolution of these two DMD modes are shown in figure \ref{fig:DMD_mode} a-b.

Based on the stability analysis of the flow field response to the single-burst actuation, one would simply assume that continuous actuation at the vicinity of $F^+=0.36$ will cause maximum energy absorbing from the freestream and hence, the maximum lift increment. However, this is not true due to the strong burst-burst interaction observed by \cite{an2016modeling}. The multi-burst actuation will be studied next.   

\FloatBarrier

\FloatBarrier

\FloatBarrier

\section{Multi-burst Actuation}\label{sec:multi}
In common applications of dynamic active flow control, a continuous sequence of bursts is amplitude modulated to provide variable strength control.  For example, dynamic stall vortex control on a pitching airfoil will require a time-varying burst amplitude during the airfoil maneuver.  If linear dynamics governed the burst-mode response, then a convolution integral using the single-pulse response as a kernel in the integral should be sufficient to predict the lift response to an arbitrary actuator input. We expect this will be the case when the bursts are sufficiently far apart in time that interactions between single pulses are small.  On the other hand, as the time between bursts decreases, then we expect interactions to occur between the current burst and the previous bursts, which will cause deviations from the convolution integral prediction.  In this section the interactions between sequences of 10 bursts with different time intervals between their individual bursts are examined  to determine the boundaries between linear and nonlinear behavior.

\begin{figure}
\centering
\includegraphics[width=.8\textwidth]{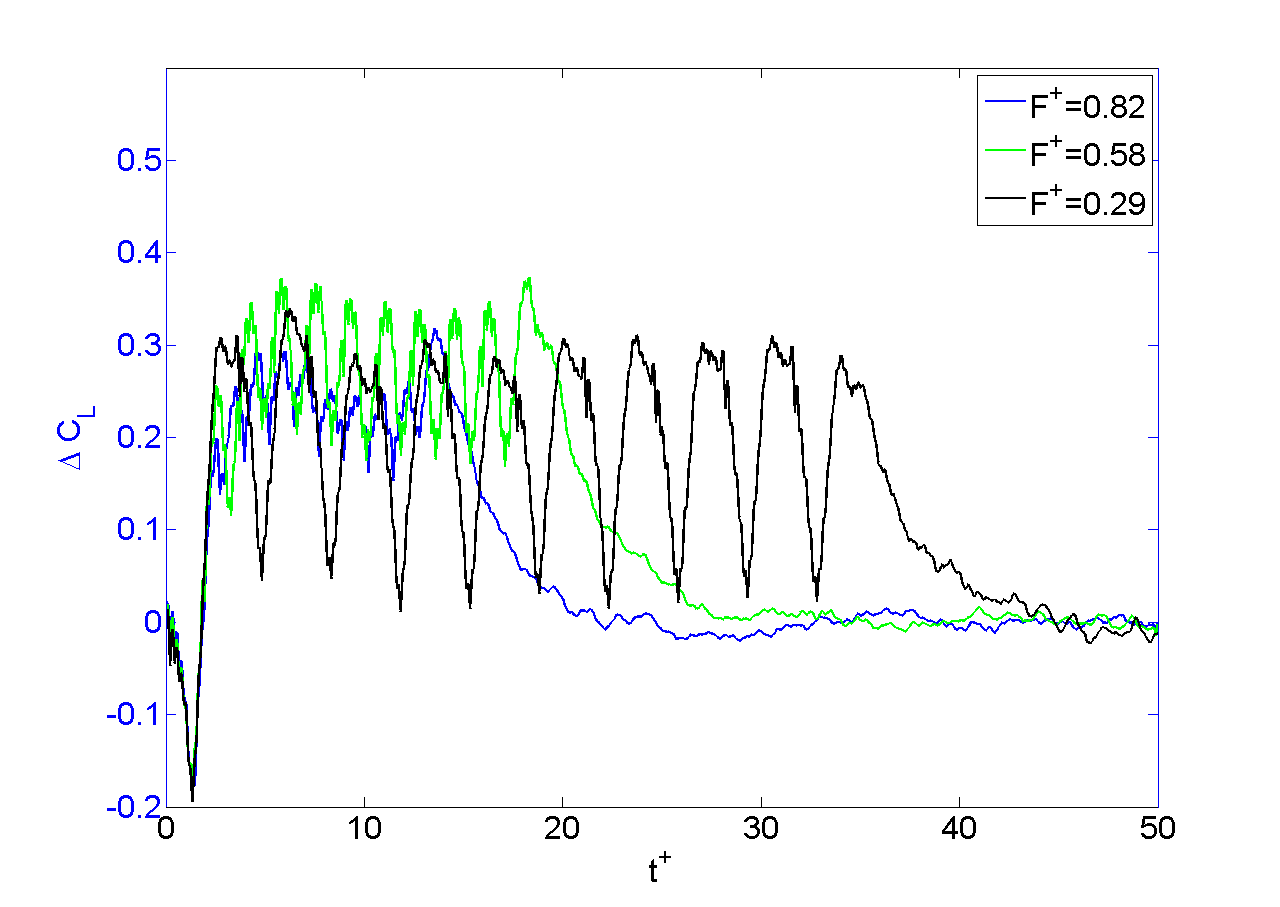}
\caption{A ten-burst actuation sequence with different actuation frequencies.}
\label{fig:multi_pulse}
\end{figure}

The 10 burst sequence is long enough to allow a steady state lift response to be reached.  Five burst frequencies ($F^+=0.145, 0.29, 0.58, 0.65, 0.82$) were investigated. 
The lift coefficient increment  $\Delta C_L$ following a sequence of 10 bursts is shown in figure \ref{fig:multi_pulse} for three frequencies, $F^+= 0.29, 0.58, 0.82$.  The initial lift reversal is the same for the first burst in all three cases. After the initial burst, the lift increment quickly reaches a new steady state value (local mean) that is superposed with large amplitude oscillations at the burst frequency. From the lift coefficient increment data in figure \ref{fig:multi_pulse}, one observes that both the local mean and the amplitude of the $\Delta C_L$ oscillations depend on the burst frequency.  The largest frequency $F^+ = 0.82$ has the lowest lift fluctuation level, and the lowest frequency $F^+ = 0.29$ has the largest lift fluctuation levels.  The intermediate burst frequency $F^+ = 0.58$ has the largest average lift increment $\Delta C_L = 0.28$.  The lift recovery after the final ($10^{th}$) burst is the same for all three cases shown.  

\begin{figure}
	\centering
	
		\begin{subfigure}{0.45\textwidth}
		        \includegraphics[width=2.6in]{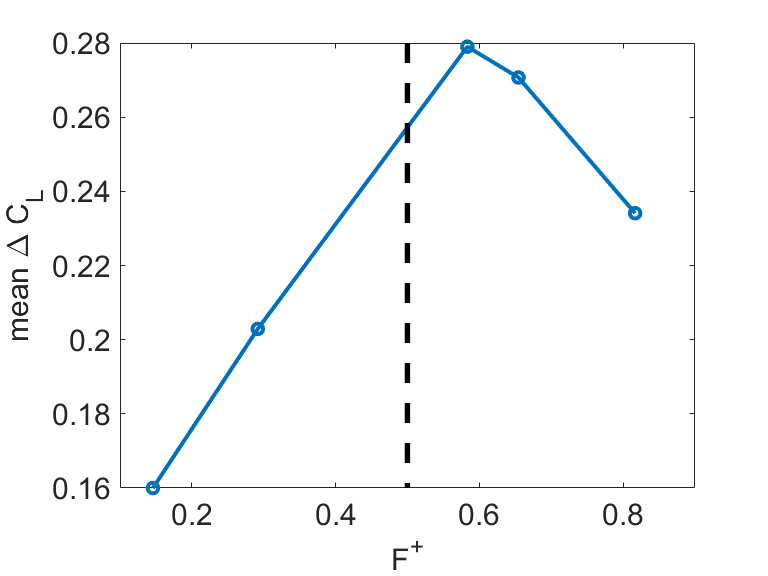}
		          \caption{Mean $\Delta C_L$ vs. $F^+$}
		          \label{fig:mean}
		\end{subfigure}	
~
		\begin{subfigure}{0.45\textwidth}
		        \includegraphics[width=2.6in]{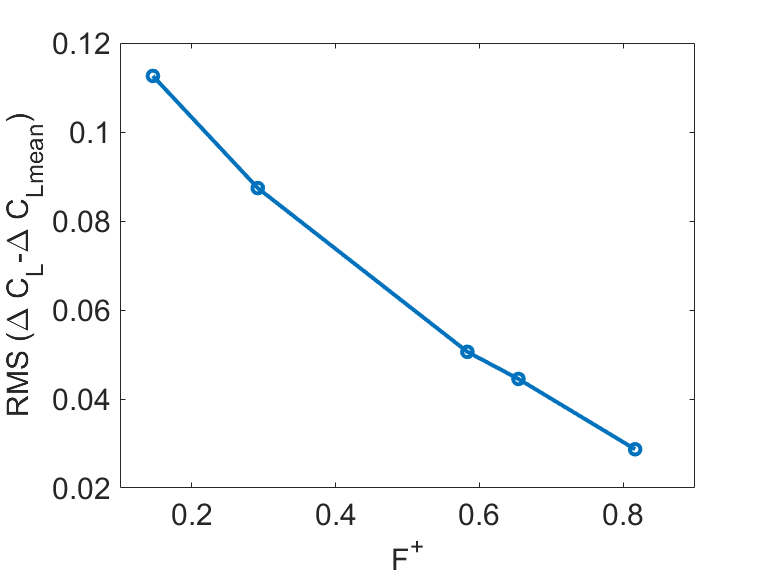}
		          \caption{Oscillation amplitude (RMS) vs. $F^+$}
		          \label{fig:rms}
		\end{subfigure}	
	\caption{Actuation frequency effect on the mean $\Delta C_L$ and the oscillation amplitude (RMS), the black dashed line in (a) indicates the first subharmonic of the separation bubble frequency $f_{sep}^+$.}
    \label{fig:mean_rms}		
\end{figure}

The local mean and the root mean square (rms) value of the $\Delta C_L$ response are plotted in figure \ref{fig:mean_rms}. The dependence of the local mean value of $\Delta C_L$ on the burst frequency $F^+$ is plotted in figure \ref{fig:mean}. The local mean value of $\Delta C_L$ is calculated by averaging the $\Delta C_L$ between the fourth and the ninth burst, so that it is not affected by the initial or final transients. The rms of $\Delta C_L$ (with the mean subtracted) is plotted in figure \ref{fig:rms}. This value is also calculated between the fourth and ninth bursts to avoid the transient region on the burst-burst interaction.   
  
Figure \ref{fig:mean} shows that the maximum local mean of $\Delta C_L$ occurs in the vicinity of $F^+\approx0.58$.  \citet{raju2008dynamics} reported that maximum lift increment occurs when the actuation frequency is close to the subharmonic of the separation bubble frequency $f_{sep}^+$. In the current research, $f_{sep}^+\approx U_{\infty}/L_{sep}\cdot c/U_{\infty}$, where $L_{sep}$ is the separation bubble length. Since the baseline flow is fully separated, we assume $L_{sep} = c$, so $f_{sep}^+\approx 1$. Hence, the burst frequency, $F^+\approx0.58$, corresponding to the mean $\Delta C_L$ peak is close to the first subharmonic of the separation bubble frequency (figure \ref{fig:mean}), which is in agreement with \citet{raju2008dynamics}. 

However, the peak mean $\Delta C_L$ frequency ($F^+\approx 0.58$) differs from the DMD growth mode frequency ($F^+\approx0.36$) from the single-burst case. This suggests that the instability triggered by the single-burst is not the only factor for the mean $\Delta C_L$ in the multi-burst actuation cases. The burst-burst interaction plays an important role as well. It is also worth of pointing out that figure \ref{fig:mean} clearly shows that the relation between the mean $\Delta C_L$ and the actuation frequency is nonlinear.   

On the other hand, within the frequency range considered, figure \ref{fig:rms} shows that the rms of $\Delta C_L$ associated with the firing of each individual burst, decreases monotonically as the burst frequency increases. 

        
\FloatBarrier

\FloatBarrier

To help clarify the differences between linear and nonlinear behavior during the multi-burst actuation, the convolution integral that uses a linear superposition of a sequence of the single-burst actuation was performed.  If the flow responds linearly to the multi-burst input, then the convolution approach will predict the lift response. Differences between the linear convolution predictions and the actual measurements are assumed to be the result of nonlinear effects. Here we use the $\Delta C_L$ response from the single-burst as the kernel for the convolution integral, which is written in discrete form as
\begin{equation} \label{eq:conv}
\Delta C_L(k)=\sum \Delta C_{L,single}(j)v(k-j).
\end{equation} 
$\Delta C_{L,single}$ is the $C_L$ response to the single-burst with its baseline value subtracted, and $v(k)$ is the input voltage to the actuator at the time step $k$. 

\begin{figure}
	\centering
	
		\begin{subfigure}{0.45\textwidth}
		        \includegraphics[width=2.6in]{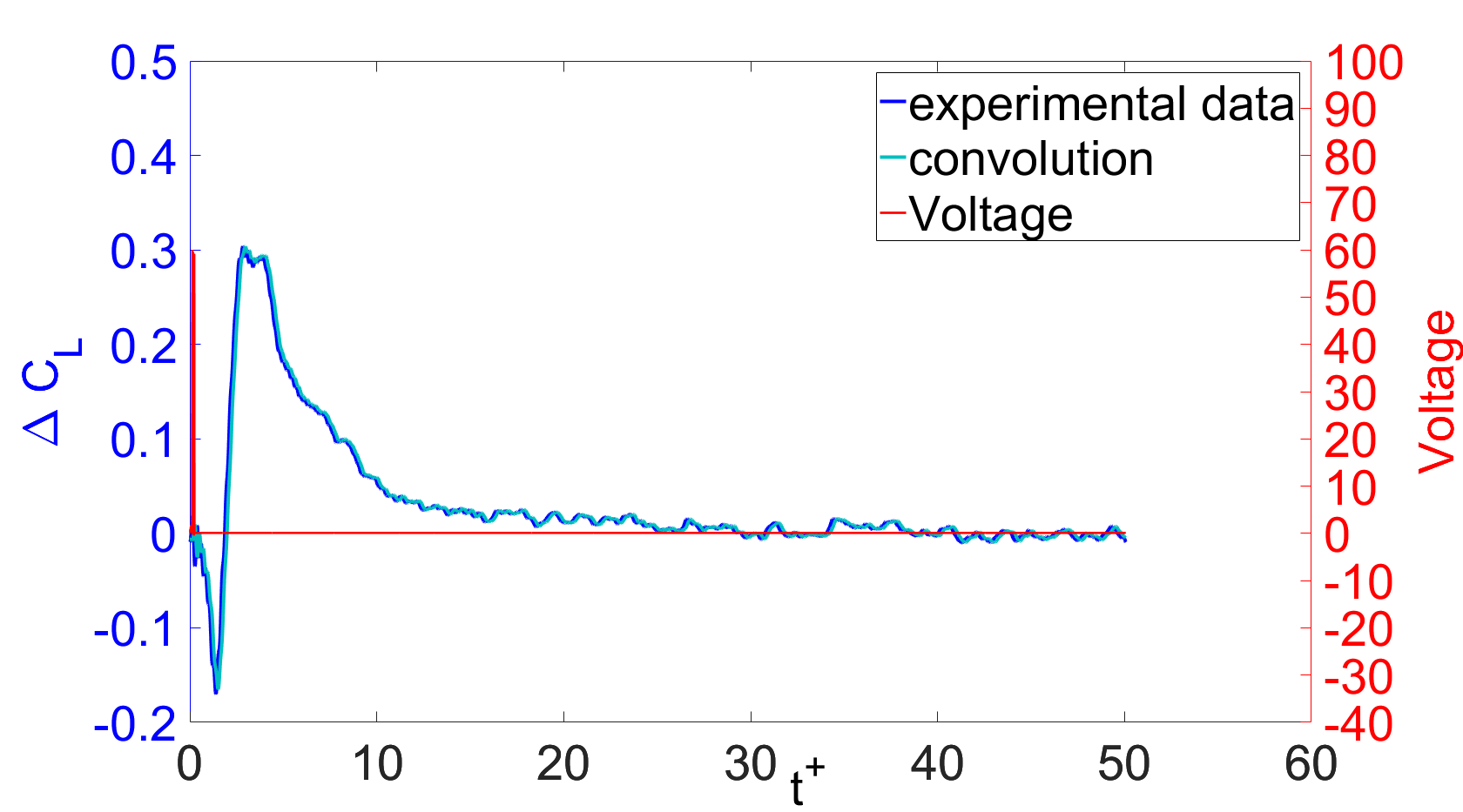}
		          \caption{Single-burst}
		          \label{fig:Single-pulse}
		\end{subfigure}	
~
		\begin{subfigure}{0.45\textwidth}
		        \includegraphics[width=2.6in]{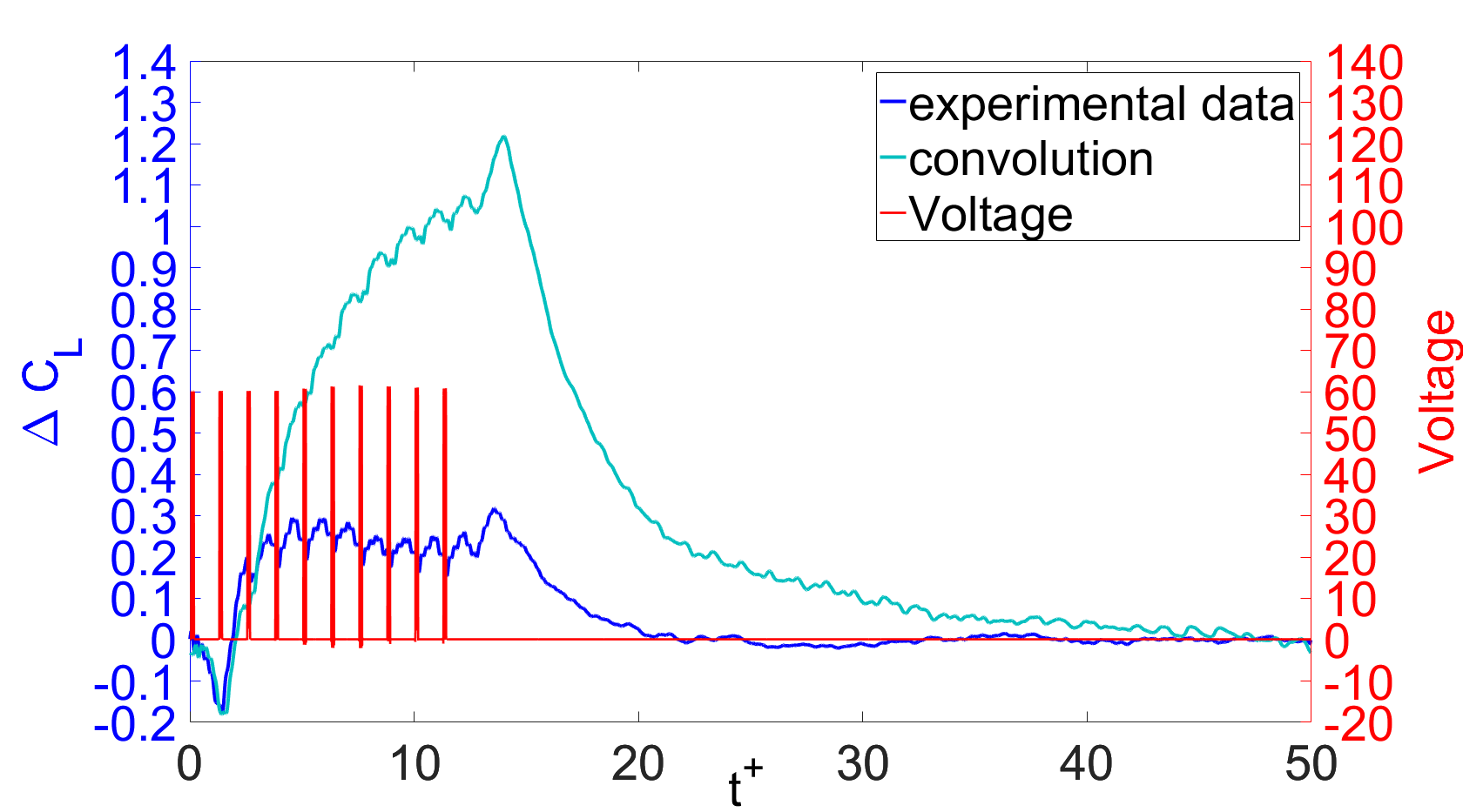}
		          \caption{Multi-burst $F^+=0.82$}
		          \label{fig:DCL_multi_1p25t+}
		\end{subfigure}	
		
		\begin{subfigure}{0.45\textwidth}
		        \includegraphics[width=2.6in]{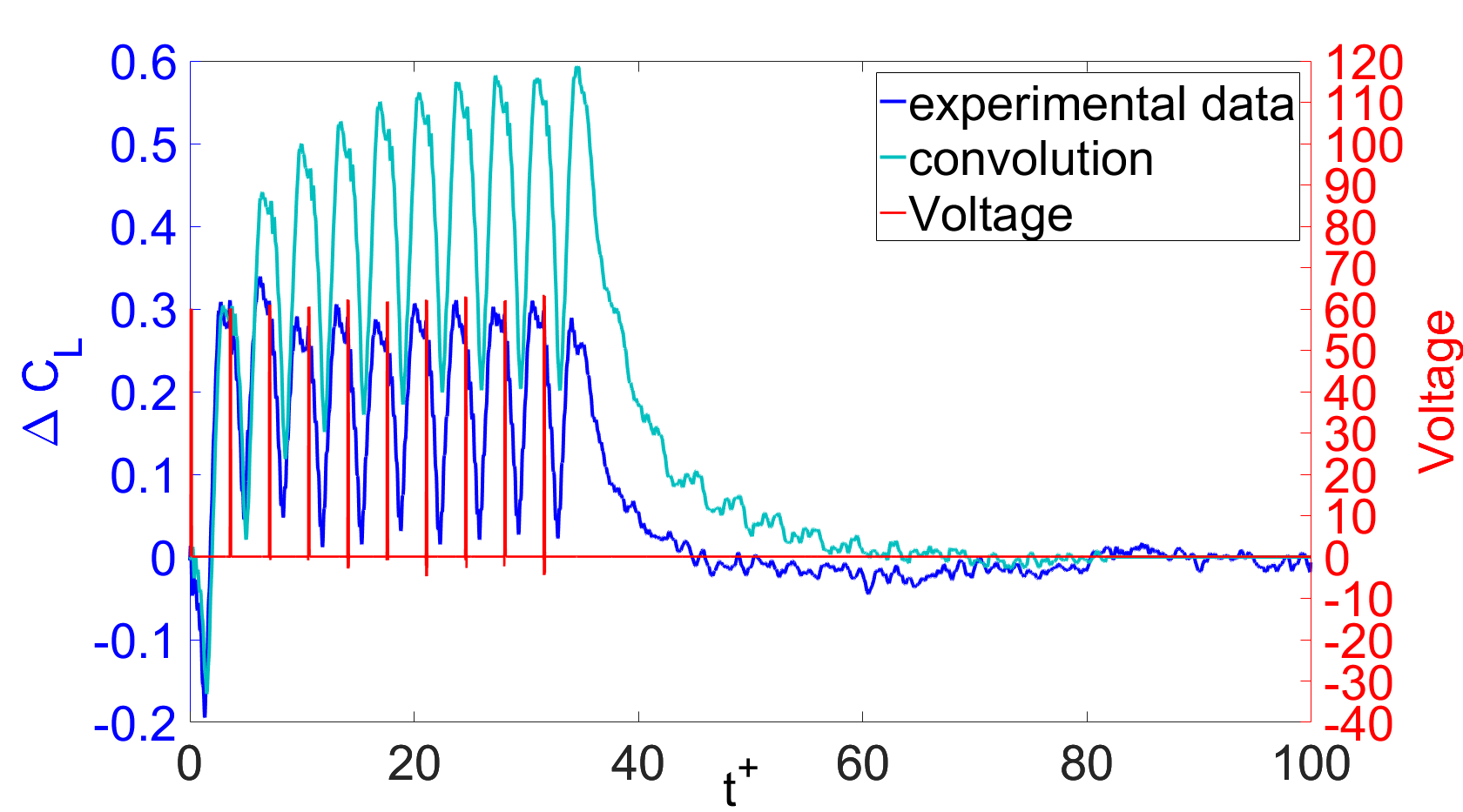}
		          \caption{Multi-burst $F^+=0.29$}
		          \label{fig:DCL_multi_3p5t+}
		\end{subfigure}	
~		
		\begin{subfigure}{0.45\textwidth}
		        \includegraphics[width=2.6in]{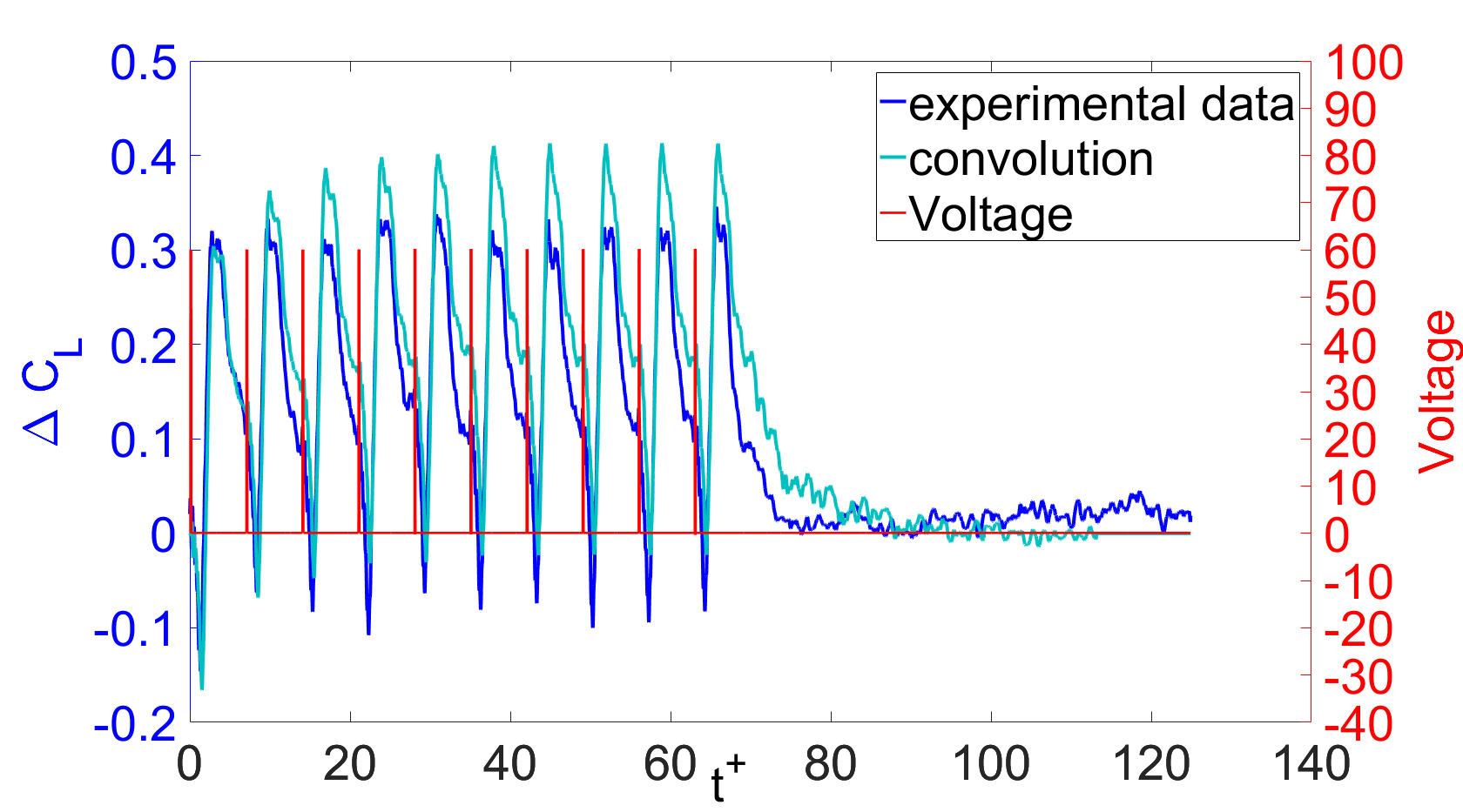}
		          \caption{Multi-burst $F^+=0.145$}
		          \label{fig:DCL_multi_7p0t+}
		\end{subfigure}				
	
    	
	\caption{$\Delta C_L$ response to single-burst and multi-burst actuation at $\alpha=12^o$. The red lines show the voltage to the actuators. a) single burst; b)10 repeating bursts at $F^+=0.82$; c) 10 repeating bursts at $F^+=0.29$; d) 10 repeating bursts at $F^+=0.145$.}
    \label{fig:multi1}		
\end{figure}

Figures \ref{fig:Single-pulse} - \ref{fig:DCL_multi_7p0t+} compare measurements with the convolution integral predictions of the lift increment response following a single burst and three multi-burst cases with different time delays between the bursts. Figure \ref{fig:Single-pulse} shows that it takes approximately 20 convective times for the lift to return to its original state. When the second burst occurs before $t^+ = 20$, then the kernel begins at a non-zero initial value, and this leads to an artificial increase in the predicted lift increment. The lift increment for the second burst begins at a higher value than the first burst.  The same effect occurs for the third burst relative to the second, and so on. In general, when the bursts are close together in time, e.g.,$1.25~t^+$ figure \ref{fig:DCL_multi_1p25t+}, then the convolution overpredicts the value of $\Delta C_L$ relative to the measured data. This phenomenon indicates that when the bursts are close to each other, the nonlinear burst-burst interaction effects become stronger. 

  In the case of figure \ref{fig:DCL_multi_1p25t+} this leads to a large over-estimation in the maximum $\Delta C_L$.  However, as the time delay between pulses increases, then the lift increment from the first pulse has decayed closer to $\Delta C_L = 0$, and the convolution model overshoot is not as large.

The high-frequency component of the $\Delta C_L$ signal that is associated with the multi-burst signal is predicted reasonably well by the convolution integral. This similarity indicates that the high-frequency oscillation is dominated by the linear dynamics. Combining the low-frequency component of the signal with the high frequency oscillations of $\Delta C_L$ dependency on the frequency mentioned above, we can assume that there are two components to the burst-burst interaction. The nonlinear component of the burst-burst interaction affects the mean value of $\Delta C_L$ and the linear burst-burst interaction corresponds to the RMS or the high-frequency oscillation of $\Delta C_L$. To further test this assumption, a more detailed investigation on the multi-burst flowfield is discussed next.     
   
\FloatBarrier

\subsection{$F^+=0.82$ multi-burst actuation}

The lift coefficient response to actuation is an integral effect from the overall flow field. To gain more insight into the mechanisms of the burst-burst interactions,   phase-averaged flow field maps over the airfoil were acquired. Inspired by the convolution integral approach on the $\Delta C_L$, we found that comparing the convolution integral simulated data with the directly measured data is an effective way of evaluating the linear-nonlinear character of the flow. 

A procedure similar to the convolution integral method is applied to the velocity field data. The kernel in Eq. \ref{eq:conv} is replaced by the flowfield evolution following the single-burst actuation. The input signal $V(k)$ is a 3-D (2 spatial dimensions and one dimension in time) variable, which means it is a spatially uniformly distributed matrix and only varying with time. For example, at the discrete time instant $k$, the flowfield is a $400\times 400$ matrix.  The input is $V(400\times 400)$ with all the elements equal to one (if the burst occurs at time $k$) or zero (if it is between the bursts). The convolution integral of the flowfield can be expressed as follows,  

\begin{equation} \label{eq:conv_flow}
\psi^{i,j,k}=\sum \psi_{single}^{i,j,m}V^{i,j,k-m}
\end{equation} 
where $\psi^{i,j,k}$ is the velocity at spatial location $i,j$ in the $k^{th}$ snapshot, $\psi_{single}^{i,j,m}$ is the velocity at spatial location $i,j$ in the $m^{th}$ snapshot for the single-burst actuation case, and $V^{i,j,k-m}$ is the input signal at spatial location $i,j$ in the $(k-m)^{th}$ snapshot.
  
\begin{figure}
\centering
    \includegraphics[width=0.2\textwidth]{figures/colorbar_conv.png}
	\centering

		\begin{subfigure}{0.3\textwidth}
		        \includegraphics[width=1.8in]{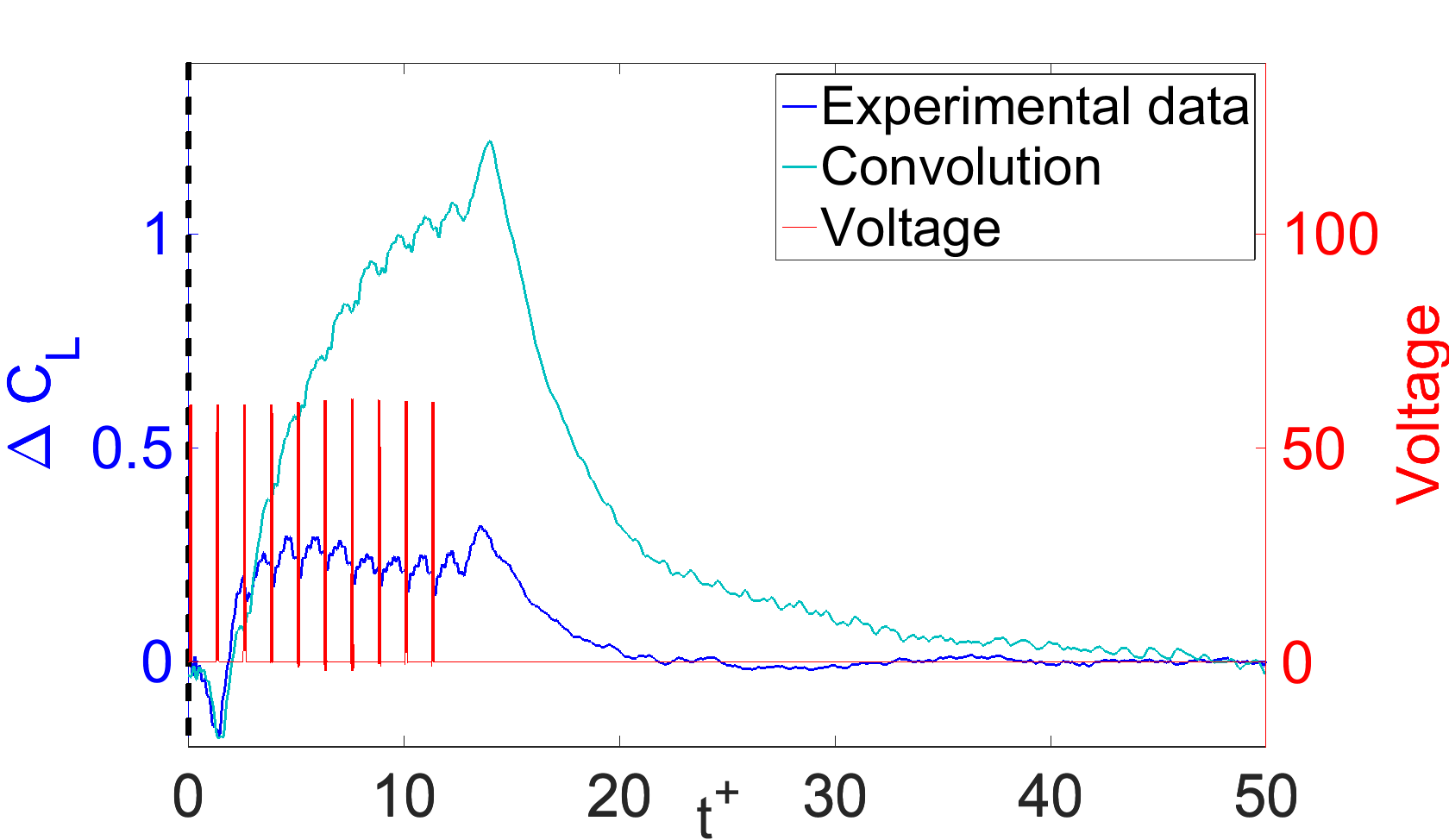}
		          \caption{$\Delta C_L$ at $0.0t^+$}
		          \label{fig:DCL_multi_1p25t+_0t+}
		\end{subfigure}	
		~
		\begin{subfigure}{0.3\textwidth}
		        \includegraphics[width=1.8in]{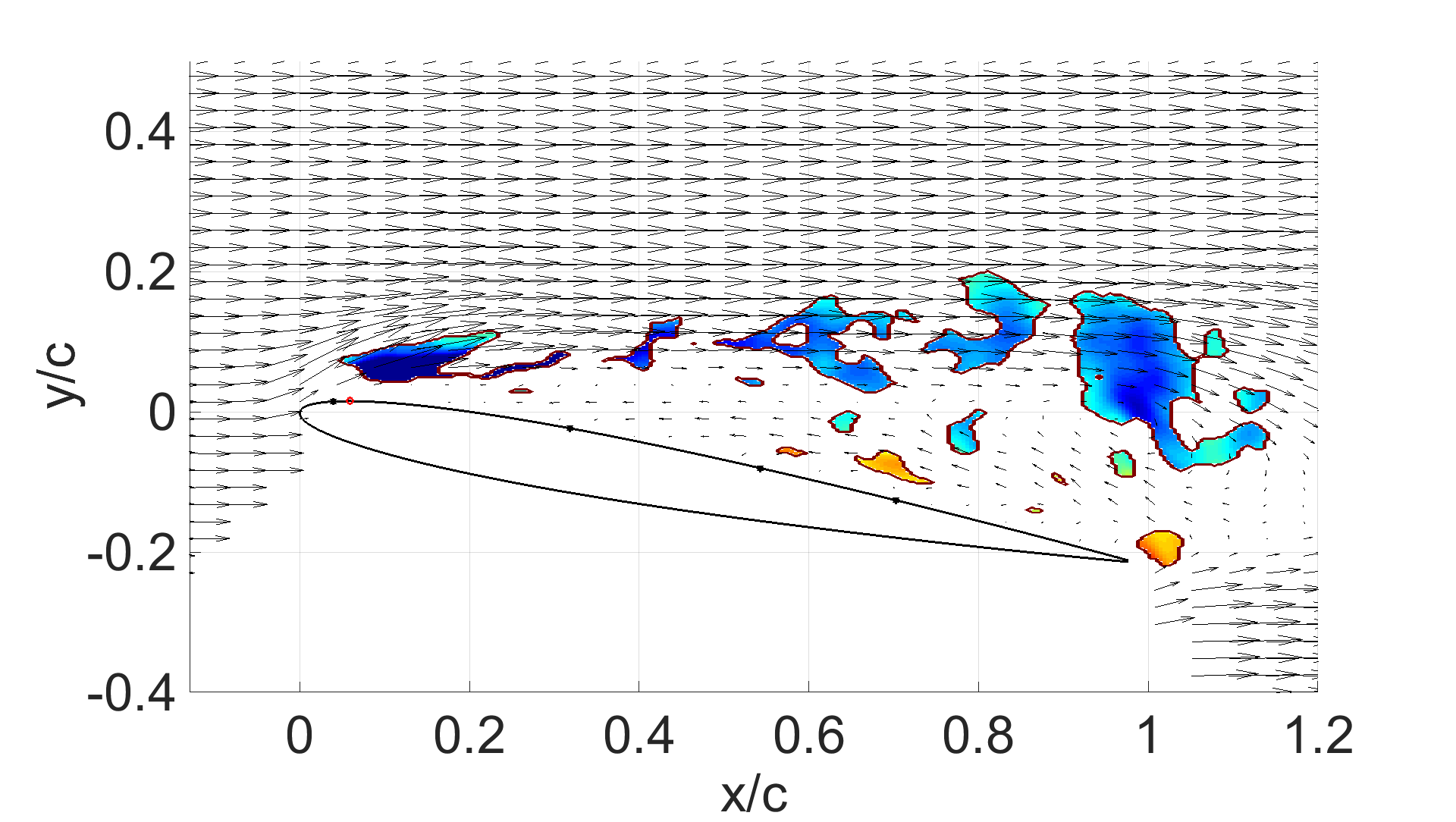}
		          \caption{DMF at $0.0t^+$}
		          \label{fig:Gamma_PIV_tplus_0_1p25}
		\end{subfigure}		
		~
		\begin{subfigure}{0.3\textwidth}
		        \includegraphics[width=1.8in]{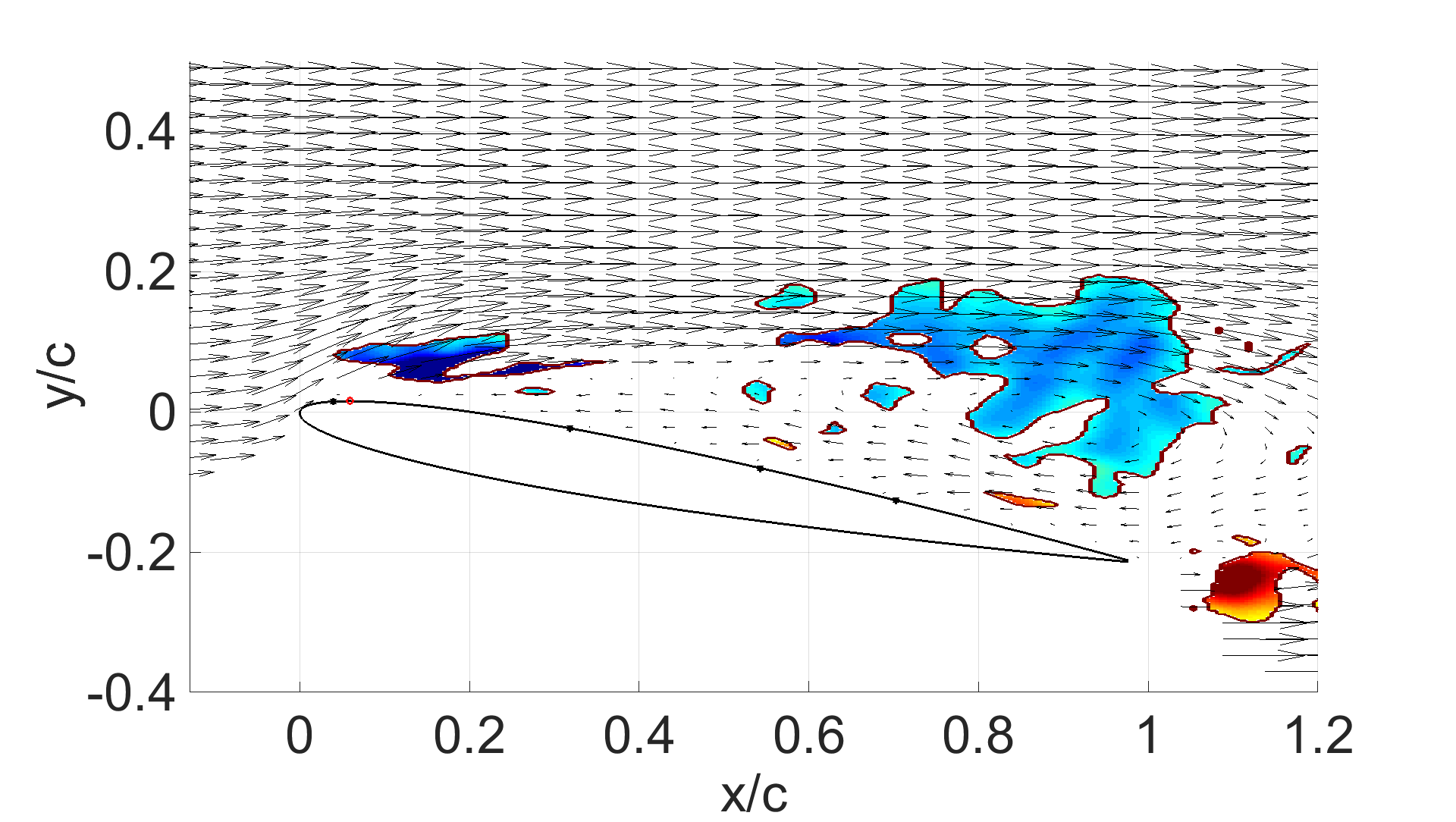}
		          \caption{CSF at $0.0t^+$}
		          \label{fig:Gamma_conv_tplus_0_1p25}
		\end{subfigure}

		\begin{subfigure}{0.3\textwidth}
		        \includegraphics[width=1.8in]{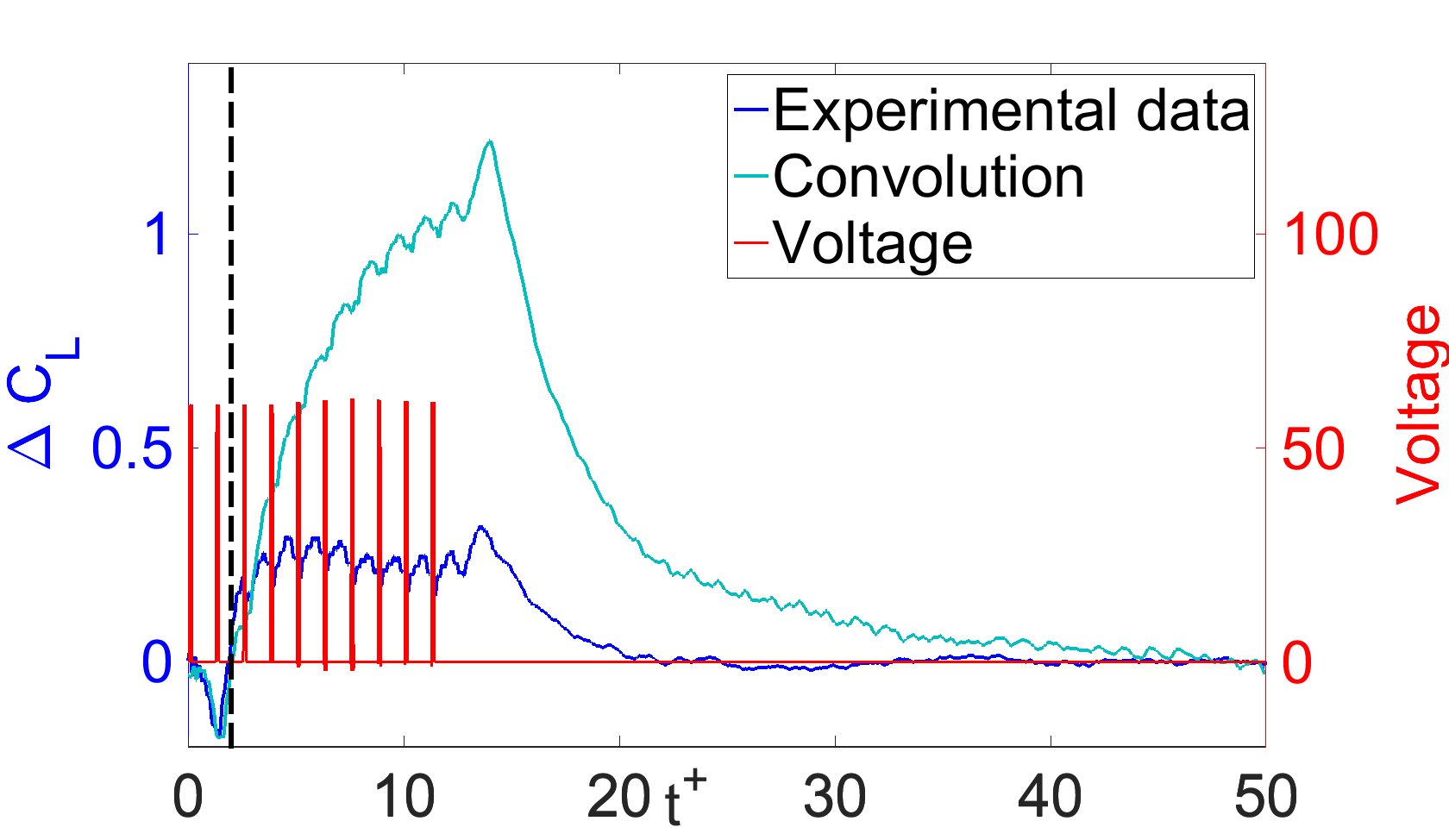}
		          \caption{$\Delta C_L$ at $2.0t^+$}
		          \label{fig:DCL_multi_1p25t+_2t+}
		\end{subfigure}	
		~
		\begin{subfigure}{0.3\textwidth}
		        \includegraphics[width=1.8in]{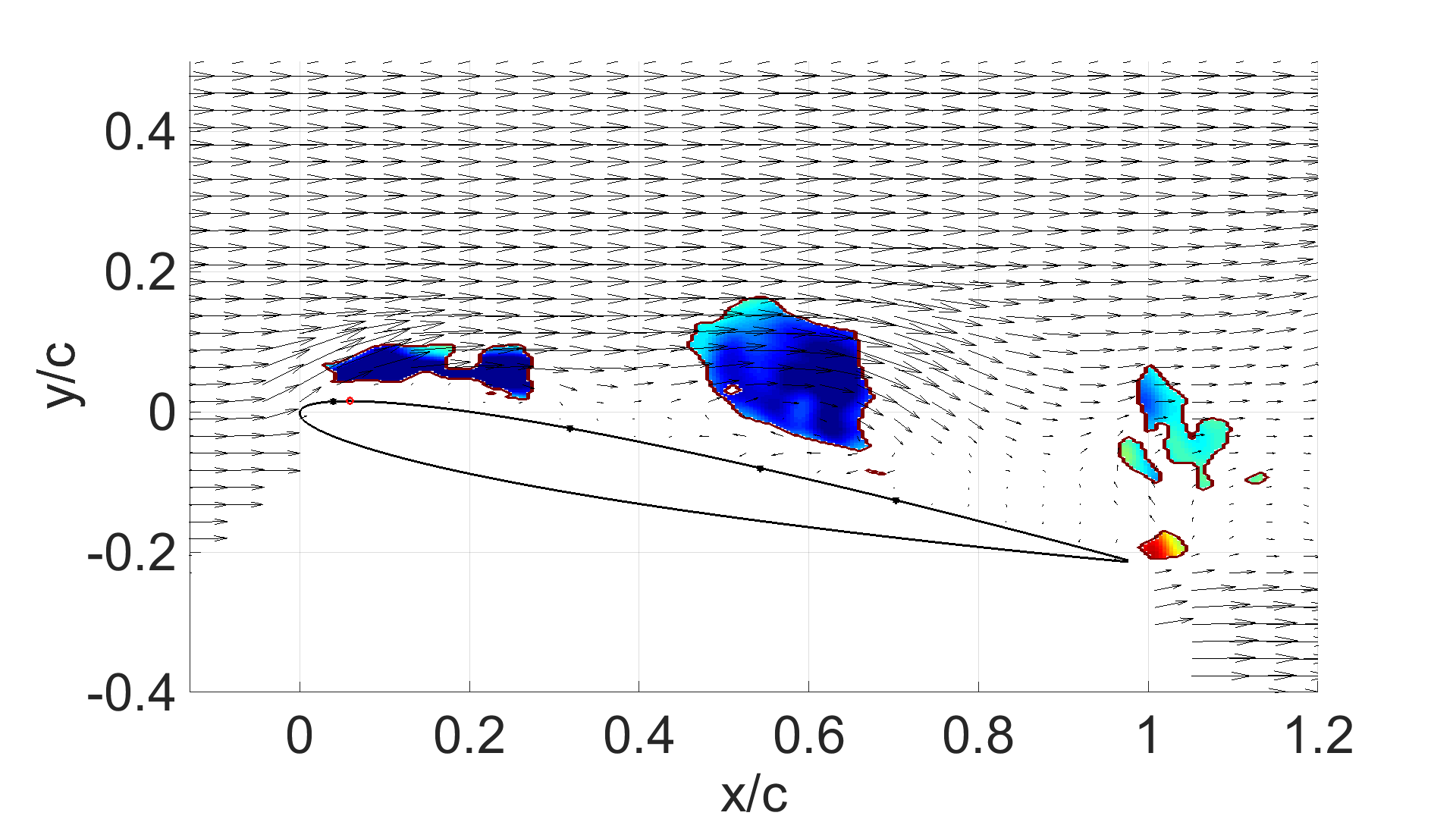}
		          \caption{DMF at $2.0t^+$}
		          \label{fig:Gamma_PIV_tplus_2_1p25}
		\end{subfigure}		
		~
		\begin{subfigure}{0.3\textwidth}
		        \includegraphics[width=1.8in]{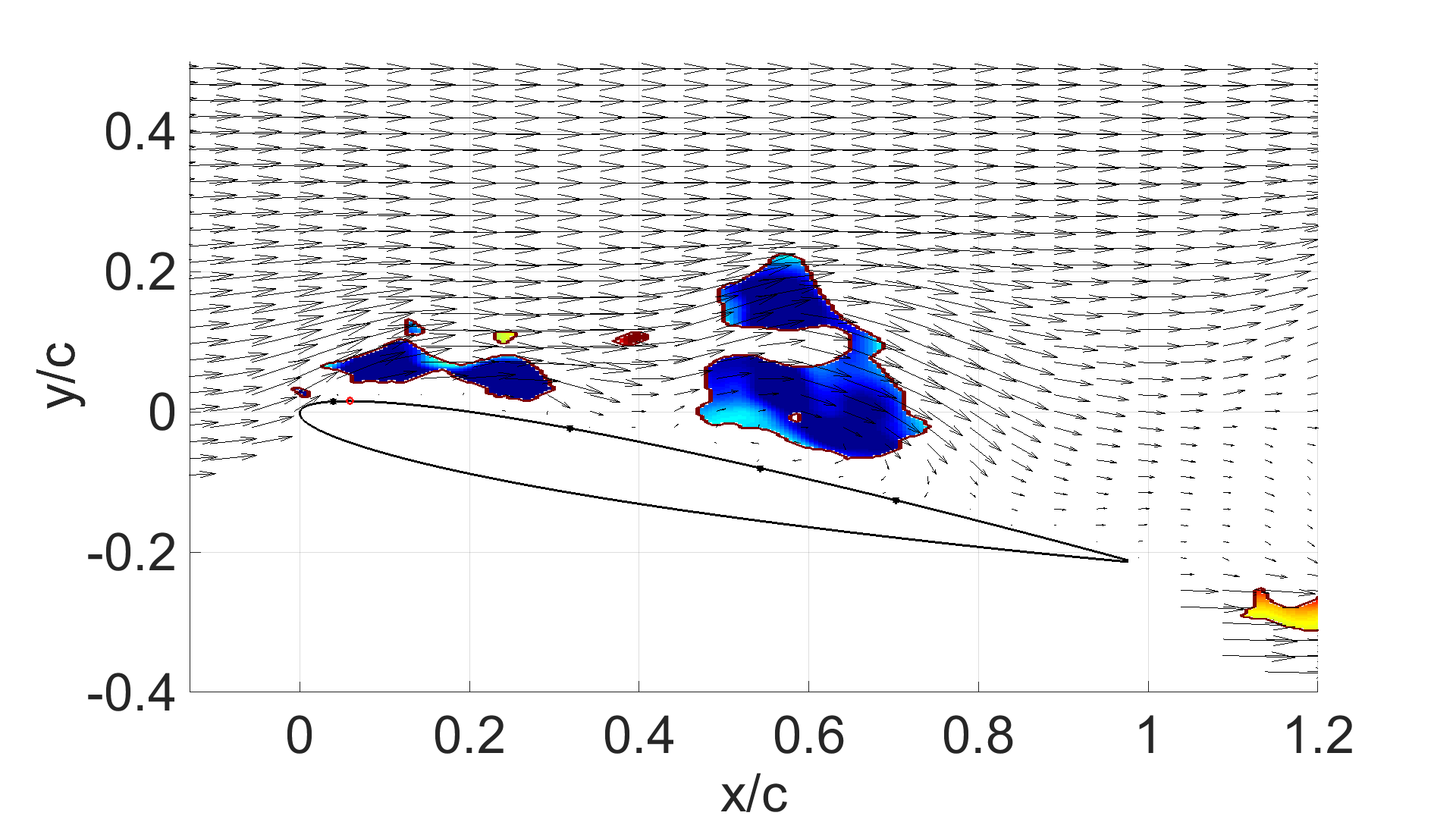}
		          \caption{CSF at $2.0t^+$}
		          \label{fig:Gamma_conv_tplus_2_1p25}
		\end{subfigure}

		\begin{subfigure}{0.3\textwidth}
		        \includegraphics[width=1.8in]{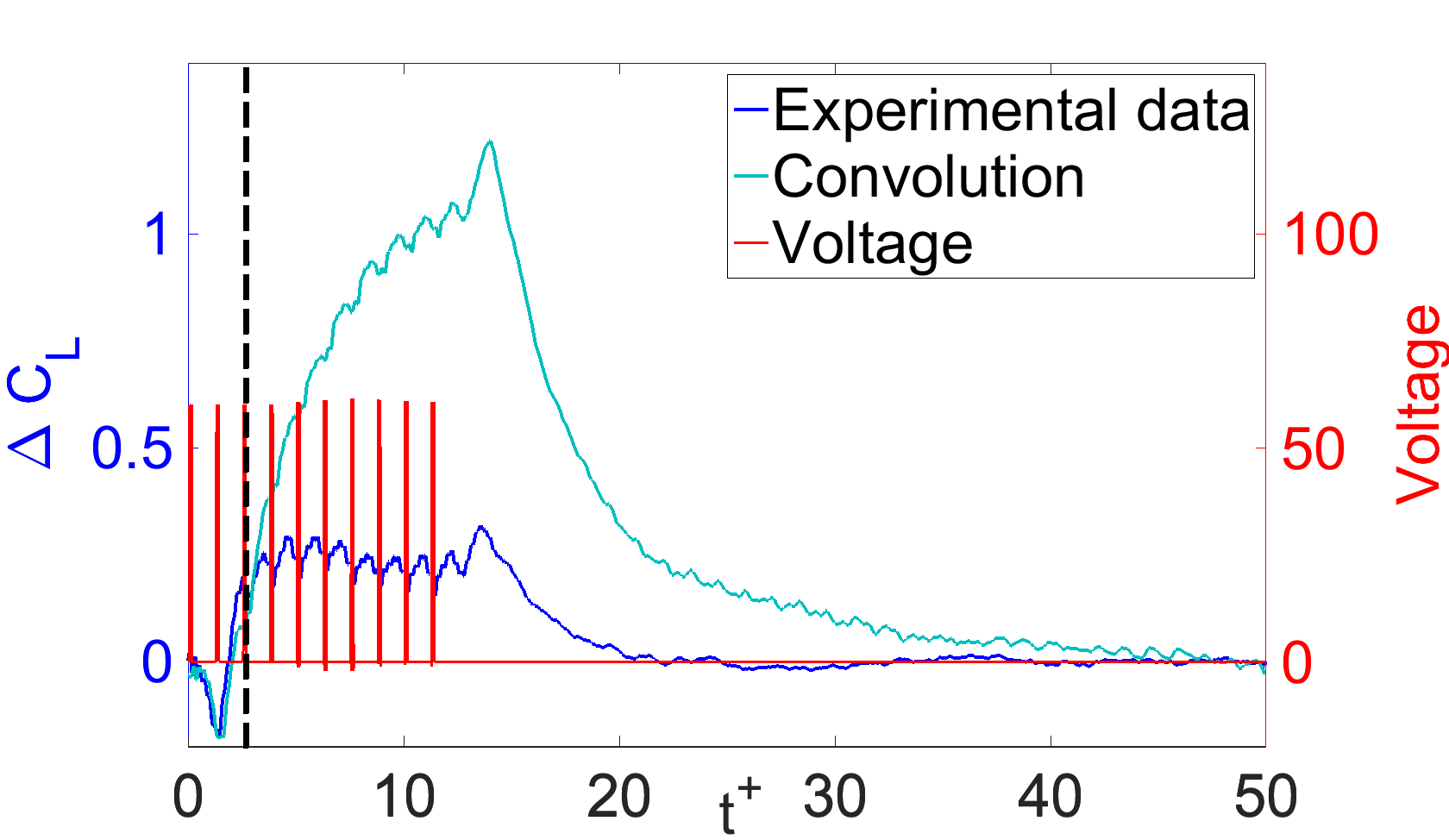}
		          \caption{$\Delta C_L$ at $2.7t^+$}
		          \label{fig:DCL_multi_1p25t+_2p7t+}
		\end{subfigure}	
		~
		\begin{subfigure}{0.3\textwidth}
		        \includegraphics[width=1.8in]{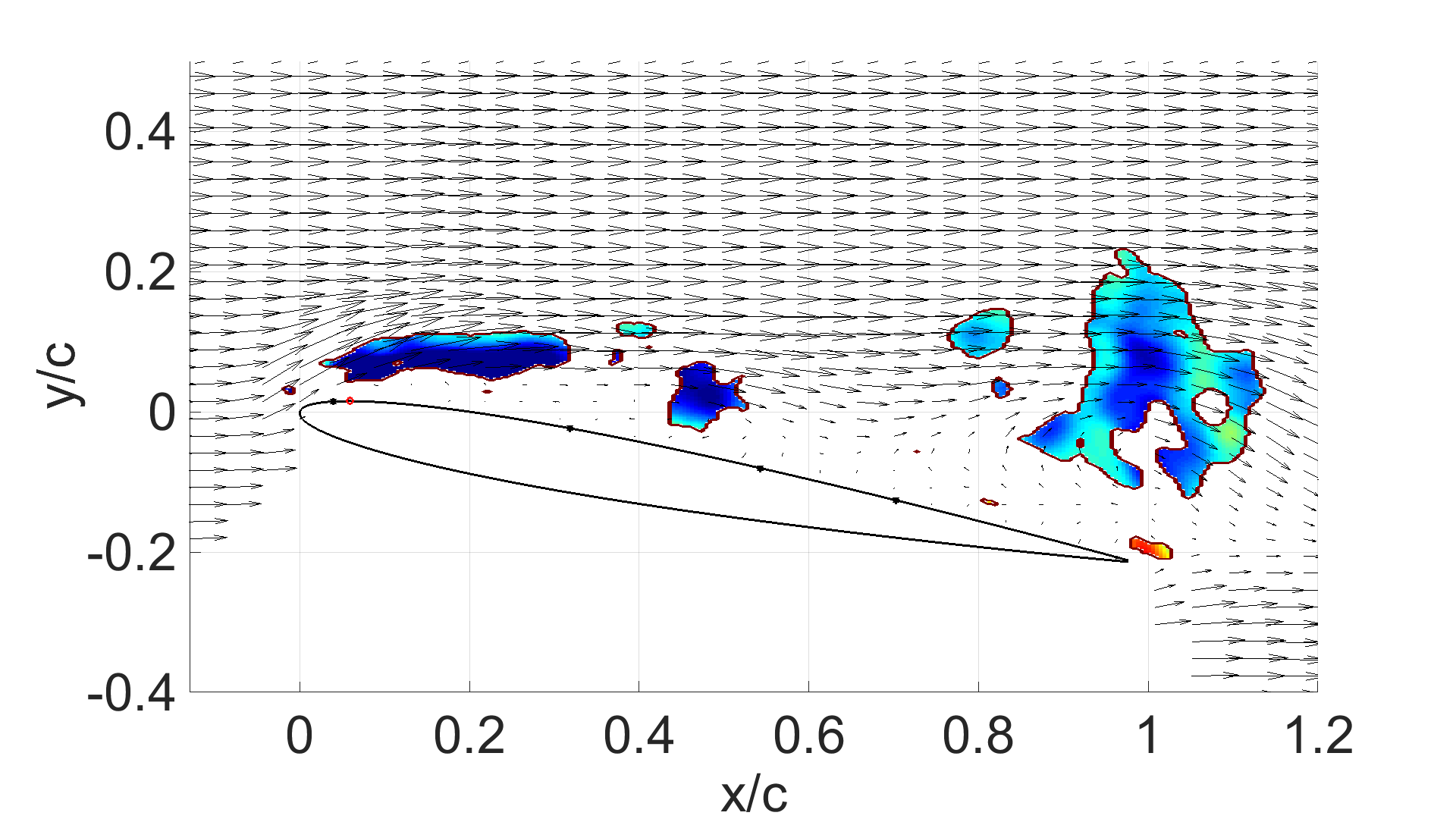}
		          \caption{DMF at $2.7t^+$}
		          \label{fig:Gamma_PIV_tplus_2p7_1p25}
		\end{subfigure}		
		~
		\begin{subfigure}{0.3\textwidth}
		        \includegraphics[width=1.8in]{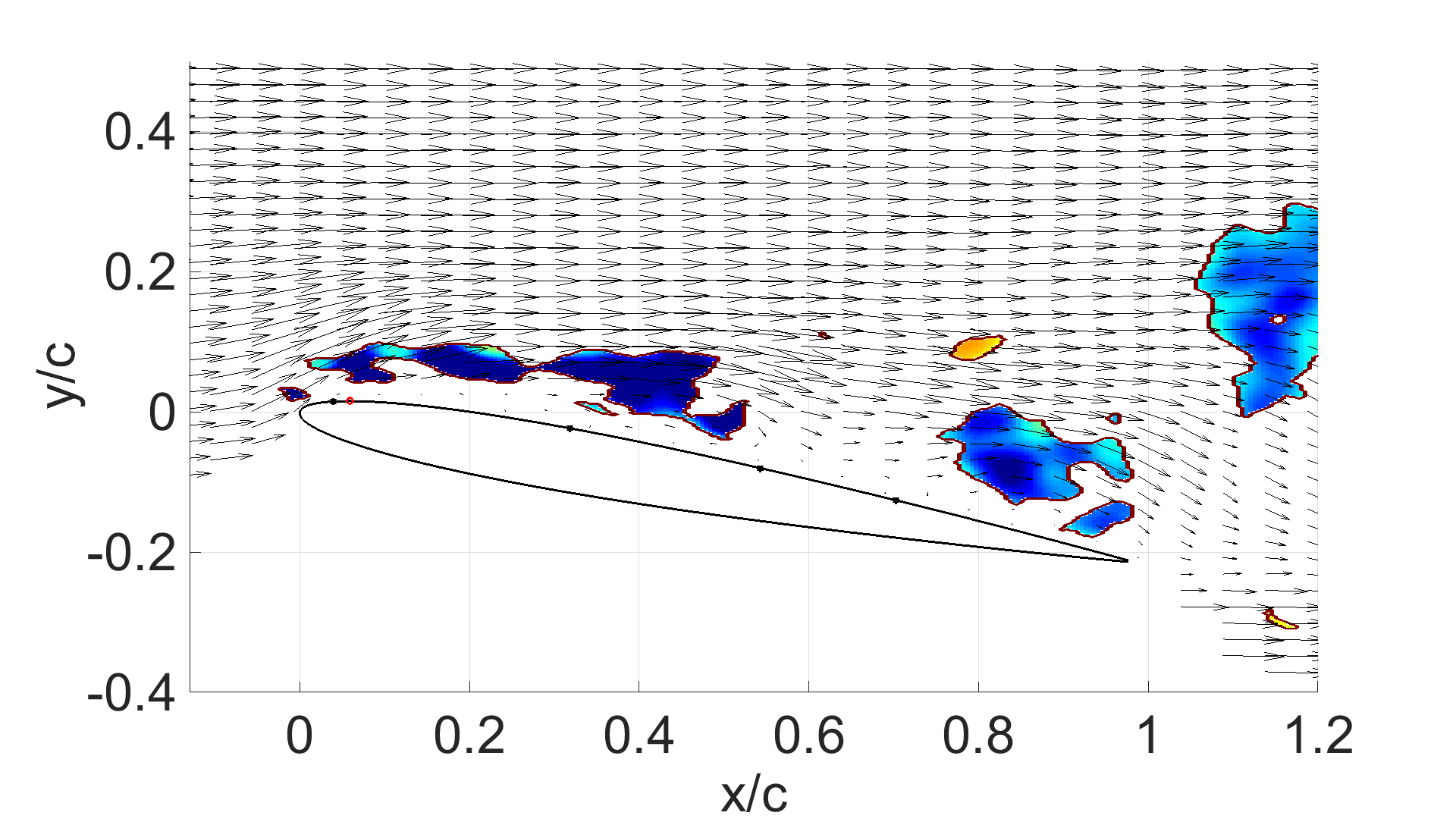}
		          \caption{CSF at $2.7t^+$}
		          \label{fig:Gamma_conv_tplus_2p7_1p25}
		\end{subfigure}

		\begin{subfigure}{0.3\textwidth}
		        \includegraphics[width=1.8in]{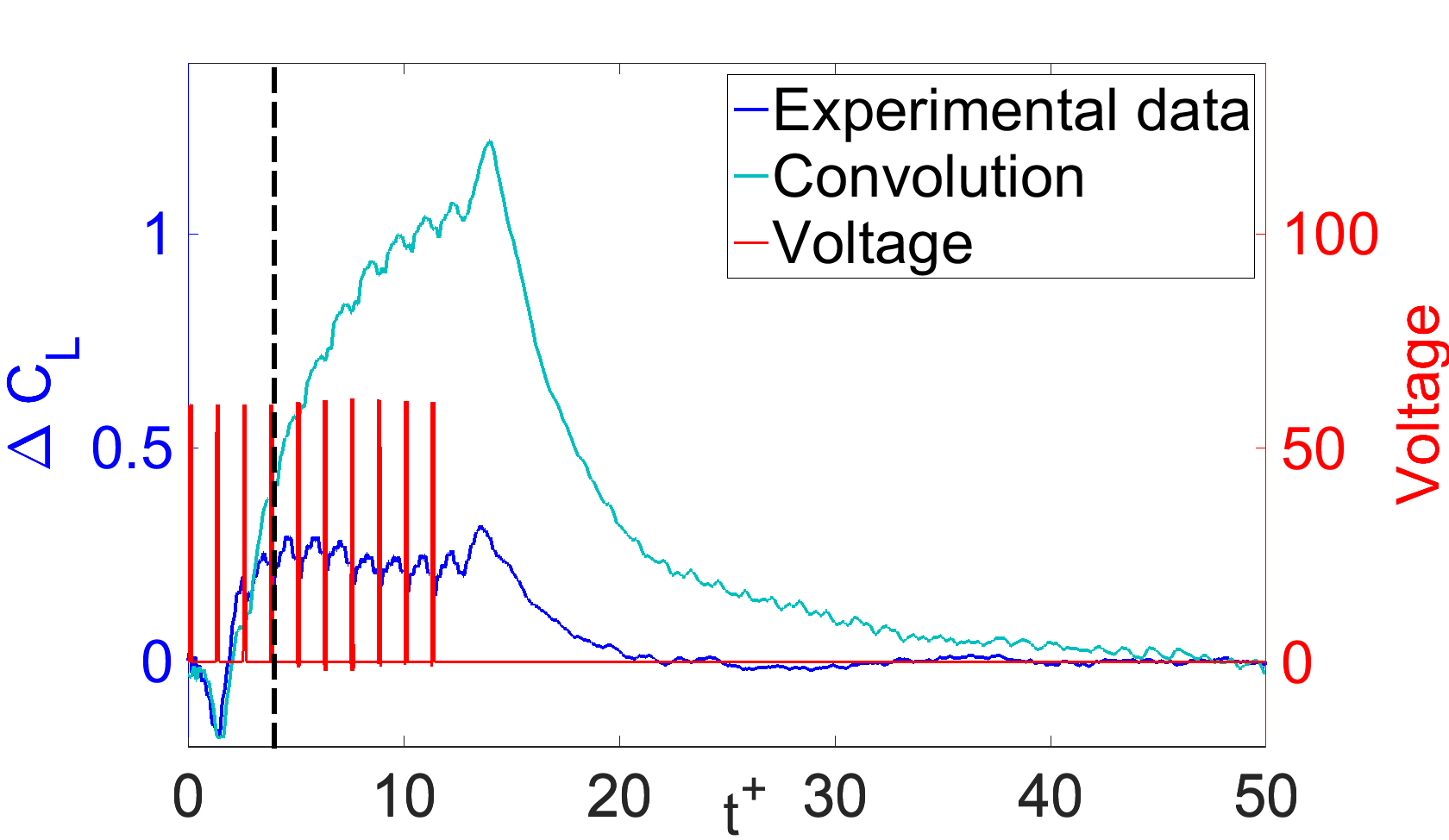}
		          \caption{$\Delta C_L$ at $4.0t^+$}
		          \label{fig:DCL_multi_1p25t+_4t+}
		\end{subfigure}	
		~
		\begin{subfigure}{0.3\textwidth}
		        \includegraphics[width=1.8in]{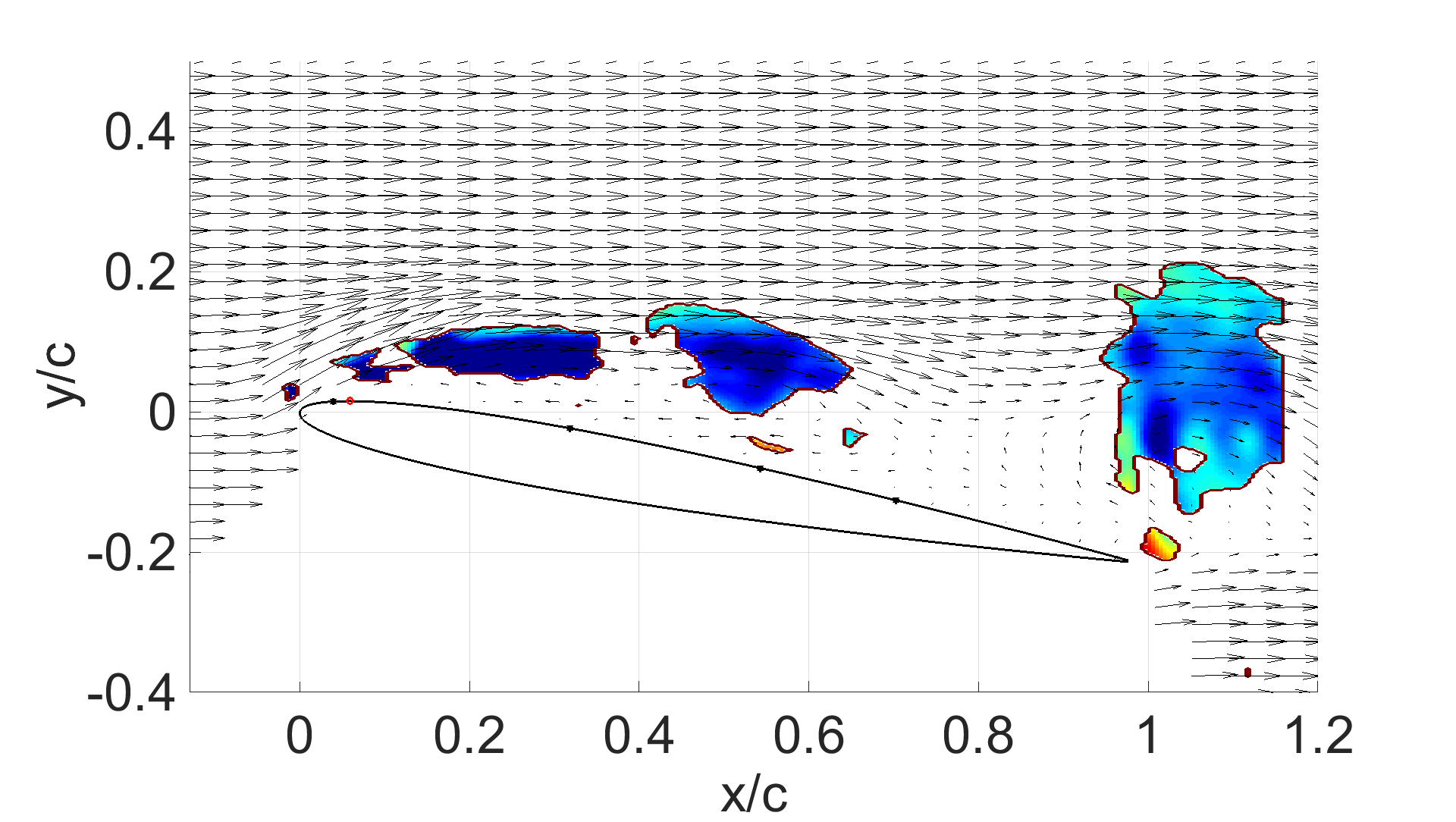}
		          \caption{DMF at $4.0t^+$}
		          \label{fig:Gamma_PIV_tplus_4_1p25}
		\end{subfigure}		
		~
		\begin{subfigure}{0.3\textwidth}
		        \includegraphics[width=1.8in]{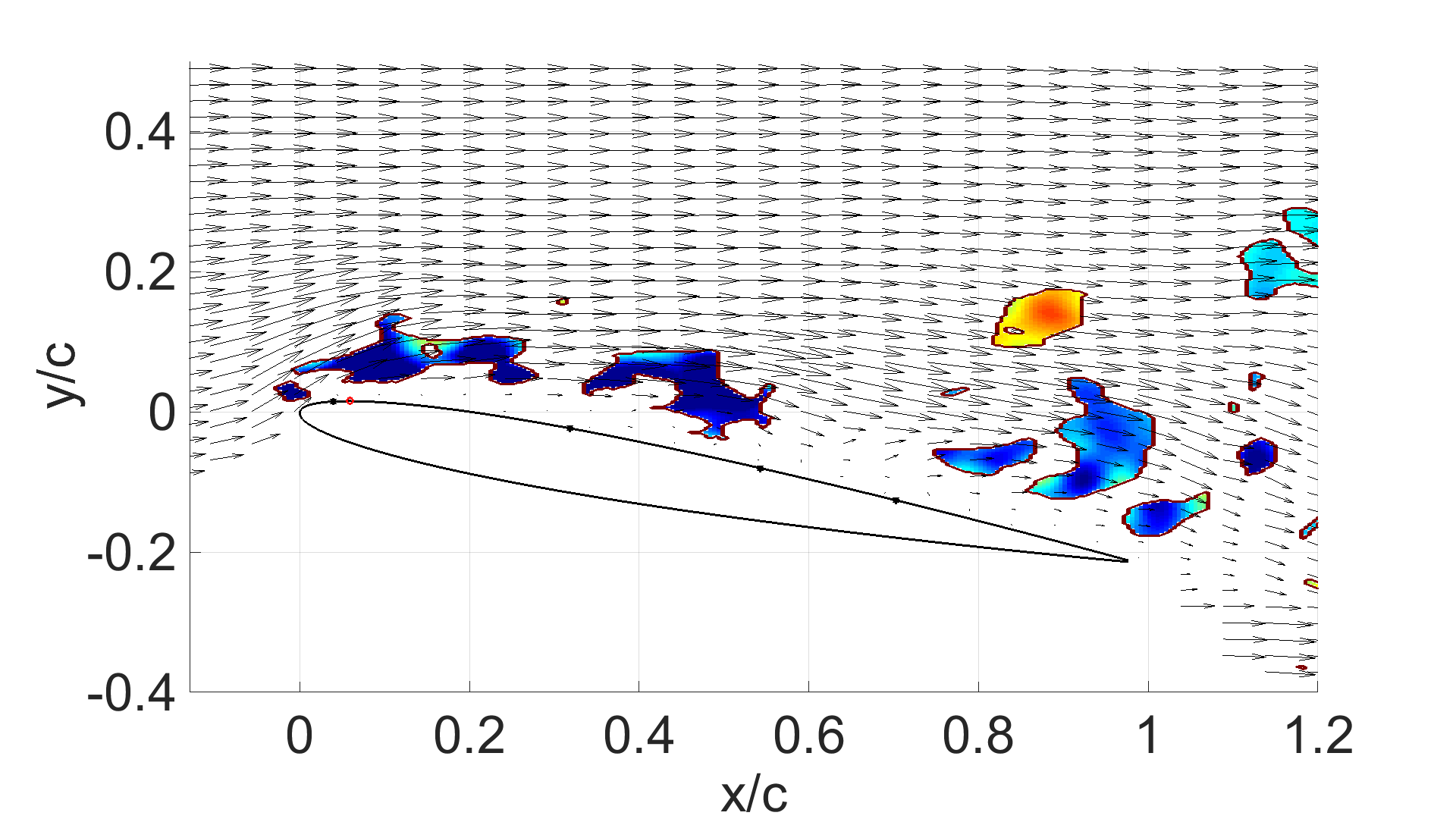}
		          \caption{CSF at $4.0t^+$}
		          \label{fig:Gamma_conv_tplus_4_1p25}
		\end{subfigure}

		\begin{subfigure}{0.3\textwidth}
		        \includegraphics[width=1.8in]{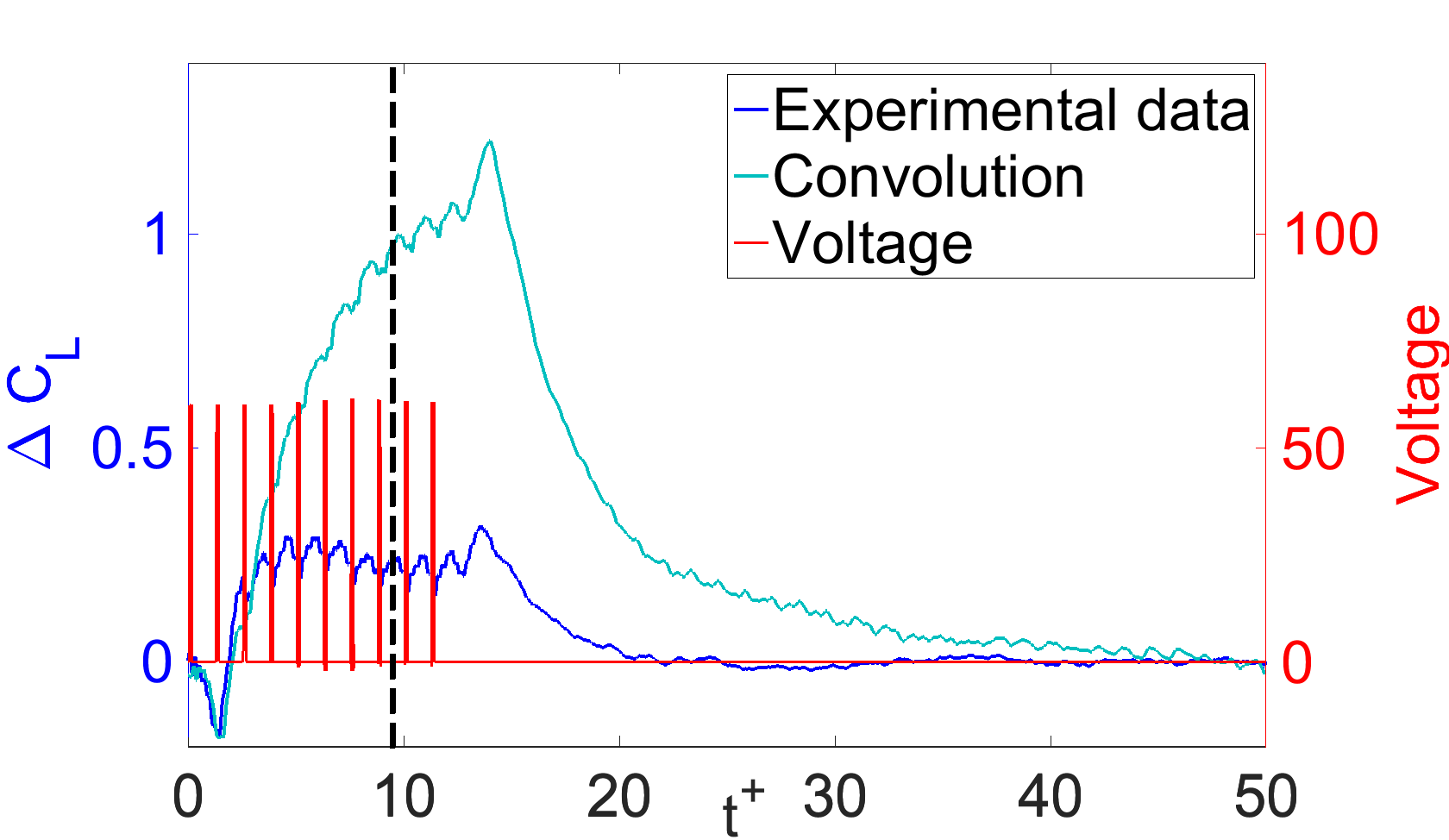}
		          \caption{$\Delta C_L$ at $9.5t^+$}
		          \label{fig:DCL_multi_1p25t+_9p5t+}
		\end{subfigure}	
		~
		\begin{subfigure}{0.3\textwidth}
		        \includegraphics[width=1.8in]{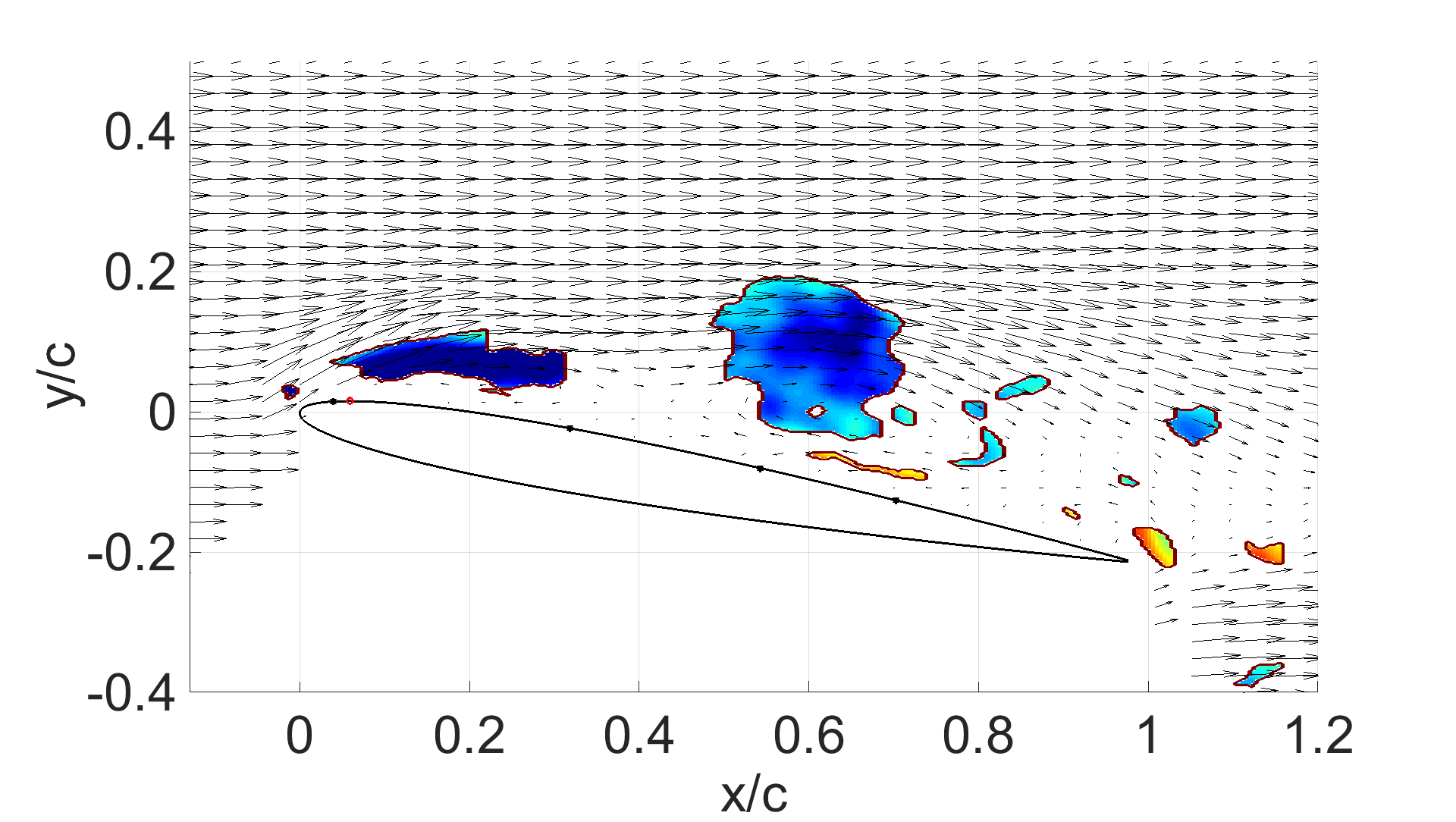}
		          \caption{DMF at $9.5t^+$}
		          \label{fig:Gamma_PIV_tplus_9p5_1p25}
		\end{subfigure}		
		~
		\begin{subfigure}{0.3\textwidth}
		        \includegraphics[width=1.8in]{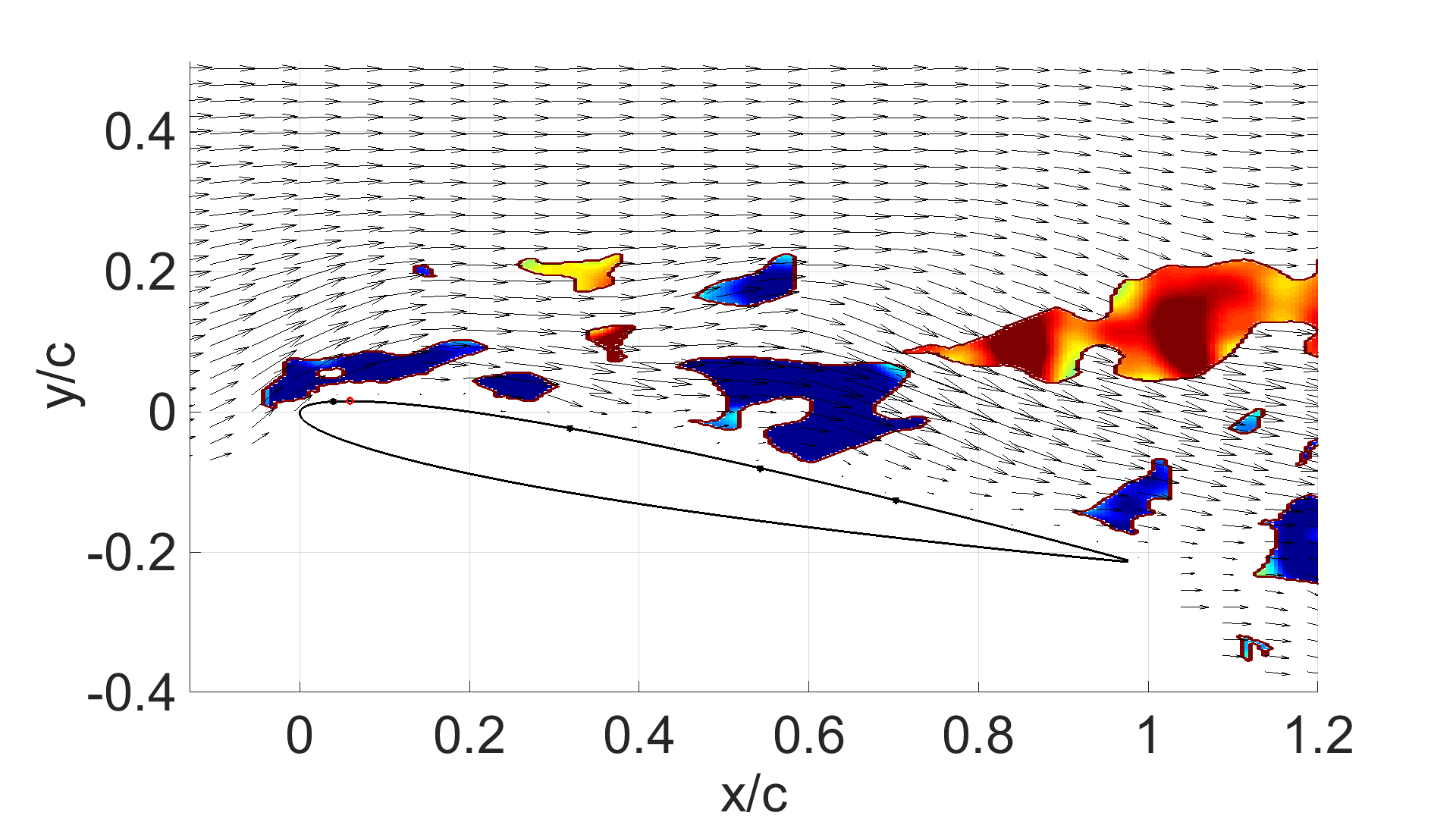}
		          \caption{CSF at $9.5t^+$}
		          \label{fig:Gamma_conv_tplus_9p5_1p25}
		\end{subfigure}

    \caption{Comparison of DMF and CSF, the burst frequency is $F^+=0.82$, which is equivalent to $1.25t^+$ burst interval. The figures on the left are the directly measured and convolution simulated $\Delta C_L$ (solid deep blue line and solid cyan), corresponding to the multi-burst input signal (solid red line). The dashed black lines indicate the instants associated with the DMF (in the middle) and the CSF (on the right), the black arrows indicate the velocity vectors and the color contour indicates the vortex strength.}
    \label{fig:conv_Gamma_1p25}		
\end{figure}

The convolution integral is applied to the $F^+=0.82$ multi-burst case, which is the actuation with the highest burst frequency.  This case shows the strongest burst-burst interaction.  The results are organized into three columns in figure \ref{fig:conv_Gamma_1p25}.  The first column contains the lift coefficient increment with a dashed vertical line marking the time corresponding to each row. The second and third columns show the PIV direct measured flowfield (DMF) and the convolution simulated flowfield (CSF), respectively.  The vortex structure is plotted on top of the velocity vectors in columns two and three. 

We expect the DMF and CSF flowfields to be similar at short times after the onset of actuation, when the measured and convolution simulated $\Delta C_L$ in figure \ref{fig:conv_Gamma_1p25} are similar, because the flowfield evolution is responsible for the $\Delta C_L$ variation. At times $0.0t^+, 2.0t^+,$ and $2.7t^+$ shown in the first three rows of figure \ref{fig:conv_Gamma_1p25}, one can observe similarities in the flow field structures of concentrated vorticity. To provide more detail, the dashed black line shown in figure \ref{fig:DCL_multi_1p25t+_0t+} denotes the instant at $0t^+$ where the $\Delta C_L$ has not been disturbed. The DMF at $0t^+$ is shown in figure \ref{fig:Gamma_PIV_tplus_0_1p25} and the CSF is shown in figure \ref{fig:Gamma_conv_tplus_0_1p25}. The DMF and CSF exhibit a very strong similarity at $0t^+$, since the burst signal was not triggered yet. At $2t^+$ after the initializing of the burst signal, the DMF still shows a strong similarity compared to the CSF despite the firing of the second burst, as it is shown in figure \ref{fig:Gamma_PIV_tplus_2_1p25} and figure \ref{fig:Gamma_conv_tplus_2_1p25}. The $\Delta C_L$ plotted at $2t^+$ (figure \ref{fig:DCL_multi_1p25t+_2t+}) exhibits the same trend as the flowfield, the direct measured and convolution simulated $\Delta C_L$ is very close. After $2.7t^+$ the CSF starts to deviate from the DMF, and so do the direct measured and the convolution simulated $\Delta C_L$ (figure \ref{fig:DCL_multi_1p25t+_2p7t+} to figure \ref{fig:Gamma_conv_tplus_9p5_1p25}).  

At the time $4.0t^+$ (fourth row) the predicted lift increment begins to deviate from the measured lift, and the directly measured and convolution simulated flow fields are also different. In particular, the measured flow field shows a well organized vortex at the trailing edge that is not reproduced by the convolution approach.


In order to better understand the differences between the directly measured flow and the convolution simulated flow, we will first try to explain the CSF. In the single-burst case there are two distinct time scales. The vortex formation and shedding that contribute to the initial lift reversal, and the following $\Delta C_L$ increment occurs in a short time scale ($O(1) t^+$) as shown in figure \ref{fig:Gamma2_0} to figure \ref{fig:Gamma2_2p8}. But the process of recovery to the original undisturbed separated flow takes much longer ($O(10) t^+$), which is shown in figure \ref{fig:Gamma2_2p8} abd figure \ref{fig:Gamma2_20}. This asymmetric $\Delta C_L$ response (in terms of time required for $C_L$ to increase and decrease) following the single-burst actuation leads to the overprediction of $C_L$ in the multi-burst case when the convolution method is used for prediction. 


In a multi-burst sequence the first burst produces the detached leading edge vortex (DLEV) which begins to convect downstream.  The second burst in the sequence will not directly affect the DLEV, because the actuation occurs upstream of the convecting DLEV. But if the second burst occurs before the flowfield has returned to its non-excited condition (which takes about $10t^+$),  then the initial condition for the lift coefficient increment in the second burst is larger than zero in the convolution integral. The process repeats for each successive burst, and the lift coefficient increment continues to increase artificially until the burst sequence is complete. 


In reality in a multi-burst sequence, the directly measured flowfield indicates that the second burst indeed has a strong influence on the reattached shear layer caused by the previous burst. It breaks the previous burst-induced reattached shear layer into two parts, and attempts to establish a new reattachment. On the other hand, in the DMF, the upstream current burst-induced vortex formation and advection will have a weaker influence on the downstream DLEV caused by the previous burst, which is similar to the CSF. 

Taking a close look at figure \ref{fig:Gamma_PIV_tplus_2p7_1p25} to figure \ref{fig:Gamma_conv_tplus_9p5_1p25}, one can observe that the clockwise rotational vortices' location and strength for both DMF and CSF show a certain level of similarity. But there is a high velocity region that covers the entire chord length above the upper surface of the wing in the CSF (figure \ref{fig:Gamma_conv_tplus_2p7_1p25}, figure \ref{fig:Gamma_conv_tplus_4_1p25} and figure \ref{fig:Gamma_conv_tplus_9p5_1p25}) which differs from the DMF (figure \ref{fig:Gamma_PIV_tplus_2p7_1p25}, figure \ref{fig:Gamma_PIV_tplus_4_1p25} and figure \ref{fig:Gamma_PIV_tplus_9p5_1p25}). This high-speed region is also in response of producing the counterclockwise rotational vortices above the trailing-edge (at $y/c\approx0.1$) in CSF. Therefore, one can speculate that the high-speed region above the wing in the CSF contributes to a higher mean lift increment than the DMF. The difference between the CSF and the DMF on this high-speed region implies the nonlinear portion of burst-burst interaction. But the similar clockwise rotating vortex structure within the DMF and CSF produces similar high-frequency periodic oscillation associated with the burst signal in $\Delta C_L$, which indicates the linear portion of burst-burst interaction. Next, to quantitatively analyze the linear and nonlinear dynamics of the multi-burst actuation, we will decompose the flowfield into different modes based on their dynamic characteristics.


\FloatBarrier

\subsection{Dynamic modal analysis for DMF and CSF}
The singular value decomposition (SVD) based dynamic mode decomposition (DMD) \citep{rowley2009spectral} \citep{schmid2010dynamic} \citep{tu2013dynamic} was employed to identify the dynamic modes for both the DMF and the CSF. In general, DMD is not suitable for the dynamic system containing nonlinear transitions. However, if the nonlinear transition only occurs once from one equilibrium state to another, then this transition may be approximated by decaying oscillating modes, whose initial value is "1" and the final value is "0". Hence, DMD was performed on the velocity field of the $F^+=0.82$ multi-burst cases from $0t^+$ to $12t^+$, it can be seen that there is only one transitional event from $0t^+$ to $12t^+$ in figure \ref{fig:DCL_multi_1p25t+}.   

The eigenvalues of the critical DMD modes of the DMF and the CSF are shown in table \ref{table:eig_PIV} and table \ref{table:eig_conv}. The critical modes of both cases are defined as those which can reconstruct the original dataset with the least number of modes. 
Table \ref{table:DMD_multi_S} shows the correlation coefficients of the spatial modes between DMF and CSF. Table \ref{table:DMD_multi_T} shows the correlation coefficients of the temporal coefficients associated with the spatial modes between DMF and CSF.

\begin{table}
\centering 
\begin{tabular}{lccc}            
{Mode 0}
   & Mode 2 & Mode 14 & Mode 18 \\ [0.5ex] 
 -0.0010+0.0000i &-7.8899+9.2895i &-3.6004-8.1792i & -0.0827-62.9156i \\              

\end{tabular}

\caption{Eigenvalues of DMF's DMD modes.}   
\label{table:eig_PIV}                
\end{table} 

\begin{table}
\centering 
\begin{tabular}{lccc}            
{Mode 0}
   & Mode 1 & Mode 8 & Mode 18 \\ [0.5ex] 
 0.0253+0.0000i &-2.5608+0.0000i &-0.2786-62.2852i & -4.6605-16.1187i \\              

\end{tabular}

\caption{Eigenvalues of CSF's DMD modes.}   
\label{table:eig_conv}                
\end{table} 
\FloatBarrier 
 
\begin{figure}
	\centering
	
		\begin{subfigure}{0.45\textwidth}
		        \includegraphics[width=2.6in]{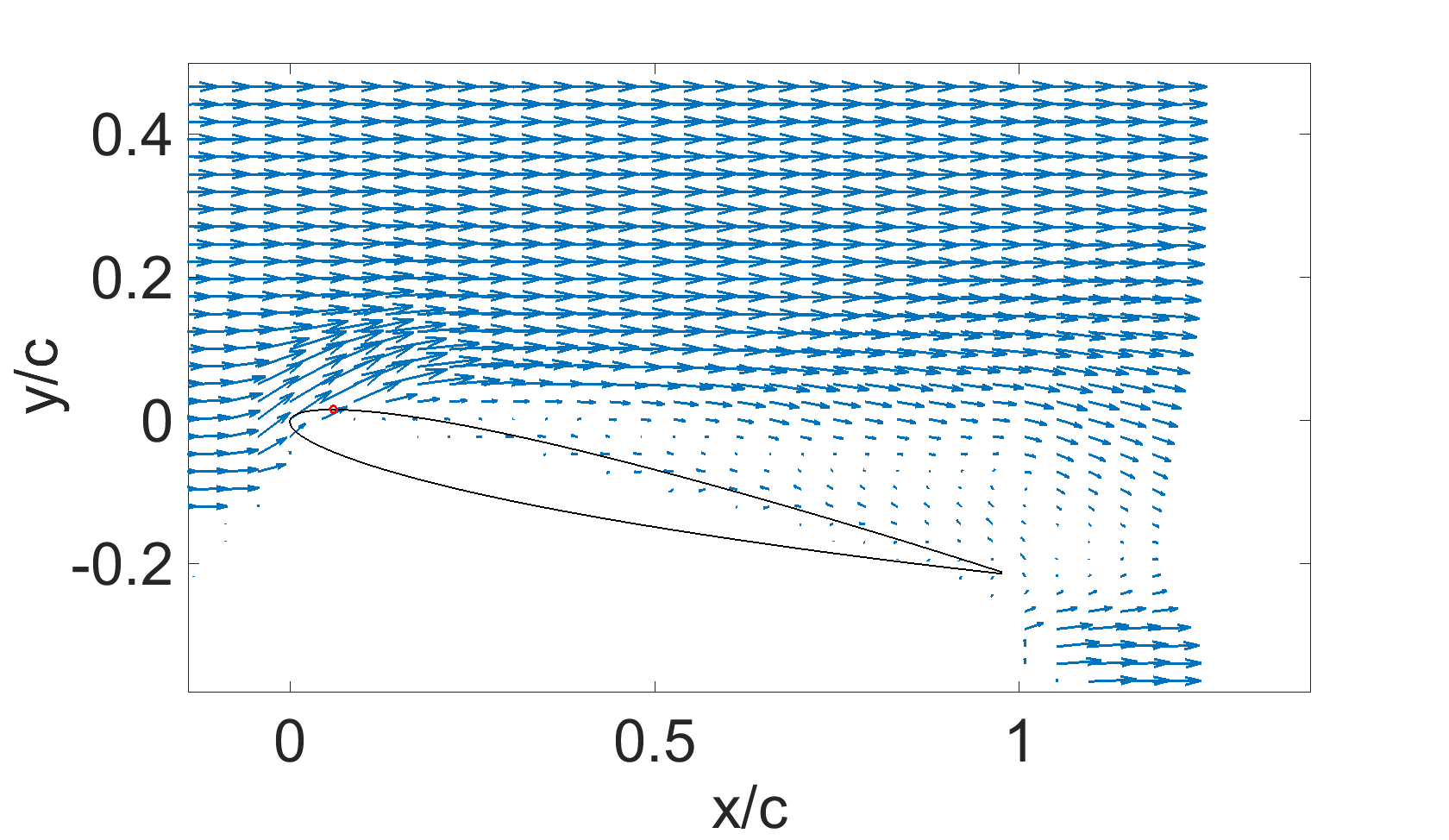}
		          \caption{Mode 0 for DMF}
		          \label{fig:mode0_S_PIV}
		\end{subfigure}	
~
		\begin{subfigure}{0.45\textwidth}
		        \includegraphics[width=2.6in]{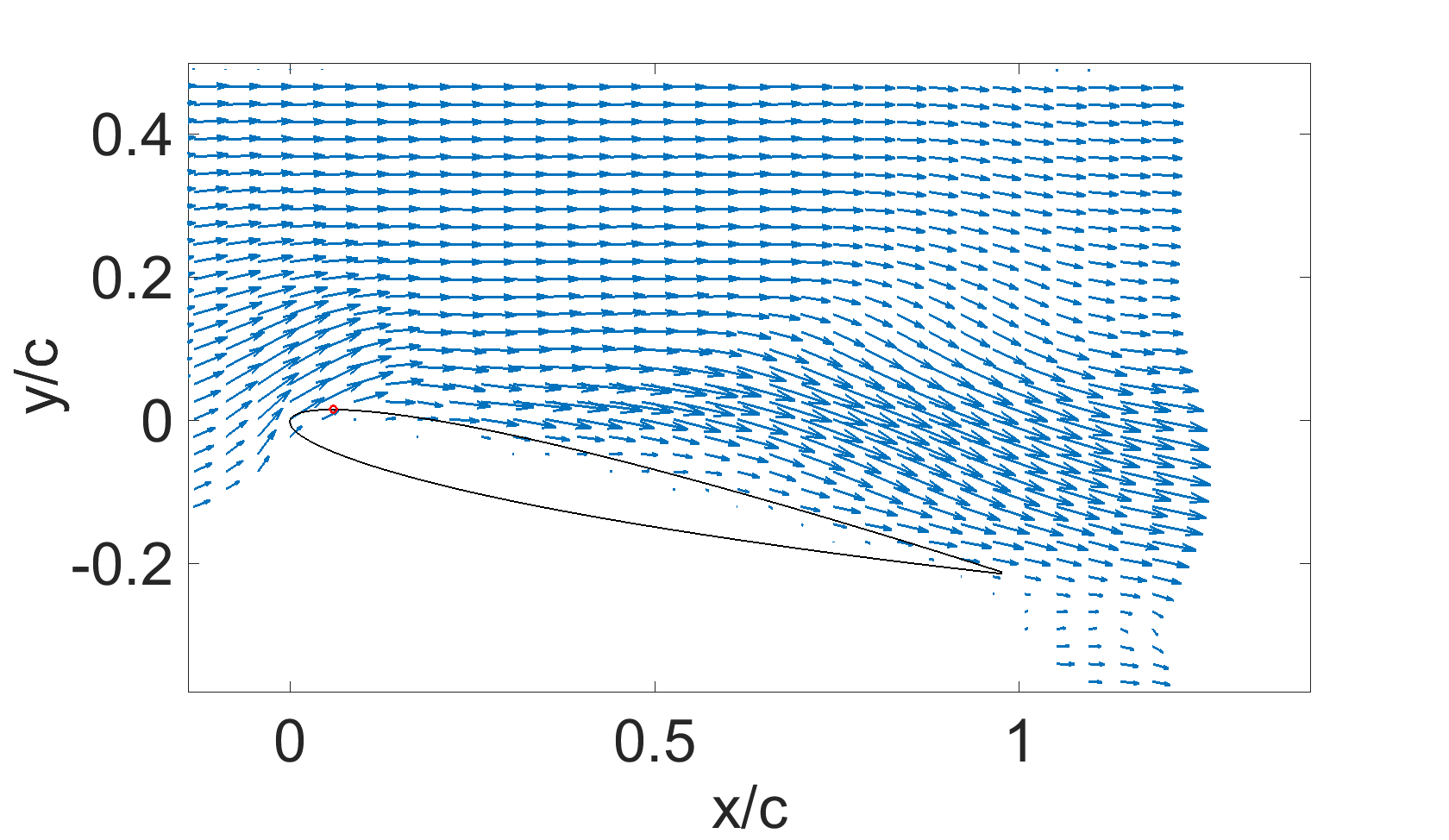}
		          \caption{Mode 0 for CSF}
		          \label{fig:mode0_S_conv}
		\end{subfigure}	
		
		\begin{subfigure}{0.45\textwidth}
		        \includegraphics[width=2.6in]{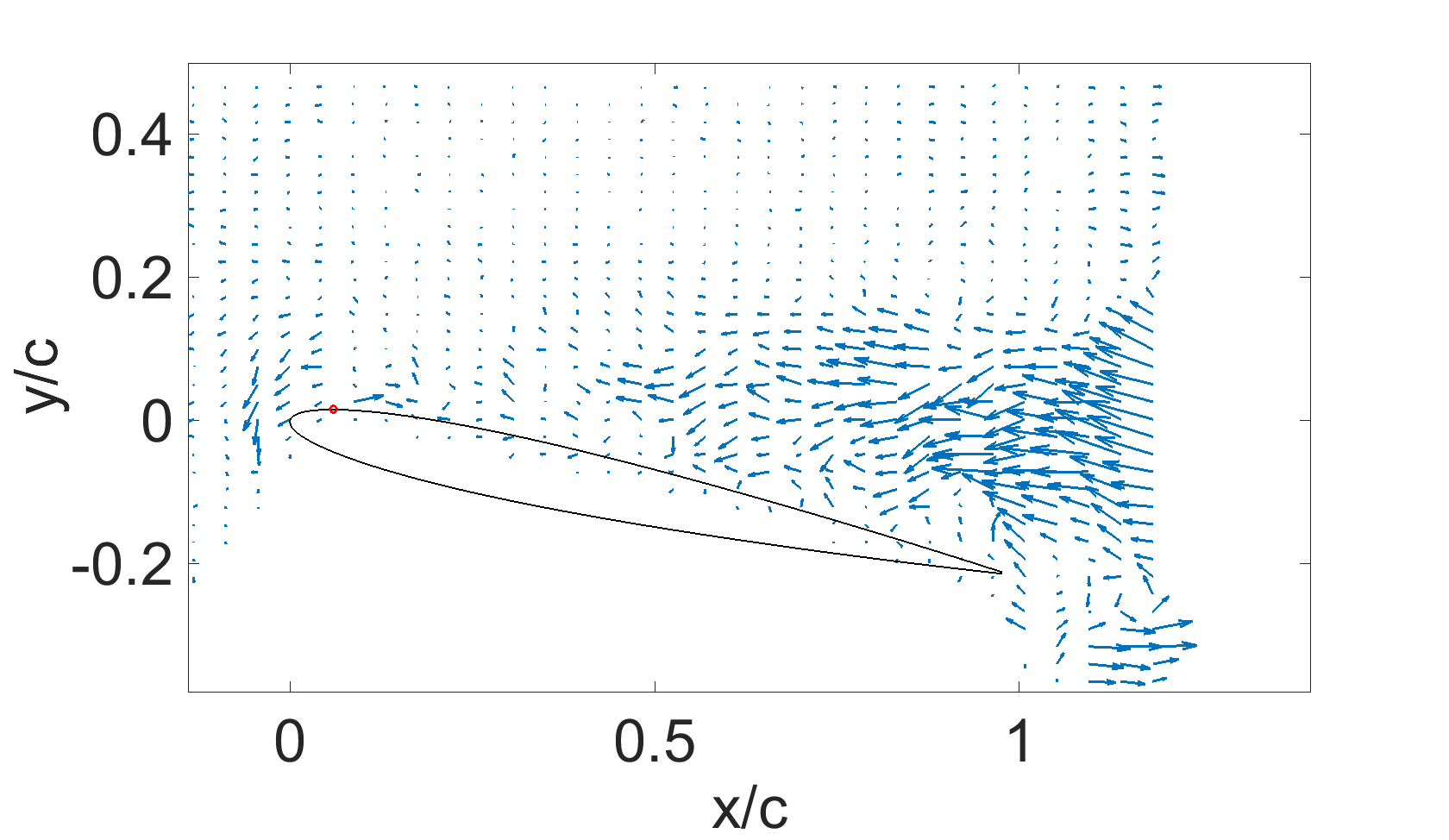}
		          \caption{Mode 2 for DMF}
		          \label{fig:mode2_S_PIV}
		\end{subfigure}	
~		
		\begin{subfigure}{0.45\textwidth}
		        \includegraphics[width=2.6in]{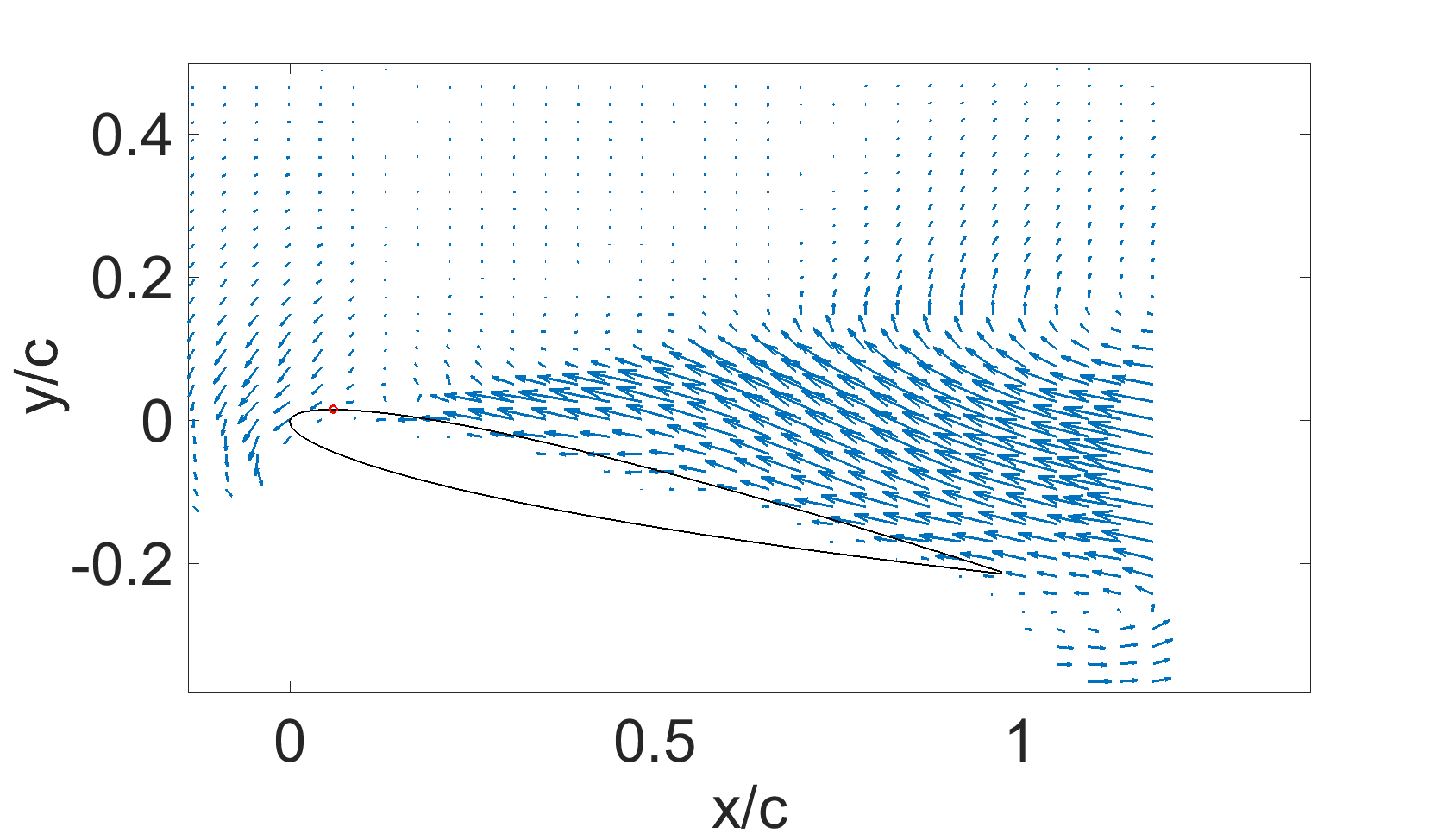}
		          \caption{Mode 1 for CSF}
		          \label{fig:mode1_S_conv}
		\end{subfigure}

		\begin{subfigure}{0.45\textwidth}
		        \includegraphics[width=2.6in]{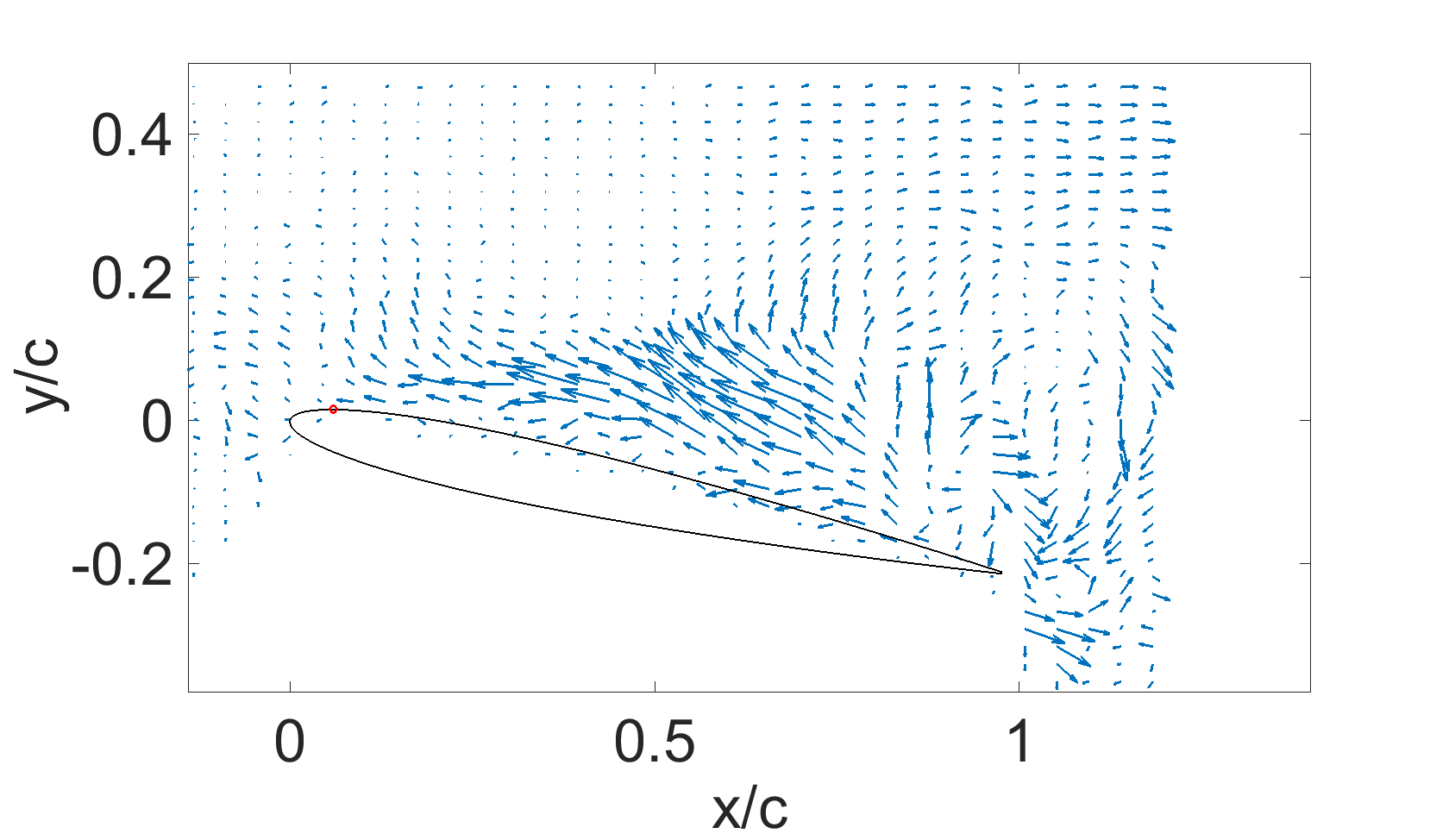}
		          \caption{Mode 14 for DMF}
		          \label{fig:mode14_S_PIV}
		\end{subfigure}	
~		
		\begin{subfigure}{0.45\textwidth}
		        \includegraphics[width=2.6in]{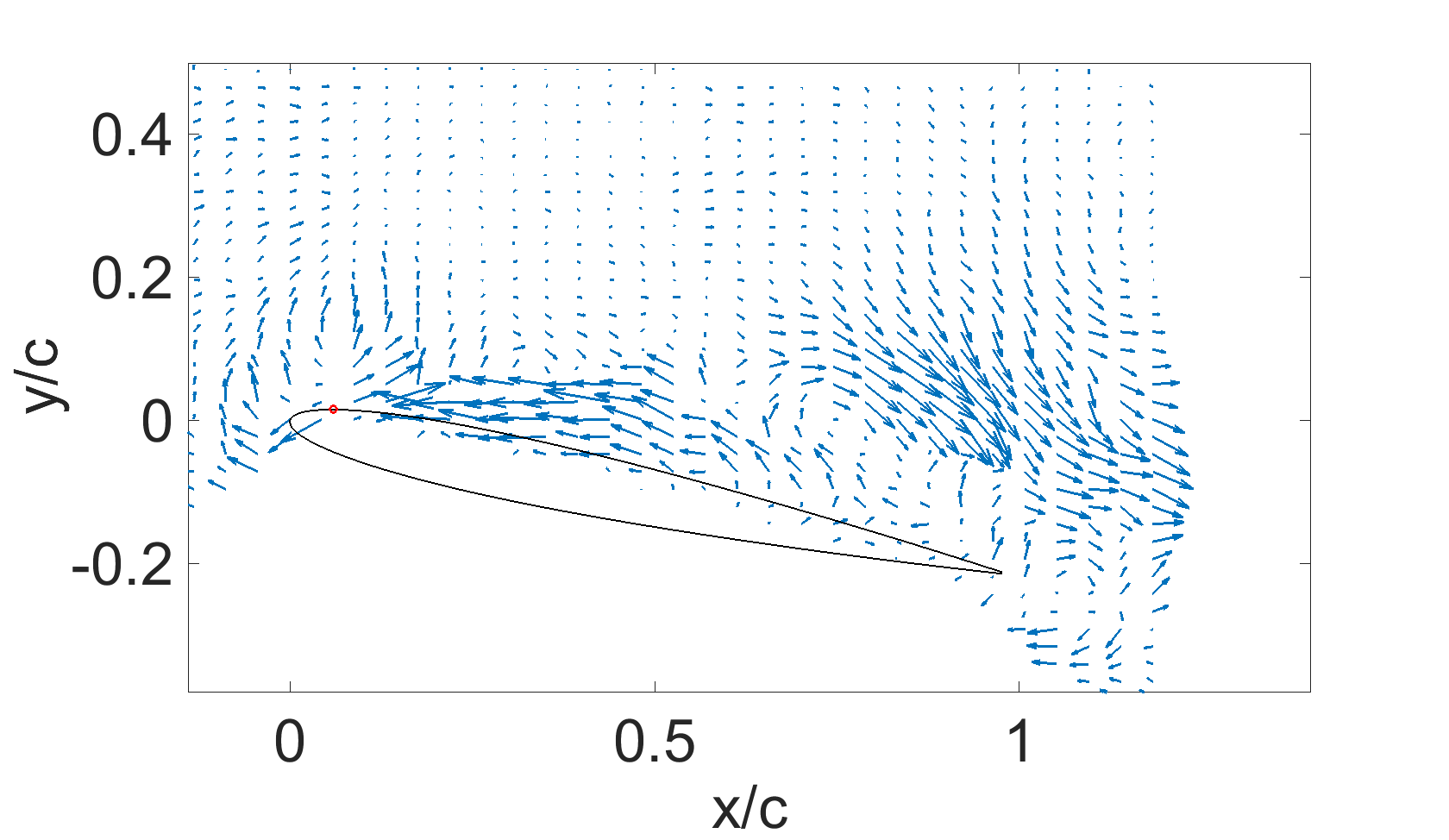}
		          \caption{Mode 18 for CSF}
		          \label{fig:mode18_S_conv}
		\end{subfigure}

		\begin{subfigure}{0.45\textwidth}
		        \includegraphics[width=2.6in]{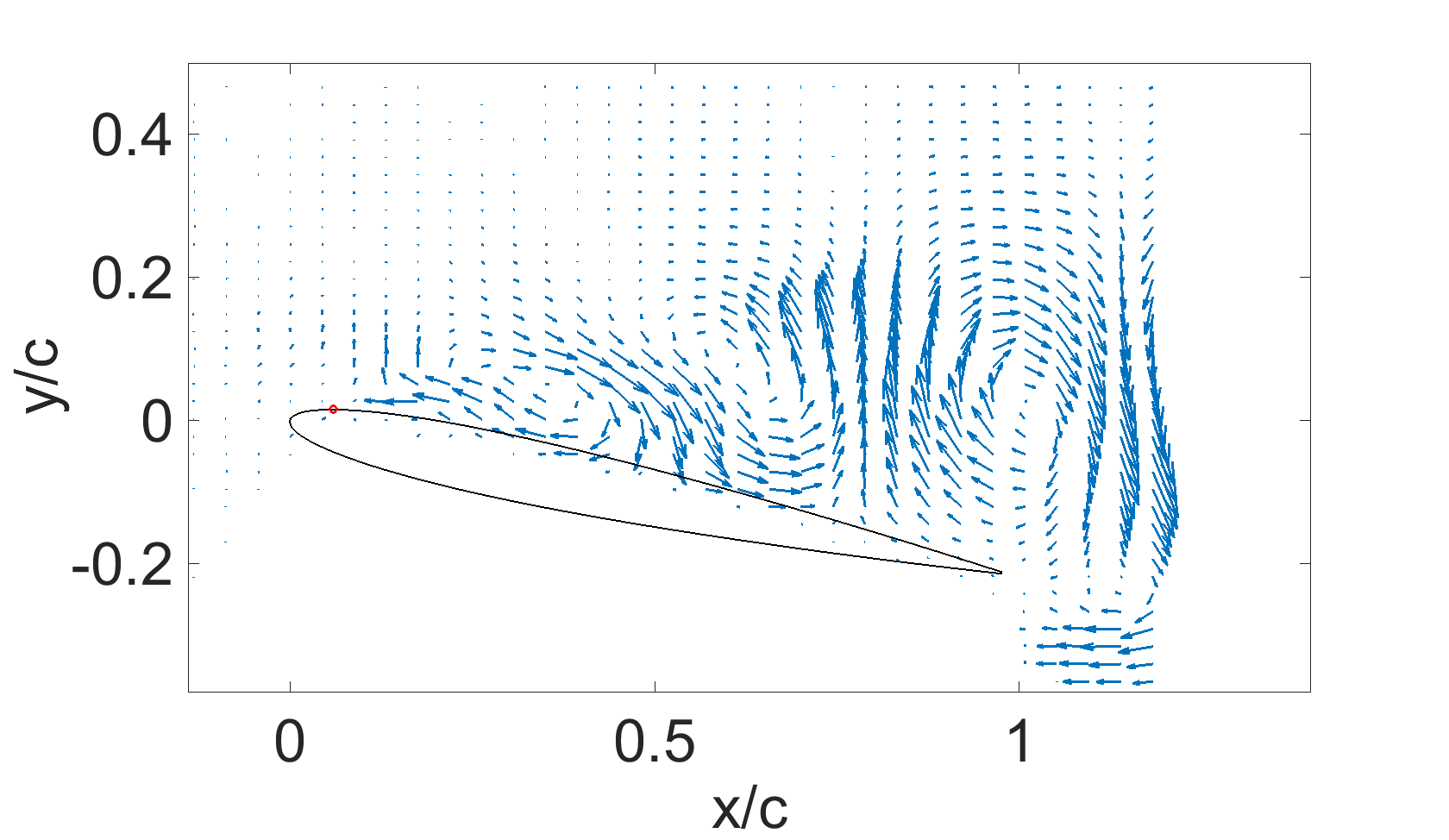}
		          \caption{Mode 18 for the DMF}
		          \label{fig:mode18_S_PIV}
		\end{subfigure}	
~				
		\begin{subfigure}{0.45\textwidth}
		        \includegraphics[width=2.6in]{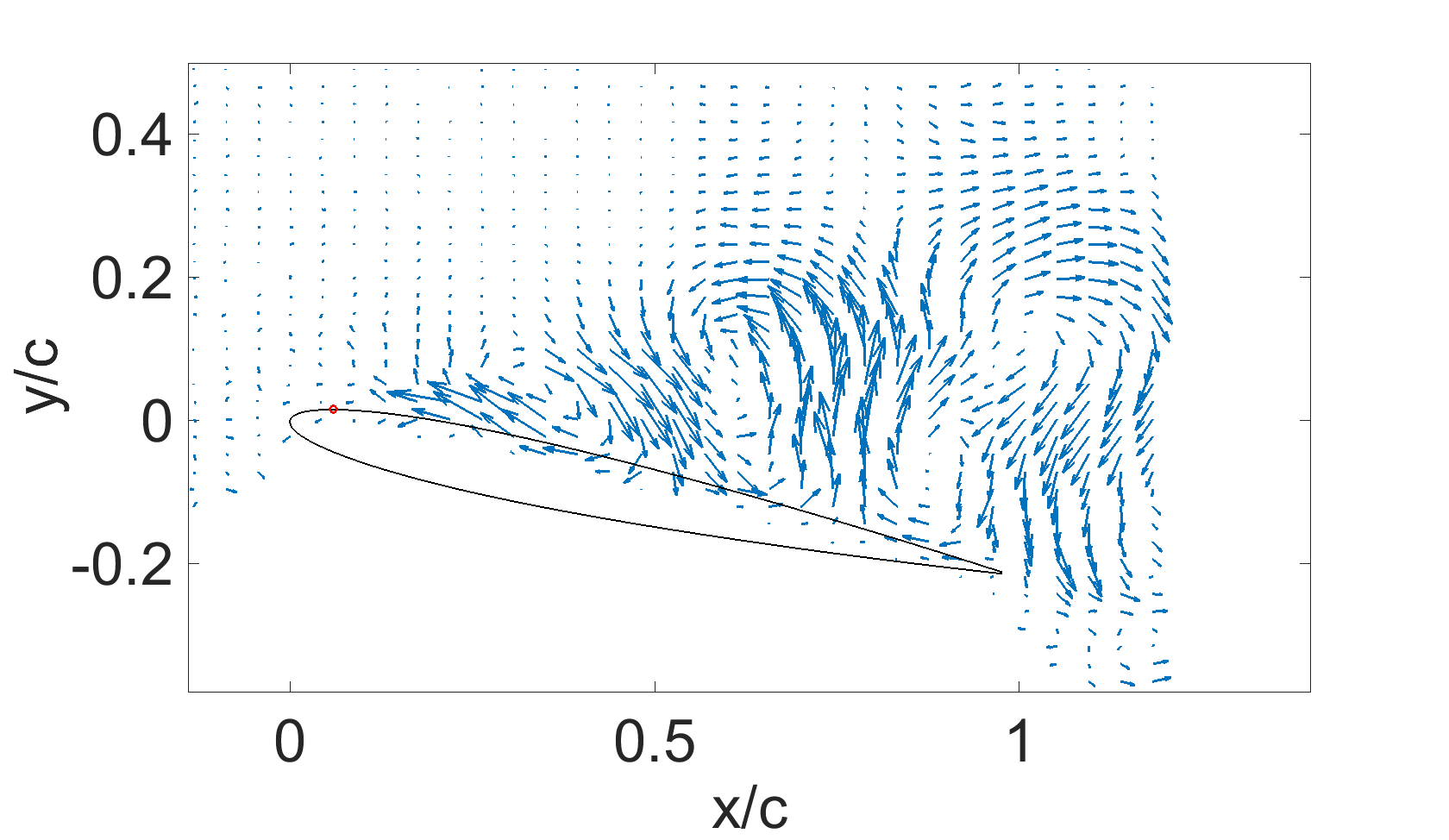}
		          \caption{Mode 8 for CSF}
		          \label{fig:mode8_S_conv}
		\end{subfigure}

	\caption{The comparison of the spatial DMD modes from the DMF and CSF. The arrows in each figure indicate the velocity magnitude and direction}
    \label{fig:DMD_multi_S}		
\end{figure}

\begin{figure}
	\centering
	
		\begin{subfigure}{0.45\textwidth}
		        \includegraphics[width=2.6in]{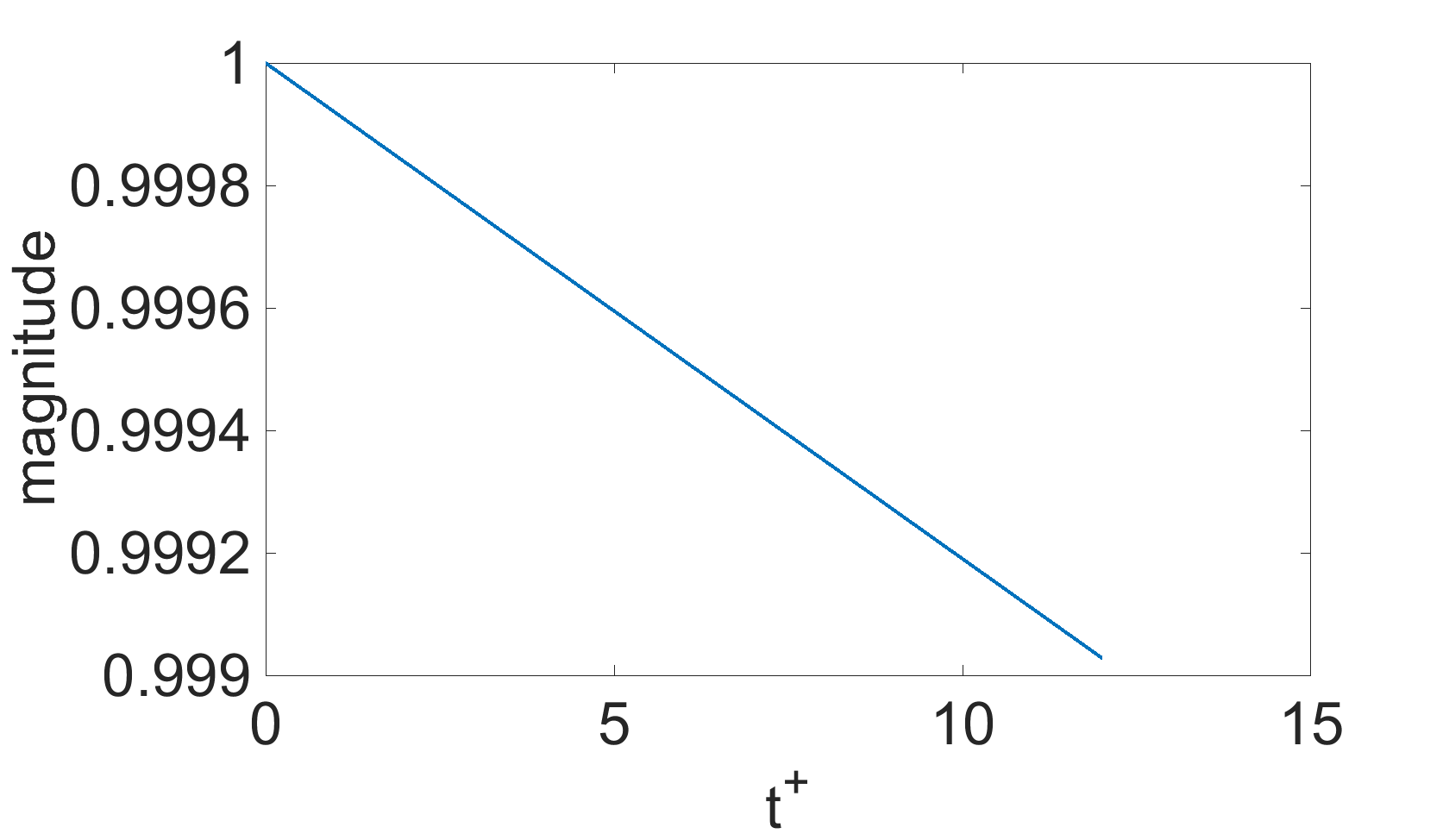}
		          \caption{Mode 0 for DMF}
		          \label{fig:mode0_T_PIV}
		\end{subfigure}	
~
		\begin{subfigure}{0.45\textwidth}
		        \includegraphics[width=2.6in]{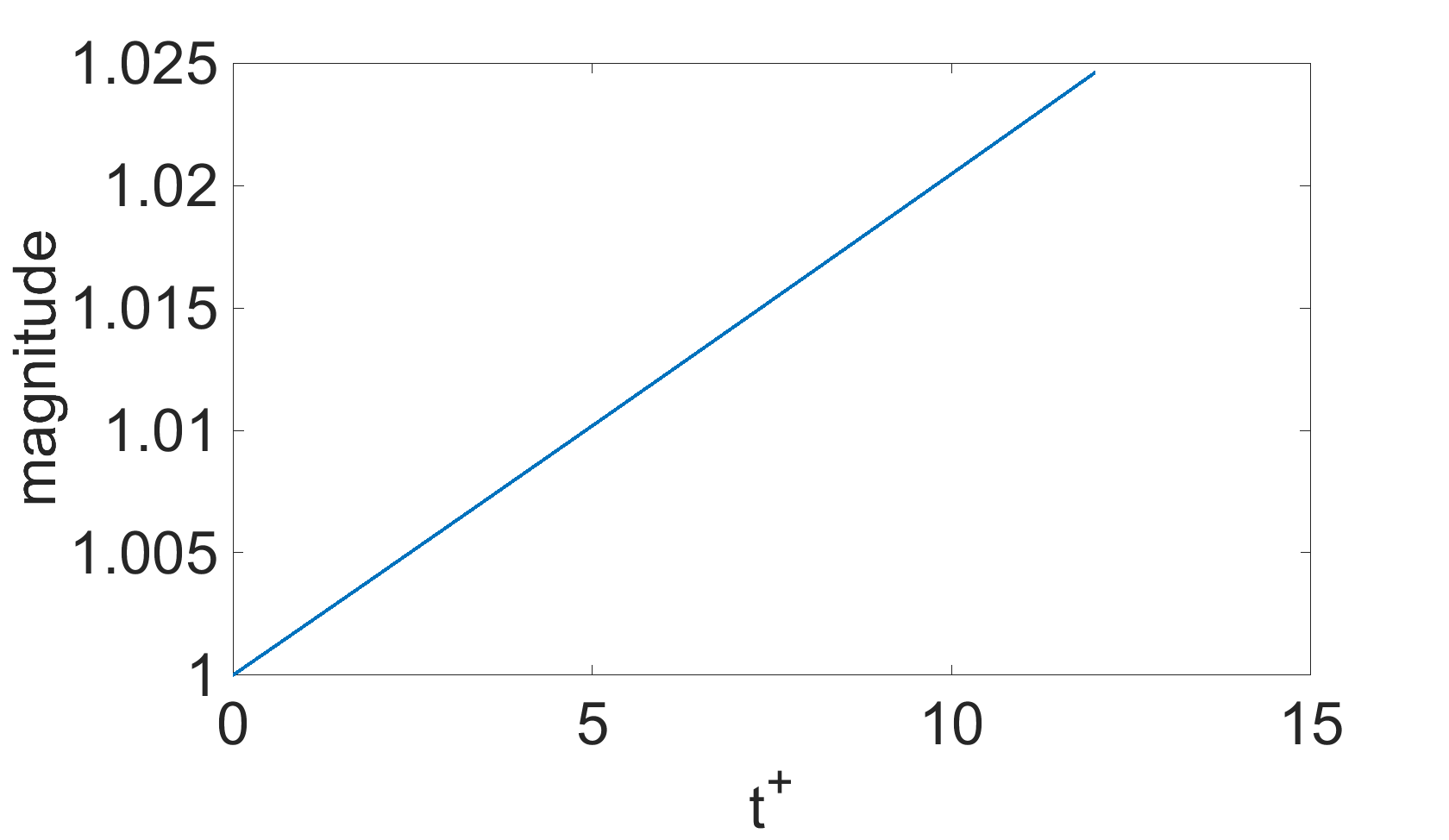}
		          \caption{Mode 0 for CSF}
		          \label{fig:mode0_T_conv}
		\end{subfigure}	
		
		\begin{subfigure}{0.45\textwidth}
		        \includegraphics[width=2.6in]{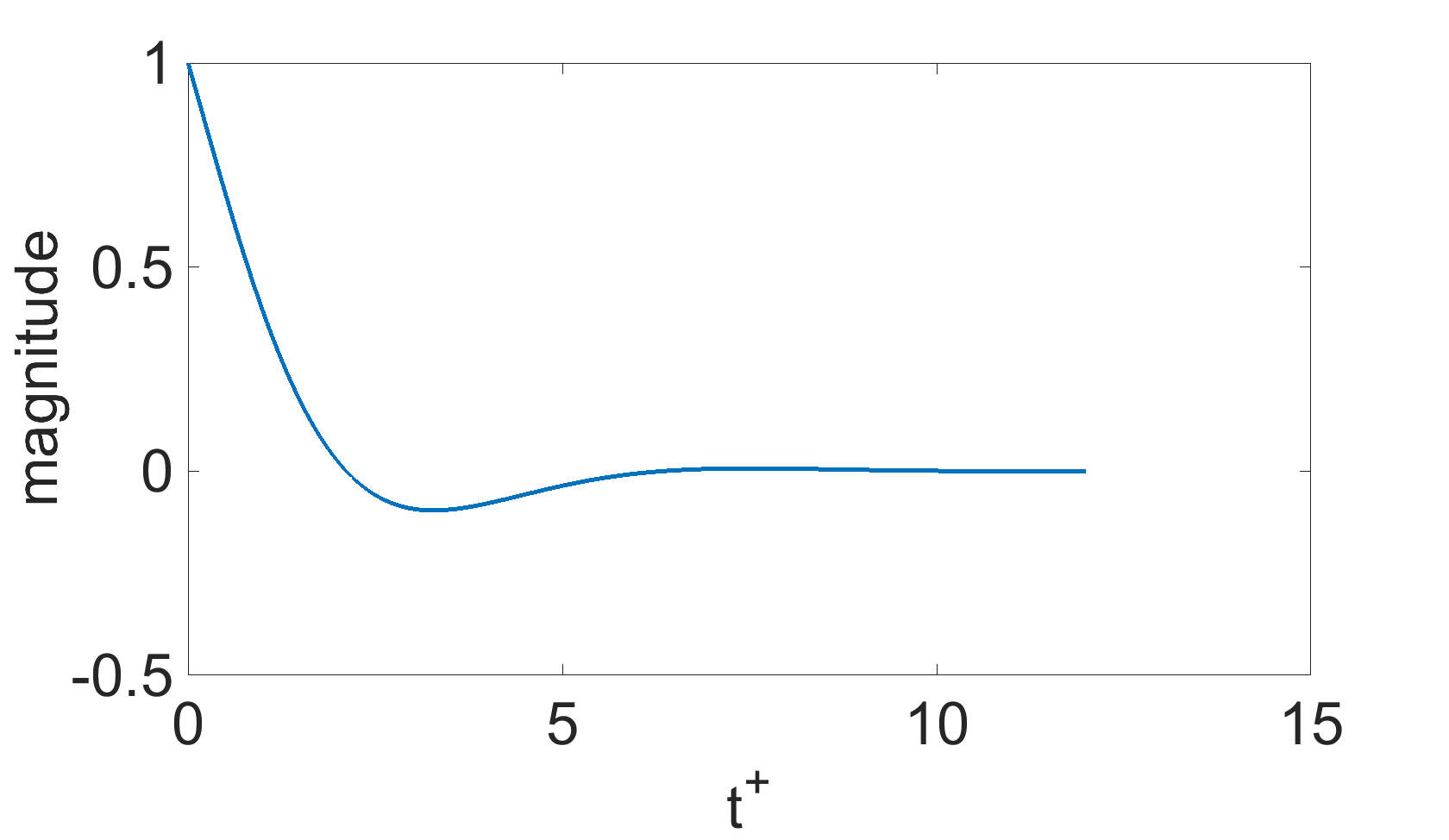}
		          \caption{Mode 2 for DMF}
		          \label{fig:mode2_T_PIV}
		\end{subfigure}	
~		
		\begin{subfigure}{0.45\textwidth}
		        \includegraphics[width=2.6in]{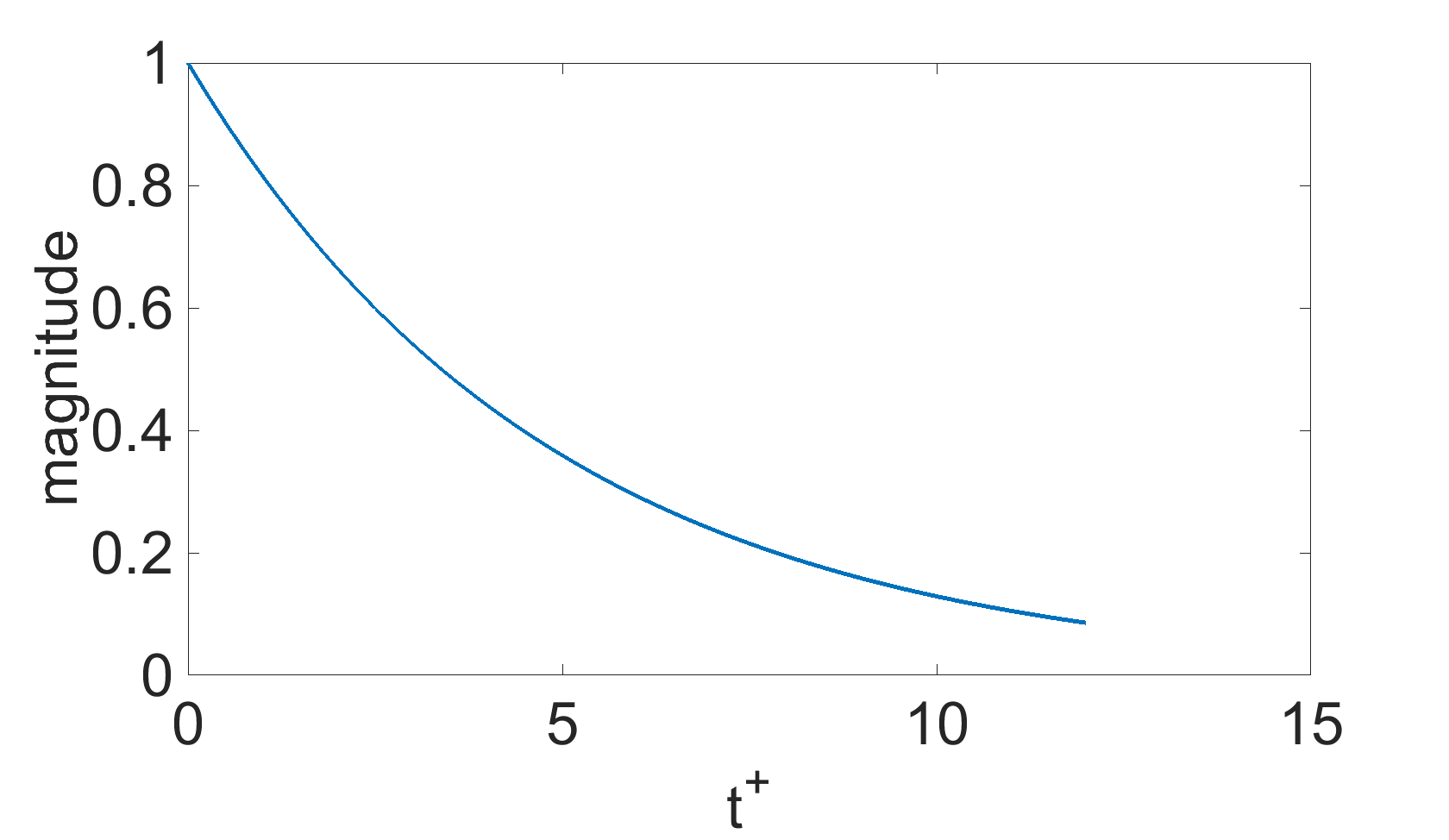}
		          \caption{Mode 1 for CSF}
		          \label{fig:mode1_T_conv}
		\end{subfigure}

		\begin{subfigure}{0.45\textwidth}
		        \includegraphics[width=2.6in]{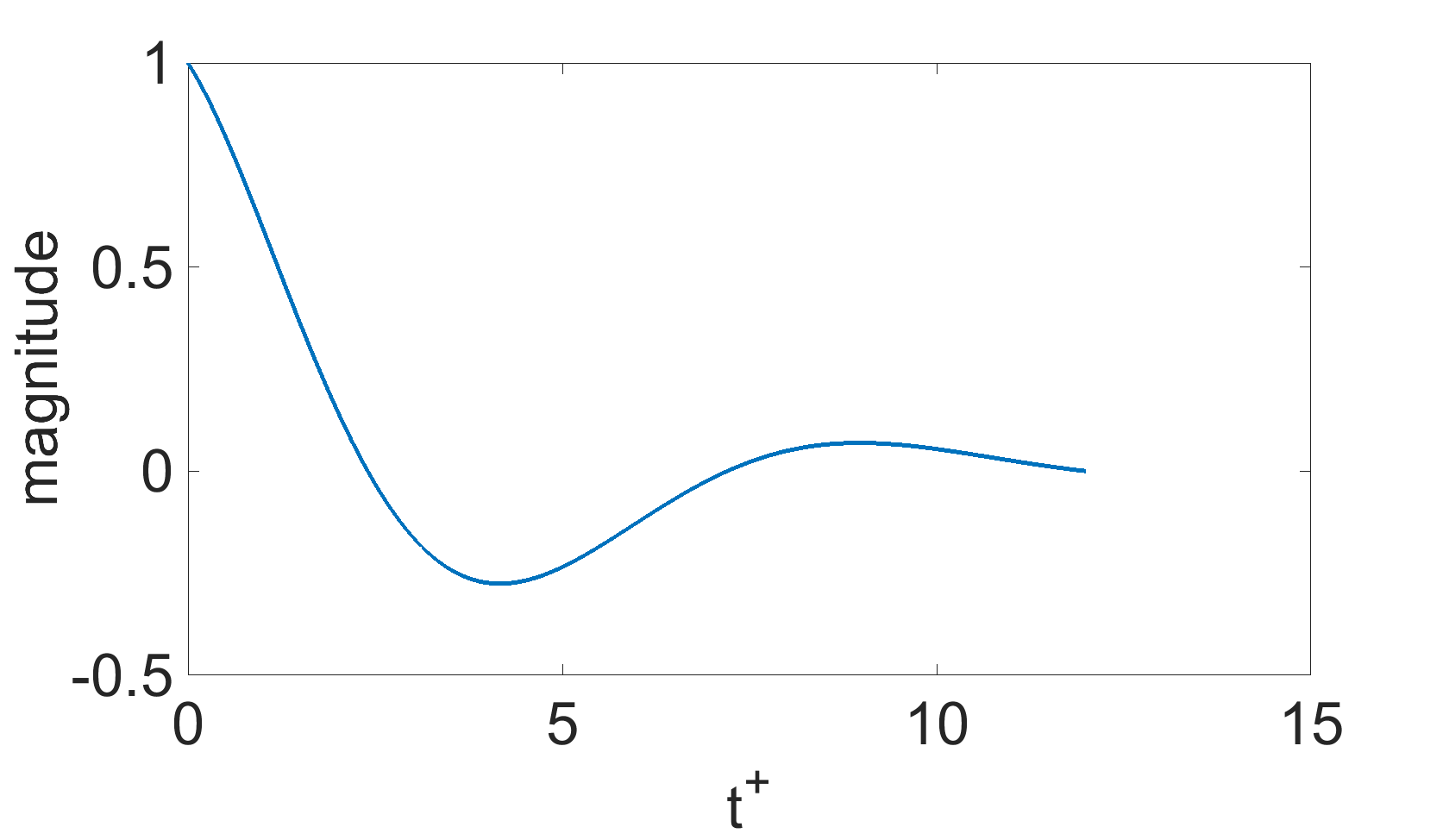}
		          \caption{Mode 14 for DMF}
		          \label{fig:mode14_T_PIV}
		\end{subfigure}	
~		
		\begin{subfigure}{0.45\textwidth}
		        \includegraphics[width=2.6in]{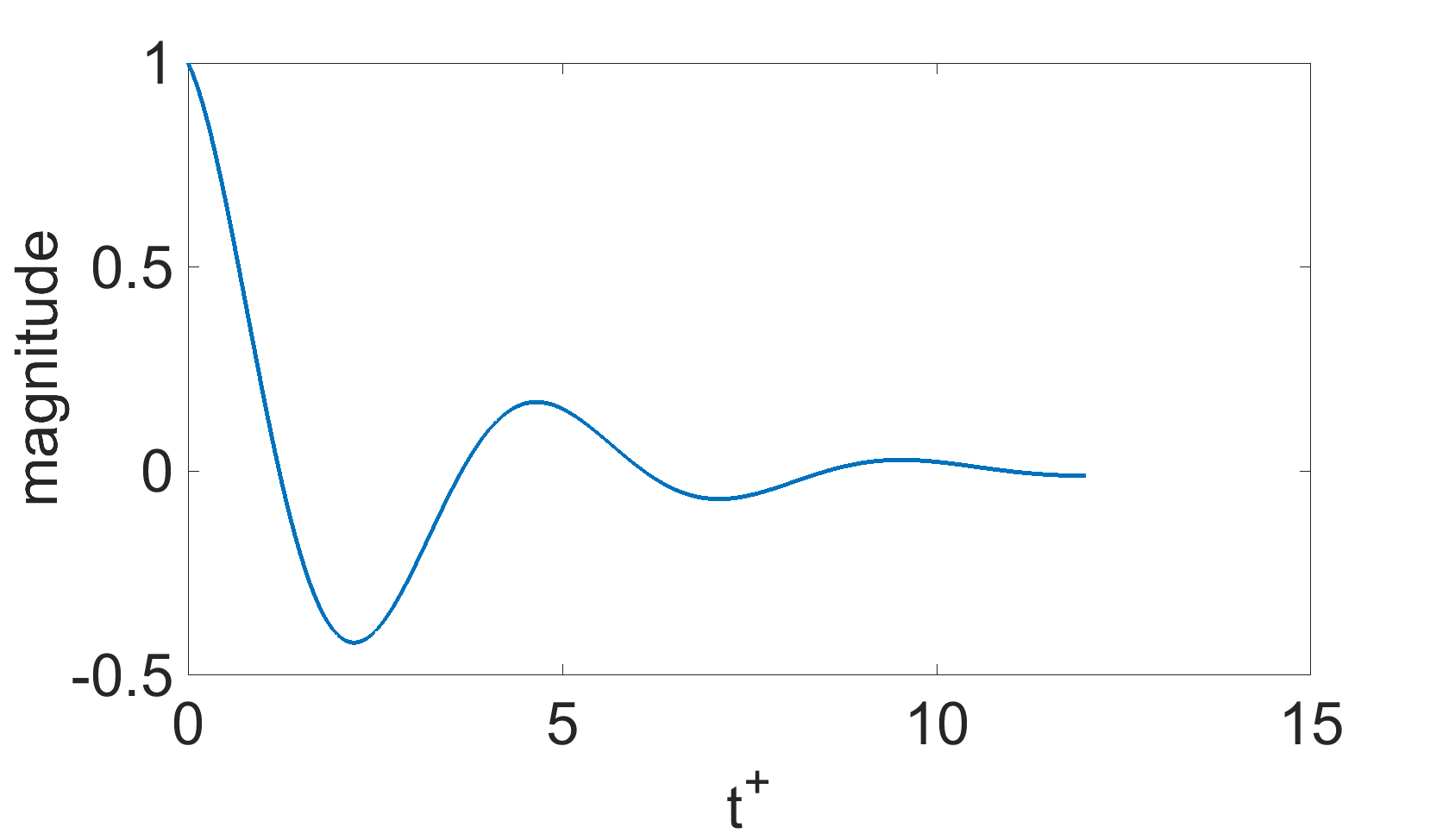}
		          \caption{Mode 18 for CSF}
		          \label{fig:mode18_T_conv}
		\end{subfigure}

		\begin{subfigure}{0.45\textwidth}
		        \includegraphics[width=2.6in]{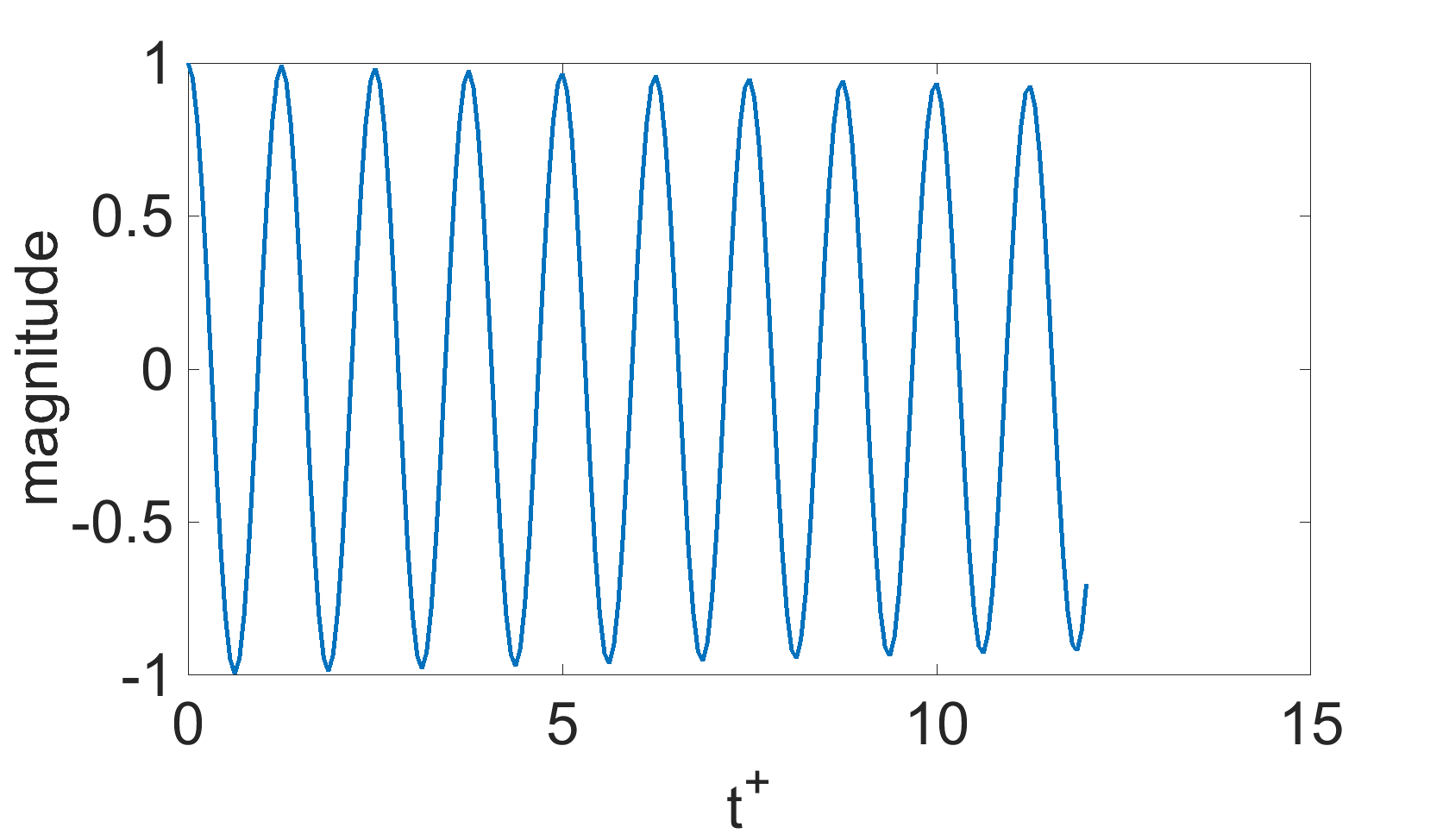}
		          \caption{Mode 18 for DMF}
		          \label{fig:mode18_T_PIV}
		\end{subfigure}	
~				
		\begin{subfigure}{0.45\textwidth}
		        \includegraphics[width=2.6in]{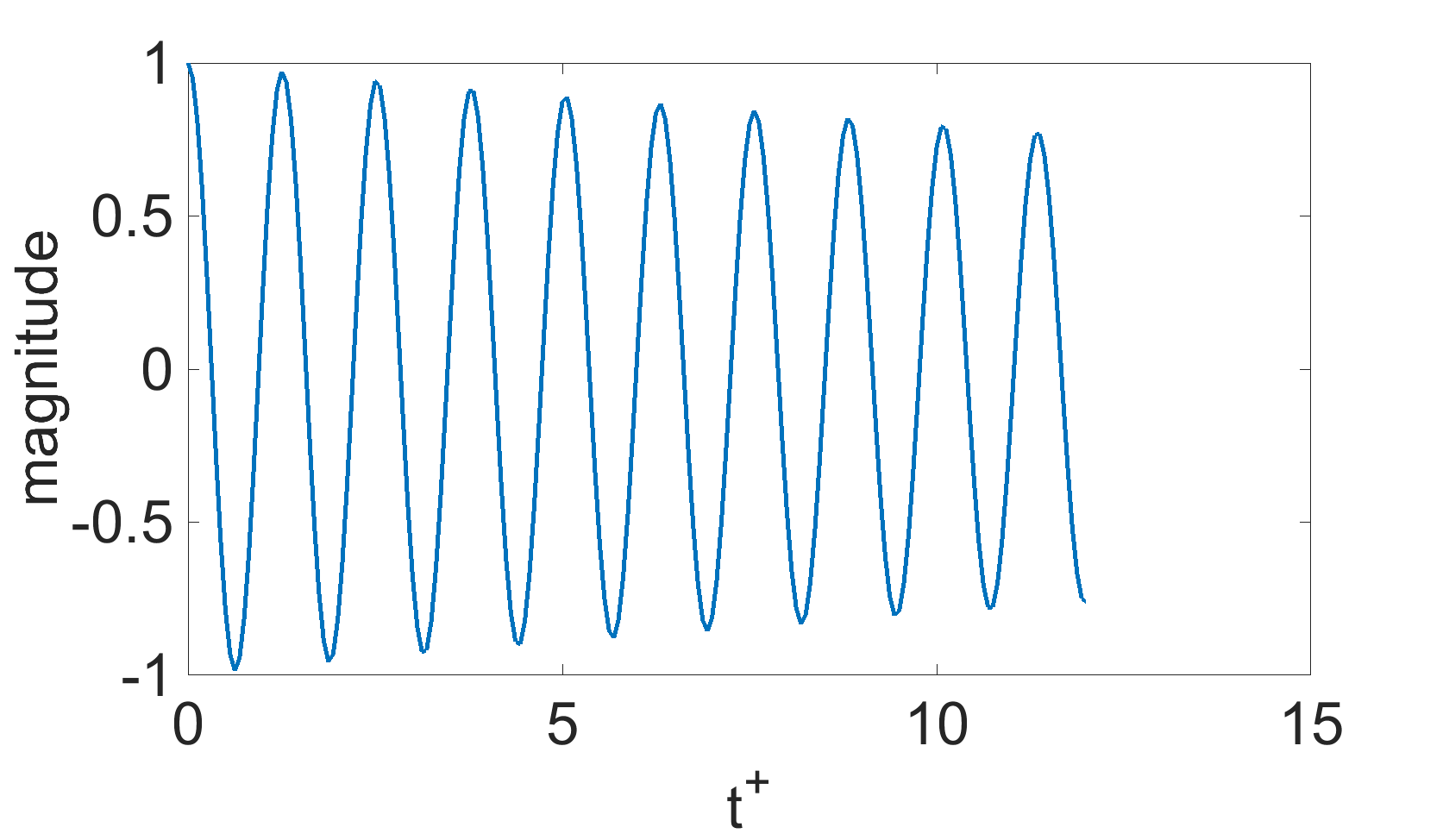}
		          \caption{Mode 8 for CSF}
		          \label{fig:mode8_T_conv}
		\end{subfigure}

	\caption{The comparison of the temporal coefficients of the DMD modes from DMF and CSF.}
    \label{fig:DMD_multi_T}		
\end{figure}

Table \ref{table:DMD_multi_S} exhibits a strong similarity between spatial mode 0 from DMF and spatial mode 0 from CSF. However, one could argue that they are different if we compare the spatial modes in figure \ref{fig:mode0_S_PIV} and figure \ref{fig:mode0_S_conv}. In fact, the high similarity is due to a large portion of these two modes that only contains the freestream flow, which is not affected by the actuation. However, the high-speed flow in CSF tends to be more attached than that in the DMF. This phenomenon partially contributes to the main trend of the $\Delta C_L$ difference for these two cases shown in figure \ref{fig:DCL_multi_1p25t+}. The correlation coefficient between the temporal coefficients of mode 0 from DMF and CSF is 1 (figure \ref{fig:mode0_T_PIV} and figure \ref{fig:mode0_T_conv}), since both of them are the linearly growth/decay modes without periodic oscillation, the temporal coefficients are linearly dependent.       
 
Table \ref{table:DMD_multi_S}, also shows that the spatial mode 1 in the CSF is similar to the spatial mode 2 in the DMF despite the velocity magnitude difference. Combining the temporal coefficients (figure \ref{fig:mode2_T_PIV} and figure \ref{fig:mode1_T_conv}) and these two spatial modes, they imply that both of these two modes from CSF and DMF represent the reduction of the reverse flow within in the separation region. However, figure \ref{fig:mode2_S_PIV} and figure \ref{fig:mode1_S_conv} exhibit that the velocity reduction magnitude of CSF mode 1 is larger than DMF mode 2. This contributes to the different $\Delta C_L$ increment magnitude in figure \ref{fig:DCL_multi_1p25t+} between the direct measurement and the convolution integral. The more interesting feature of these two modes is that their temporal coefficients approximately track the main trend of the $\Delta C_L$ in figure \ref{fig:DCL_multi_1p25t+} respectively. The magnitude difference in these two spatial modes and the trend difference in their temporal coefficients indicate the main trend of the $\Delta C_L$ variation of the multi-burst actuation is due to the nonlinear burst-burst interaction.  

Figure \ref{fig:mode14_S_PIV}, figure \ref{fig:mode14_T_PIV}, figure \ref{fig:mode18_S_conv} and figure \ref{fig:mode18_T_conv} show that both mode 18 in CSF and mode 14 in DMF are responsible for reverse flow reduction within the separation region, although the correlation between these two modes is relative weak (Table \ref{table:DMD_multi_S} and Table \ref{table:DMD_multi_T}). 


Nevertheless, the most interesting modes are mode 18 from the DMF and mode 8 from the CSF. There is a strong correlation between these two modes (Table \ref{table:DMD_multi_S} and table \ref{table:DMD_multi_T}). The oscillation frequencies of these two modes are $F^+\approx0.82$ (figure \ref{fig:mode18_T_PIV} and figure \ref{fig:mode8_T_conv}), which is the same as the burst frequency. More importantly, figure \ref{fig:mode18_S_PIV} and Fig \ref{fig:mode8_S_conv} exhibit that these two modes are responsible for the actuation induced vortex evolution. This is important evidence that the CSF is able to capture the vortex evolution in DMF, which is in response to the high-frequency ($F^+=0.82$) flowfield/$\Delta C_L$ oscillation. Hence, convolution or linear superposition of the single-burst actuation response is capable of tracking the high-frequency (burst frequency) component of the multi-burst actuation. Moreover, as it was previously mentioned, the nonlinear burst-burst interaction (mode 2 of DMF and mode 1 of CSF) has a strong connection to the main trend of the $\Delta C_L$ curve rather than the high-frequency oscillation. These important features provide us a guideline for the modeling of  $\Delta C_L$ variation related to the multi-burst or the continuous burst actuation in future research.

\begin{table}
\centering 
\begin{tabular}{ | m{5em} | m{4em}| m{4em}| m{4em}| m{4em}| }            

DMF \textbackslash CSF  & Mode 0 & Mode 1 & Mode 8 & Mode 18 \\ [0.5ex] 
Mode 0 & 0.7766 &0.1105 &0.0543 & 0.2291 \\              

Mode 2 & 0.5223& 0.7192 & 0.0425 & 0.3462  \\ 

Mode 14 & 0.2883& 0.4415 & 0.1557 & 0.1588  \\

Mode 18 & 0.0114& 0.0382 & 0.6656 & 0.0371  \\

\end{tabular}

\caption{Correlation coefficients between DMF and CSF, the spatial modes from DMD}   
\label{table:DMD_multi_S}                
\end{table} 

\begin{table}

\centering                          
\begin{tabular}{ | m{5em} | m{4em}| m{4em}| m{4em}| m{4em}| }            
 DMF \textbackslash CSF & Mode 0 & Mode 1 & Mode 8 & Mode 18 \\ [0.5ex] 
Mode 0 & 1 &0.9545 &0.0190 & 0.1798 \\              

Mode 2 & 0.4804& 0.6711 & 0.0369 & 0.7758  \\ 

Mode 14 & 0.3773& 0.6011 & 0.0209 & 0.5650  \\

Mode 18 & 0.0383& 0.0328 & 0.9451 & 0.0259  \\

\end{tabular}

\caption{Correlation coefficients between DMF and CSF, the temporal coefficients of the modes from DMD}   
\label{table:DMD_multi_T}                
\end{table} 
\FloatBarrier

\FloatBarrier

\FloatBarrier  

\section{Conclusion}\label{Sec:conc} In this work, the naturally separated flow over an NACA-0009 wing responding to single-burst and multi-burst actuation was investigated. Both the time-evolving flow structure, lift, pitching moment, and their dynamic characteristics were studied. The work is applicable to actuator design, actuation parameter selection and low-order modeling of the time-varying lift for flow control applications.     

When the separated flow is perturbed by a single-burst actuation, the streamlines of the flowfield demonstrate the reattachment process corresponding to the lift coefficient variation. There is a downwash impacting on the suction side of the wing that moves with the clockwise rotating large-scale leading edge vortex. This results in a pressure increase, and leads to an initial lift reversal before the DLEV convects into the wake. The maximum lift reversal is observed at $1.4t^+$ after initiation of the single-burst, and the maximum lift increment occurs at $2.8t^+$, when the flow is reattached. The POD analysis shows that most of the disturbed kinetic energy is stored in mode 1 and mode 2. The combined temporal coefficients of POD mode 1 and mode 2 track the negative lift coefficient curve, especially during the lift reversal. Unlike the lift coefficient curve, there is no obvious positive increment observed in the pitching moment coefficient, and it is closely tracked by the temporal coefficient of the POD mode 2. Stability analysis on the kinetic energy density field following the single-burst actuation shows that the DMD modes at $F^+\approx0.36$ are responsible for the kinetic energy growth within the measurement window.  

The investigations of the multi-burst actuation showed that the maximum mean lift gain occurs at the burst frequency $F^+\approx0.58$, which is close to the first subharmonic of the separation bubble frequency ($F^+\approx 1$), but differs from the growth DMD mode of the single-burst actuation ($F^+\approx0.36$). This is important evidence that the maximum lift gain is determined by a combination of single-pulse triggered instability and the burst-burst interaction. The mean $\Delta C_L$ dependency on the burst frequency is nonlinear process, while the RMS or the high oscillation amplitude (corresponding to the firing of the bursts) of $\Delta C_L$ dependency on the burst frequency can be modelled with a linear convolution method within the frequency range that has been investigated.     

The convolution integral on the multi-burst actuation flowfield was performed to investigate the interactions occurring between the bursts. The comparison of DMD modes between DMF and CSF indicates that the nonlinear burst-burst interaction has a strong influence on the main trend of the time-varying flowfield/lift. On the other hand, the high-frequency oscillations associated with the burst frequency strongly depend on the linear burst-burst interaction. These important features provide a guideline for the modeling of  $\Delta C_L$ variation related to the multi-burst or the continuous burst actuation in future research.

\bibliographystyle{jfm}
\bibliography{jfmbib}

\end{document}